\documentclass[prb,a4paper,twocolumn]{revtex4}
\usepackage{amssymb}
\usepackage{t1enc}
\usepackage{amsmath}
\usepackage{amsmath}
\usepackage{epsfig}
\usepackage{amsfonts}
\usepackage{amssymb}

\begin{document}
%
%
%
%
%

\providecommand{\eq}{\begin{eqnarray}}

\providecommand{\qe}{\end{eqnarray}}

\providecommand{\e}{\varepsilon}

\providecommand{\h}{\hslash}

\providecommand{\p}{\partial}

\providecommand{\s}{\textrm{sign}}

\providecommand{\Q}{\widetilde{Q}}

\providecommand{\muTr}{\mu_{\textrm{Tr}}}

\providecommand{\al}{\alpha_{\textsf{l}}}

\providecommand{\aq}{\alpha_{\textsf{qu}}}

\providecommand{\F}{F_i(q,\omega)}

\providecommand{\e}{\varepsilon}

\providecommand{\eF}{\varepsilon_{\textrm{F}}^{}}

\providecommand{\kf}{k_{\textrm{F}}^{}}

\providecommand{\qTF}{q_{\textrm{TF}}^{}}

\providecommand{\s}{\textrm{sign}}

\newcommand{\dd}{\textrm{d}} 

\newcommand{\kb}{k_{\textrm{B}}^{}}

\newcommand{\om}{\omega}

\newcommand{\TF}{T_{\textrm{F}}}

\newcommand{\acc}{a_{\textsc{c}^{\rule{1mm}{0.2mm}}\textsc{c}}}

\newcommand{\at}{\tilde{a}}

\newcommand{\N}{\mathfrak{n}}

\newcommand{\m}{\mathfrak{m}}

\newcommand{\LL}{\mathfrak{L}}

\newcommand{\kk}{\mathfrak{K}}

\newcommand{\J}{\mathcal{J}}

\title{Intershell resistance in multiwall carbon nanotubes: A Coulomb drag study}
\author{Anders Mathias Lunde$^{\dag,\ddag,\ast}$, Karsten Flensberg$^{\dag}$ and Antti-Pekka Jauho$^\ddag$}
\affiliation{$^{\dag}${\O}rsted Laboratory, Niels Bohr Institute,
University of Copenhagen, DK-2100 Copenhagen,
Denmark\\
$^{\ddag}$MIC-Department of Micro and Nanotechnology, Technical
University of Denmark, {\O}rsteds Plads, Bldg.~345~east, DK-2800
Kgs.~Lyngby, Denmark}

\date{\today}
\pacs{}

\begin{abstract}

We calculate the intershell resistance $R_{21}$ in a multiwall
carbon na\-no\-tu\-be as a function of temperature $T$ and Fermi
level $\eF$ (e.g.~a gate voltage), varying the chirality of the
inner and outer tubes. This is done in a so-called Coulomb drag
setup, where a current $I_1$ in one shell induces a voltage drop
$V_2$ in another shell by the screened Coulomb interaction between
the shells neglecting the intershell tunnelling. We provide
benchmark results for $R_{21}=V_2/I_1$ within the Fermi liquid
theory using Boltzmann equations. The band structure gives rise to
strongly chirality dependent suppression effects for the Coulomb
drag between different tubes due to selection rules combined with
mismatching of wave vector and crystal angular momentum
conservation near the Fermi level. This gives rise to orders of
magnitude changes in $R_{21}$ and even the sign of $R_{21}$ can
change depending on the chirality of the inner and outer tube and
misalignment of inner and outer tube Fermi levels. However for
\emph{any} tube combination, we predict a dip (or peak) in
$R_{21}$ as a function of gate voltage, since $R_{21}$ vanishes at
the electron-hole symmetry point. As a byproduct, we classified
\emph{all} metallic tubes into either zigzag-like or
armchair-like, which have two different non-zero crystal angular
momenta $\m_a$, $\m_b$ and only zero angular momentum,
respectively.


\end{abstract}

\maketitle

\section{Introduction}
\subsection{General considerations on nanotubes}

Carbon nanotubes are widely recognized as being among the most
promising materials for future na\-no\-tech\-no\-lo\-gy
applications. Furthermore, they are of fundamental scientific
interest due to several unique electronic, mechanical and thermal
properties.\cite{Reich-et-al-Book-CNT-2003} These properties often
depend on the microscopic details of their composition, e.g.~the
way the graphene sheets are rolled into tubes and whether one has
a single or multiwall carbon nanotube or a rope or bundle of
these.
Electrical transport measurements have shown a tendency for
ballistic transport in individual singlewall carbon
nanotubes\cite{Bachtold-ballistic-SWNT-diffusiv-MWNT-PRL-2000,Liang-ballistic-experiment-SWNT-nature-2001,diffusive-MWCNT-ballistic-SWCNT-experiment-McEuen-PRL-2000}
(SWCNT) and diffusive transport in multiwall carbon
nanotubes\cite{Bachtold-Aharonov-Bohm-nanotubes-nature-1999,gatevoltage-doping-schonenberger-condmat-2001,Martel-diffusiv-experiment-SWNT-APL-1998,diffusive-MWCNT-ballistic-SWCNT-experiment-McEuen-PRL-2000}
(MWCNT), but this issue is not completely settled
yet\cite{frank-Ballistic-MWNT-1998} and seems to depend crucially
on the contacts to the tubes and the amount of defects and
impurities in and near the tube.
%
Many
experiments\cite{Coulomb-blockade-in-CNT-science-Bockrath-1997,Coulomb-blockade-in-CNT-nature-Tans-1997,Coulomb-blockade-in-CNT-nygaard-PRL-2002,Kondo-in-CNT-nygaard-nature-2000,electron-hole-symmetry-in-semicond-CNT-eksperiment-Herrero-nature-2004,review-transport-bl-a-Coulomb-blockade-nygaard-1999,pseudospin-Mceuen-PRL-1999}
have explored the Coulomb blockade regime, where the tube can be
treated as a quantum dot, due to poor electric contact. More
recently, better electrical contacts have been
achieved,\cite{Liang-ballistic-experiment-SWNT-nature-2001,Ballistic-4ee/h-good-contacts-Kong-PRL-2001,Ballistic-CNT-good-contacts-Pd-Javey-nature-2003,Ballistic-CNT-good-contacts-Pd-Mann-nanolett-2003}
which gives larger conductance, approaching the predicted
$\frac{4e^2}{h}$, and a coherent (or Landauer-B\"{u}ttiker-like)
regime is thereby reached. Palladium seems to be a promising
candidate for good future ohmic
contacts.\cite{Ballistic-CNT-good-contacts-Pd-Javey-nature-2003,Ballistic-CNT-good-contacts-Pd-Mann-nanolett-2003}
Another interesting feature of carbon nanotubes is their
one-dimensional nature, which may have profound consequences on
the basic physical phenomenology for their description:
SWCNT's have been predicted to be Luttinger
liquids\cite{Egger-luttinger-liquids-teori-SWCNT-prl-1997,Egger-Luttinger-liquid-SWCNT-lang-udgave-Euro-phys-Jour-B-1998}
and some experimental evidence
exists\cite{Luttinger-liquids-experiment-nature-mceuen-1999,Luttinger-liquids-experiment-in-CNT-nature-ishil-2003}
even though other interpretations have been
suggested.\cite{Luttinger-liquid-effects-could-be-Coulomb-blockade-effect-superledning-Kasumov-PRB-2003}
Whether MWCNT's are Fermi or Luttinger Liquids has been
investigated extensively
experimentally\cite{luttinger-liquids-kontra-fermi-liquids-MWNT-Tarkiainen-PRB,luttinger-liquids-absent-in-doped-MWNT-PRB-Krstic-2003,luttinger-liquids-kontra-fermi-liquids-MWNT-PRB-Xie}
and theoretically\cite{Egger-Luttinger-liquids-in-MWNT-PRL-1999}
and seems to depend on the situation, but is still subject to
debate. Also in ropes the situation is not clear
yet.\cite{Luttinger-liquid-and-Contact-barriers-Hunger-PRB-2004}

The structure of this paper is as follows. We begin by introducing
the intershell resistance problem in MWCNT's and our approach to
it in section
\ref{sec:Introduction-to-intershell-resistance-in-MWCNT}. In
section \ref{sec:qualitative-features} we review the basic
qualitative features of our theory of the intershell resistance
using a Coulomb drag setup. Sections
\ref{sec:energy-bands-of-CNT-main-text} and \ref{sec:Coulomb-int}
are devoted to a summary of the band structure and a calculation
of the screened Coulomb matrix element including the important
suppression rules for backscattering in metallic tubes, and in
section \ref{sec:transresistivity-model} we indicate how the
standard transresistance formulae are modified in the nanotube
configuration. Sections
\ref{sec:Electron-hole-symmetry-and-Coulomb-drag},
\ref{sec:Coulomb-drag-between-metallic-tubes} and
\ref{sec:Comment-on-the-drag-between-semiconducting-tubes} give
our results for several different nanotube combinations. Details
of the nanotube band structure and the screening model including
the band structure are found in the appendices.

\subsection{Intershell resistance in
MWCNT's}\label{sec:Introduction-to-intershell-resistance-in-MWCNT}
\begin{center}
\begin{figure}
\epsfig{file=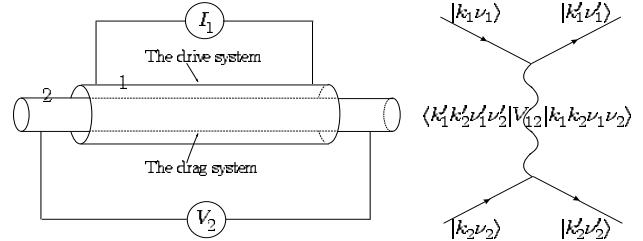,width=0.46\textwidth}
\caption{\footnotesize (Left): The experimental setup to directly
measure the Coulomb drag effect in a MWCNT. The intershell
resistance is $R_{21}=V_2/I_1$. (Right): The basic mechanism in
the intershell resistance in a drag configuration: the intershell
$e\!-\!e$ interaction and thereby momentum transfer to induce the
voltage drop $V_2$.} \label{fig:drag-setup-med-tubes}
\end{figure}
\end{center}
%
Let us now consider electron interaction and transport in the
concentric tubes (or shells) in a MWCNT. Yoon~\emph{et
al.}\cite{no-intershell-tunnelling-MWCNT-theory-Yoon-PRB-2002}~have
argued theoretically that the intershell tunnelling of electrons
is vanishingly small between both commensurable and
incommensurable long defect-free MWCNT's.
Lack of intertube tunnelling is also expected in nanotube
ropes.\cite{Maarouf-small-tunnelling-between-tubes-PRB-2000}
Furthermore, Aharonov-Bohm
experiments\cite{Bachtold-Aharonov-Bohm-nanotubes-nature-1999}
indicate that current only flows in the outer tube in a MWCNT.
Another experiment by Collins \emph{et
al.}\cite{collins-shell-remove-science-2001}~supports this picture
and finds no leakage between the shells in the low-bias limit.
This is concluded by removing the shells in a MWCNT one by one and
measuring the gate voltage response of the remaining MWCNT after
each shell removal. Other shell removing experiments has also been
reported.\cite{cumings-shell-remove-nature-2000,Dohn-shell-remove-masterthesis-2003,Colling-shell-remove-Appl-Phys-2002,Li-shell-remove-China-2004}
Furthermore, Cumings \emph{et
al.}\cite{cuming-oscillator-in-MWCNT-eksperiment-2000-science}~have
demonstrated relative motion between the inner and outer shells in
a MWCNT indicating that the shell are weakly coupled by van der
Waals forces. In addition to Yoon~\emph{et
al.}\cite{no-intershell-tunnelling-MWCNT-theory-Yoon-PRB-2002}
also other theoretical papers have calculated the intershell
resistance
using tunnelling as the \emph{only}
mechanism.\cite{Roche-intershell-tunnelling-PRB-2001,Roche-intershell-tunnelling-Phys-lett-A-2001,Intershell-conductance-MWCNT-Roche-PRB-2004,Intershell-tunneling-MWCNT-Ahn-Physica-E-2004,hansson-intershell-resistance-PRB-2003,sanvito-intershell-resistance-PRL-2003,Tomanek-intershell-resistance-PRB-2002,Uryu-matrix-elementer-mellem-tubes-PRB-2004,Kim-telescoping-MWCNT-teori-PRB-2002}
For example Roche~\emph{et
al.}\cite{Roche-intershell-tunnelling-PRB-2001,Roche-intershell-tunnelling-Phys-lett-A-2001,Intershell-conductance-MWCNT-Roche-PRB-2004}~have
considered the time evolution of a wave packet initially on the
outer tube in a disorder-free MWCNT including tunnelling in a
tight-binding approximation. This is not in contrast to
[\onlinecite{no-intershell-tunnelling-MWCNT-theory-Yoon-PRB-2002}]
due to the localization of the wave packet of Refs.
[\onlinecite{Roche-intershell-tunnelling-PRB-2001,Roche-intershell-tunnelling-Phys-lett-A-2001,Intershell-conductance-MWCNT-Roche-PRB-2004}]\cite{no-intershell-tunnelling-MWCNT-theory-Yoon-PRB-2002}.
Using density functional theory (DFT), Hansson \emph{et
al.}\cite{hansson-intershell-resistance-PRB-2003}~consider
concentric armchair tubes and find no essential change in the
conductance steps for a ballistic MWCNT, when the intershell
tunnelling is turned on and off.
Ref.~[\onlinecite{sanvito-intershell-resistance-PRL-2003,Tomanek-intershell-resistance-PRB-2002}]
also model intershell tunnelling by DFT. Very recently,
experiments with a MWCNT with 11 contacts on the outer tube, where
a current is driven though some of the tube and a voltage drop is
measured elsewhere on the tube, have been
published.\cite{Bacdtold-intershell-resistance-measurement-condmat-2004}
Using a transmission line model, information about the intershell
conductance is deduced.

In the present paper, we approach the intershell resistance in a
MWCNT from a different point of view: \emph{We consider the
intershell resistance $R_{21}$ from the electron-electron
($e\!-\!e$) interaction between the shells neglecting tunnelling,
i.e.~in a Coulomb drag configuration.} In general, Coulomb
drag\cite{Pogrebinskii-original-drag-forslag-1977,Price-original-drag-forslag-1983}
means that moving charges in one subsystem (the drive subsystem)
exchange momentum (and other quantum numbers) with carriers in a
nearby subsystem (the probe or drag subsystem) thus exerting a
drag force on the probe, inducing a current, or a voltage, in the
probe (see Fig.~\ref{fig:drag-setup-med-tubes}). Here the
intershell or transresistance $R_{21}=V_2/I_1$ is found as a
function of gate voltage (i.e.~Fermi level $\eF$) and temperature
$T$, varying the chirality of the inner and outer tubes. Once the
chiralities of the tubes are chosen, our theory has no remaining
free parameters.
Coulomb drag is a unique transport measurement in the sense that
the $R_{21}$ is dominated by the intershell Coulomb
interaction.\cite{footnote-other-interactions-than-Coluomb}
%
%
Therefore serious attention to the intershell Coulomb matrix
element and the use of proper Bloch states of the individual tubes
is necessary. As will be seen below, the effects of including the
band structure (and the underlying symmetries of the constituent
nanotubes) are absolutely crucial, leading to orders-of-magnitude
changes in the intershell resistance, occasionally also reversing
its sign.
Furthermore, the present work also gives a new source of
friction against relative motion of concentric tubes, which could
be considered in the context of using MWCNT as GHz nano-mechanical
oscillators.\cite{MWCNT-as-GHz-oscillatorer-PRL-2002}

A direct measurement of the intershell resistance in a Coulomb
drag setup (ig.~\ref{fig:drag-setup-med-tubes}) requires
independent contacts on an inner and an outer tube, a difficult
but possible technological achievement\cite{private-comm-nygard}
in the light of the resent shell removal
experiments.\cite{collins-shell-remove-science-2001,cumings-shell-remove-nature-2000,Dohn-shell-remove-masterthesis-2003,Colling-shell-remove-Appl-Phys-2002,Li-shell-remove-China-2004}
As a model, we consider two shells, but our considerations can be
extended for many shells. Also, a direct growth of double wall
tubes seems feasible.\cite{Sugai-gro-DWCNT-NL-2003}

Coulomb drag has been an extremely successful tool in studying
interactions in coupled quantum
wells\cite{Rojo-review-af-drag-mellem-2DEG-1999,Gramila-first-experiment-drag-mellem-2DEG-PRL-1991,Antti-smith-original-drag-PRB-93,karsten-ben-PRL-plasmon-1994,Hill-Flensberg-palsmon-peak-i-drag-PRL-1997,Boensager-phonon-drag-PRB-1998}
(notably in the quantum Hall
regimes\cite{Rojo-review-af-drag-mellem-2DEG-1999,Gornyi-drag-i-lave-B-felter-condmat-2004}),
and indeed it was realized very early that Coulomb drag between
Luttinger liquids would be an important object to
study\cite{crossed-nanotubbes-egger-PRL-1998,crossed-luttinger-liquids-Flensberg-PRL-1998,Luttinger-drag-Nazarov-PRL-1998,Luttinger-drag-Ponomarenko-PRL-2000,Luttinger-drag-Klesse--PRB-2000,crossed-nanotubbes-egger-EPJB-2000,Luttinger-drag-Trauzettel-RPL-2002}
These studies focused on Coulomb drag on either crossed or
adjacent subsystems, and used very simple models for the Coulomb
interaction. Several interesting theoretical predictions emerged
from these papers, some of which may have been confirmed
experimentally.\cite{crossed-SWCNT-Luttinger-liquid-experiment-Egger-Bachtold-PRL-2004}
%
%
We work in the Fermi-liquid framework using Boltzmann equations.
We think that it is important to establish a clear picture of what
one expects within this simple model before turning to strongly
interacting theories. Note that our approach also gives valuable
information about drag between parallel tubes.

\subsection{Nanotube Coulomb drag - qualitative features of the
theory}\label{sec:qualitative-features}

As explained in detail in subsequent sections, the
trans\-re\-sis\-tan\-ce or intershell resistance $R_{21}$ is
computed from the expression
\begin{equation}
R_{21}\propto\int\sum (SR)^2|V_{12}|^2
\mathcal{A}(T)F^{(1)}F^{(2)},\label{eq:drag-qualitative}
\end{equation}
where the integration is taken over transferred momentum and
energy in the intershell interaction, and the summation includes
all involved bands and other quantum numbers required to specify
the states. $\mathcal{A}(T)$ is a thermal factor, $V_{12}$ is the
screened intershell Coulomb interaction, and the $F$-functions for
the two subsystems account for the available phase-space for
electronic scattering. Of crucial importance is the factor $SR$
accounting for the selection rules (or rather suppression rules)
stemming from the intershell Coulomb matrix element between the
Bloch states. (In the final formula some $SR$ is incorporated into
the $F$-functions). As known from
experimental\cite{pseudospin-Mceuen-PRL-1999} and
theoretical\cite{backscattering-Ando-JPJ-1998-1,backscattering-Ando-JPJ-1998-2,Klesse-PRB-2002-Coulomb-matrix-element-MWNT}
studies, backscattering between the linear bands in metallic tubes
by impurities with slowly varying potentials are strongly
suppressed leading to very long mean free paths. The selection
rules for intershell Coulomb interaction lead to a similar
suppression, which depends strongly on the inner and outer tubes'
chirality. A detailed analysis of these effects is one of the
central tasks of the present article. The structure of
Eq.~(\ref{eq:drag-qualitative}) is much richer than its
counterparts' for coupled quantum wells due to the rather
complicated band structure combinations of the various MWCNT's.

\section{Carbon nanotube band structure}\label{sec:energy-bands-of-CNT-main-text}

In appendix \ref{sec:energy-bands-of-CNT-Appendix}, we give a
detailed account for the band structure of a SWCNT with chirality
$(n,m)$, since it turns out to be of crucial importance to the
intershell Coulomb matrix element and thereby also for the drag.
Here we only outline the important points of the band structure
used later.

The carbon nanotube band structure can be found by applying
periodic boundary conditions to the band structure of a single
graphite layer (graphene). Graphene has two atoms in the primitive
unit cell, so the tight-binding state (or Wannier decomposition)
have two components with weights $\alpha$ and $\beta$ (see
Eq.~(\ref{eq:Bloch-state-graphene})). When applying the periodic
boundary condition the wave vector component around the tube $k_c$
becomes quantized into a discrete values,
$k_c=\frac{2\pi}{|\mathbf{C}|}n_c$. However, it is important to
realize that $n_c$ is \emph{not} the crystal angular momentum $\m$
stemming from the rotation symmetry, but only related to it by
$n_c=\m$ (mod $\N$). (Here $\mathbf{C}$ is the chiral vector and
$\N$ is the greatest common divisor of $(n,m)$, $\N=\gcd(n,m)$).
This is due to the non-primitive (large) nanotube unit cell, when
using translational symmetry instead of helical
symmetry.\cite{White-1993-RPB,Mintmire-1995-carbon}

\begin{center}
\begin{figure}
\epsfig{file=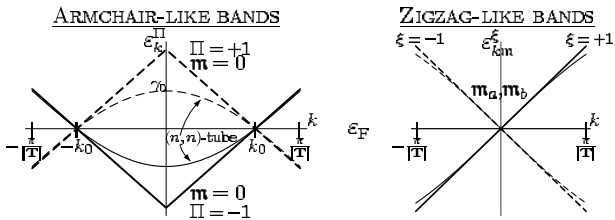,width=0.45\textwidth}
\caption{\footnotesize The two categories of metallic nanotubes:
Armchair-like (AL, left) and zigzag-like (ZL, right). The AL bands
near $\eF=0$ have zero crystal angular momentum $\m=0$ and
$\Pi=\pm1$, where $k_0\equiv\frac{2\pi}{3|\mathbf{T}|}$. The ZL
tubes have doubly degenerate bands crossing $\eF=0$, i.e.~for each
$\xi=\pm1$ we have either $\m_a=\frac{2n+m}{3}\ (\textrm{mod}\
\N)$ or $\m_b=\frac{2m+n}{3}\ (\textrm{mod}\ \N)$, where $\m_a\neq
\m_b$, ($\N=\gcd(n,m)$). The thin lines are the
tight-binding bands near $\eF=0$ for a $(n,n)$ tube (with
$|\mathbf{T}|=a$) and a $(3m,0)$ tube.}
\label{fig:two-types-of-tubes}
\end{figure}
\end{center}

Linearizing the tight-binding band structure around the Fermi
level $\eF=0$ the states and bands for metallic tubes near $\eF$
become
\begin{align}
\e_{\kk_T^{}}^{\xi} &=\xi\h v_0 \kk_T \quad \textrm{and}
\label{eq:general-metallic-nanotube-bands}
\\
\left(\!\!\begin{array}{c} \alpha \\ \beta
\end{array}\!\!\right)_{\xi,\varsigma}&=
\frac{1}{\sqrt{2}} \left(\!\begin{array}{c} -\xi
\frac{i(m-n)-\varsigma\sqrt{3}(n+m)}{2\sqrt{n^2+m^2+mn}}
\\ 1
\end{array}\!\right),
\label{eq:general-metallic-nanotube-states}
\end{align}
where $\kk_T$ is the wave vector along the tube \emph{measured}
from the point, where the band crosses
$\eF=0$,($\kk_T=k-k_{\eF=0}$), $\xi=\pm1$ is the sign of the
velocity in the band and $\varsigma=\pm1$ describes which
$\mathbf{K}_{\varsigma}$ point of graphene the linear band
originate from. Here $v_0=\frac{\sqrt{3}\gamma_0 a}{2\h}$ with
$\gamma_0\simeq 3$eV and $a\equiv\sqrt{3}\acc$ ($\acc=0.142$nm).
The metallic states can thus described by $(k,\xi,\varsigma)$.
Using this, we can \emph{classify all metallic tubes into two
categories: zigzag-like (ZL) and armchair-like (AL) tubes}, with
the following bands near the Fermi level (shown on
Fig.~\ref{fig:two-types-of-tubes}):
\begin{align}
\textrm{Zigzag-like:}& \quad \e_{k\m}^{\xi}=\xi \h v_0 k, \quad
\m\in \{\m_a,
\m_b\} \label{eq:zigzag-like-bands} \\
\textrm{Armchair-like:}&  \quad \e_{k}^{\Pi}=-\Pi \h v_0
(|k|-k_0), \ \ \textrm{($\m=0$)}. \label{eq:armchair-like-bands}
\end{align}
Here $k\in]-\frac{\pi}{|\mathbf{T}|},\frac{\pi}{|\mathbf{T}|}]$ is
the wave vector along the tube, $k_0=\frac{2\pi}{3|\mathbf{T}|}$
and $\mathbf{T}$ is the translational vector generating the
translational symmetry. Note that two different tubes can have
different $|\mathbf{T}|$ even though they belong to the same
category. $\Pi=\pm1$ originates from
$\xi$, but does not give the sign of the velocity, 
and for a $(n,n)$ tube $\Pi$ is the parity in the cylindrical
angle.\cite{Vukovic-2002-PRB-bloch-states,Damnjanovic-1999-symmetri-og-potentialer-NT}
%
%
The linearity of the bands near the Fermi level is, of course,
well known, but it is important to recognize the entirely
different angular momentum quantum numbers $\m$ that characterize
the AL and ZL bands crossing the Fermi level.  Specifically, for
AL tubes it always holds that $\m=0$, while for the ZL tube one
has $\m_a=\frac{2n+m}{3}\ (\textrm{mod}\ \N)$ or
$\m_b=\frac{2m+n}{3}\ (\textrm{mod}\ \N)$ (note that $\m_a\neq
\m_b$ and $\m_a$, $\m_b$ are never zero).
There is a one-to-one correspondence between $\varsigma=\pm1$ and
the crystal angular momentum of the linear bands.
We note that the most commonly studied metallic zigzag and
armchair tubes, with indices $(3n,0)$ and $(n,n)$, are of course
special cases of ZL and AL tubes, respectively.

\section{Intershell Coulomb interaction}\label{sec:Coulomb-int}

We next consider the Coulomb interaction
between Bloch states $|kn_c \rangle$ 
for electrons in different shells in a MWCNT. Before calculating
the Coulomb matrix element involving
products\cite{footnote-exchange-interaction}
of Bloch states it is useful to consider the less complicated
problem of the impurity matrix element $\langle
k^{\prime}_{}n^{\prime}_{c}|V(\mathbf{r})|k n_c\rangle$. The
essential assumption that we use in calculating both the impurity
and Coulomb matrix element is that \emph{the potential is slowly
varying on the scale of the interatomic distance $\acc$}. In the
case of impurity scattering this is a fair assumption for an
impurity held on the tube by Van der Waals forces as is often the
case.\cite{doping-lee-PRB-2000,doping-zhou-sceince-2000} For
Coulomb interaction between different shells it is also a good
assumption, since the electrons do not get close enough to
experience the $1/r$ singularity.

The impurity matrix element $\langle
\mathbf{k}^{\prime}|V(\mathbf{r})|\mathbf{k}\rangle$ between the
two component Bloch states $\psi_{\mathbf{k}}(\mathbf{r})$
Eq.~(\ref{eq:Bloch-state-graphene}) (before applying periodic
boundary conditions) is
\begin{align}
& \langle\mathbf{k}^{\prime}| V(\mathbf{r})|\mathbf{k}\rangle
=
\frac{1}{N} \sum_{\mathbf{R},\mathbf{R}^{\prime}}
e^{-i\mathbf{k}^{\prime}\cdot\mathbf{R}^{\prime}+i\mathbf{k}\cdot\mathbf{R}}\nonumber\\
& \times \Bigg[
\alpha^{\ast}_{\mathbf{k}^{\prime}} \alpha_{\mathbf{k}}  \int \!\!
\dd \mathbf{r}
\Psi^{\ast}(\mathbf{r}-\mathbf{R}^{\prime})V(\mathbf{r})
\Psi(\mathbf{r}-\mathbf{R})  \nonumber \\
%
&+\alpha^{\ast}_{\mathbf{k}^{\prime}}\beta_{\mathbf{k}} 
 \int \!\! \dd \mathbf{r}
\Psi^{\ast}(\mathbf{r}-\mathbf{R}^{\prime})V(\mathbf{r})
\Psi(\mathbf{r}-\mathbf{R}-\mathbf{d})  \nonumber \\
%
&+\beta^{\ast}_{\mathbf{k}^{\prime}}\alpha_{\mathbf{k}} 
\int \!\! \dd \mathbf{r}
\Psi^{\ast}(\mathbf{r}-\mathbf{R}^{\prime}-\mathbf{d})V(\mathbf{r})
\Psi(\mathbf{r}-\mathbf{R})  \nonumber \\
&+\beta^{\ast}_{\mathbf{k}^{\prime}}\beta_{\mathbf{k}}
 \int \!\! \dd \mathbf{r}
\Psi^{\ast}(\mathbf{r}-\mathbf{R}^{\prime}-\mathbf{d})V(\mathbf{r})
\Psi(\mathbf{r}-\mathbf{R}-\mathbf{d}) \Bigg].
\end{align}
By using the assumption of slow variation of $V(\mathbf{r})$ we
can take the potential outside the integrals. The first and last
term in the square bracket become
$\delta_{\mathbf{R}^{\prime},\mathbf{R}}V(\mathbf{R})(\alpha^{\ast}_{\mathbf{k}^{\prime}}
\alpha_{\mathbf{k}}+\beta^{\ast}_{\mathbf{k}^{\prime}}\beta_{\mathbf{k}})$
and the second and third term are found (including a sum) by
summing over the nearest neighbors to be $\propto
s_0\Big(\alpha^{\ast}_{\mathbf{k}^{\prime}}\beta_{\mathbf{k}}
\Upsilon(\mathbf{k}^{\prime})+
\alpha_{\mathbf{k}}\beta^{\ast}_{\mathbf{k}^{\prime}}
\Upsilon^{\ast}(\mathbf{k}) \Big)$.
Eq.~(\ref{eq:H-AB-matrix-element}) defines $\Upsilon(\mathbf{k})$.
Introducing the Fourier transform of the potential $V(\mathbf{k})$
and the reciprocal lattice vector $\mathbf{G}$ we find:
\begin{align}
\langle\mathbf{k}^{\prime}|
V(\mathbf{r})|\mathbf{k}\rangle 
&= g(\mathbf{k},\mathbf{k}^{\prime}) \frac{1}{\mathcal{A}}
\sum_{\mathbf{G}}V(\mathbf{k}^{\prime}-\mathbf{k}+\mathbf{G})
\label{eq:scattering-matrix-element-graphene},
\end{align}
where $\mathcal{A}$ is the surface area and the $g$-factor is
\begin{align}
\hspace{-3mm}g(\mathbf{k},\mathbf{k}^{\prime})&\equiv
\alpha_{\mathbf{k}}\alpha^{\ast}_{\mathbf{k}^{\prime}}
+\beta_{\mathbf{k}}\beta_{\mathbf{k}^{\prime}}^{\ast}\nonumber\\
&\quad
+s_0\Big(\alpha_{\mathbf{k}^{\prime}}^{\ast}\beta_{\mathbf{k}}
\Upsilon(\mathbf{k}^{\prime})+
\alpha_{\mathbf{k}}\beta_{\mathbf{k}^{\prime}}^{\ast}
\Upsilon^{\ast}(\mathbf{k}) \Big), \label{eq:g-factor-graphene}
\end{align}
i.e.~the matrix element is essentially the plane wave result
\emph{times a band structure factor, which we will refer to as the
$g$-factor}.

To obtain the matrix element for the screened Coulomb interaction
$V(\mathbf{r}_1,\mathbf{r}_2)$ (suppressing the frequency argument
$\omega$ in the notation) we note that
$\langle \mathbf{k}_1^{\prime}\mathbf{k}_2^{\prime}|
V(\mathbf{r}_1,\mathbf{r}_2)|\mathbf{k}_1^{}\mathbf{k}_2^{}\rangle=
\langle \mathbf{k}_2^{\prime}| \langle \mathbf{k}_1^{\prime}|
V(\mathbf{r}_1,\mathbf{r}_2)|\mathbf{k}_1^{} \rangle
|\mathbf{k}_2^{}\rangle$,
%
where $i=1,2$ labels the outer/inner tube, respectively. Therefore
we can use the impurity potential result
Eq.~(\ref{eq:scattering-matrix-element-graphene}) to obtain:
\begin{multline}
\langle \mathbf{k}_1^{\prime}\mathbf{k}_2^{\prime}|
V(\mathbf{r}_1,\mathbf{r}_2)
|\mathbf{k}_1^{}\mathbf{k}_2^{}\rangle=
g_1(\mathbf{k}_1,\mathbf{k}_1^{\prime})
g_2(\mathbf{k}_2,\mathbf{k}_2^{\prime})\\
\times \frac{1}{\mathcal{A}_1\mathcal{A}_2}
\sum_{\mathbf{G}_1,\mathbf{G}_2}
V(\mathbf{k}_1^{\prime}-\mathbf{k}_1+\mathbf{G}_1,\mathbf{k}_2^{\prime}-\mathbf{k}_2+\mathbf{G}_2),
\label{eq:Coulomb-int-matrix-element-graphene-tight-bin-screened}
\end{multline}
where we have a $g$-factor for each system and the screened
potential is Fourier transformed separately in both $\mathbf{r}_1$
and $\mathbf{r}_2$.

For a $(n_2,m_2)$ tube inside a $(n_1,m_1)$ tube the screened
Coulomb matrix element is found using cylindrical coordinates
$\mathbf{r}=(r,\theta,z)$ to be
\begin{widetext}
\begin{align}
&\langle k_1^{\prime}n_{c_1}^{\prime}k_2^{\prime}n_{c_2}^{\prime}|
V(\mathbf{r}_1,\mathbf{r}_2)|k_1^{}n_{c_1}^{}k_2^{}n_{c_2}^{}\rangle=
\frac{1}{(2\pi L)^2}
g_1(k_1^{}n_{c_1}^{},k_1^{\prime}n_{c_1}^{\prime})
g_2(k_2^{}n_{c_2}^{},k_2^{\prime}n_{c_2}^{\prime})\nonumber\\
&\hspace{3.7cm}\times \sum_{G_1,G_2} \sum_{u_1,u_2\in\mathbb{Z}}
V(k_1^{\prime}-k_1+G_1,\m_1^{\prime}-\m_1^{}+\N_1^{}u_1^{},
k_2^{\prime}-k_2+G_2,\m_2^{\prime}-\m_2^{}+\N_2^{}u_2^{},r_1,r_2),
\label{eq:Coulomb-int-matrix-element-trans-unitcell-tight-bin-screened}
%
\end{align}
\end{widetext}
where $L$ is the length of tubes, $\N_i=\gcd(n_i,m_i)$,
$G_i=\frac{2\pi}{|\mathbf{T}_i|}s$ ($s\in\mathbb{Z}$), $r_i$ is
the radius\cite{footnote-2D-til-3D} of tube $i$.
%
%

We will also need the unscreened Coulomb matrix element $V^0$,
which is a function of the interparticle distance
$|\mathbf{r}_1-\mathbf{r}_2|$, i.e.~a function of $z_1-z_2$,
$\theta_1-\theta_2$, $r_1$ and $r_2$, so we Fourier transform in
the differences $z_1-z_2$ and $\theta_1-\theta_2$. Therefore the
matrix element is:
\begin{align}
&\langle k_1^{\prime}n_{c_1}^{\prime}k_2^{\prime}n_{c_2}^{\prime}|
V^0(|\mathbf{r}_1-\mathbf{r}_2|)|k_1^{}n_{c_1}^{}k_2^{}n_{c_2}^{}\rangle=\nonumber \\
&\frac{1}{2\pi L}
g_1(k_1^{}n_{c_1}^{},k_1^{\prime}n_{c_1}^{\prime})
g_2(k_2^{}n_{c_2}^{},k_2^{\prime}n_{c_2}^{\prime})\nonumber\\
& \sum_{G_1,G_2} \sum_{u_1,u_2\in\mathbb{Z}}
V^0(k_1^{\prime}-k_1+G_1,\m_1^{\prime}-\m_1^{}+\N_1^{}u_1^{},r_1,r_2)\nonumber\\
&
\times\delta_{k_1^{}+k_2^{},k_1^{\prime}+k_2^{\prime}+G_1^{}+G_2^{}}
\delta_{\m_1^{\prime}+\m_2^{\prime}+\N_1^{}u_1^{},
\m_1^{}+\m_2^{}+\N_2^{}u_2^{}}.
\label{eq:Coulomb-int-matrix-element-trans-unitcell-tight-bin-unscreened}
\end{align}
Note that the $\pm$ in the states
Eq.~(\ref{eq:energi-mulige-near-EF}) (the $\xi$ index for metallic
states
Eq.~(\ref{eq:general-metallic-nanotube-bands}-\ref{eq:general-metallic-nanotube-states}))
is suppressed in the notation and that \emph{this index only
appears in the $g$-factors} in both
Eq.~(\ref{eq:Coulomb-int-matrix-element-trans-unitcell-tight-bin-screened})
and
Eq.~(\ref{eq:Coulomb-int-matrix-element-trans-unitcell-tight-bin-unscreened}).
Here we have used the crystal angular momentum difference in the
Fourier transforms instead of the $n_c$ difference, since this is
the physical (crystal) angular momentum being
transferred.\cite{footnote-explicit-periodisk}
%
%
%
Note that we have included Umklapp scattering and that the
unscreened interaction
Eq.~(\ref{eq:Coulomb-int-matrix-element-trans-unitcell-tight-bin-unscreened})
has crystal (angular) momentum conservation. Similar matrix
element were considered by S.
Uryu.\cite{Uryu-matrix-elementer-mellem-tubes-PRB-2004}

\subsection{The $g$-factor and backscattering in metallic
tubes}\label{subsec:g-factor}

We now consider the $g$-factors and show that they contain
essential information about the electronic scattering.
The $g$-factor for any $(n,m)$ metallic tube for the scattering
 process $(k,\xi,\varsigma)\rightarrow
(k^{\prime},\xi^{\prime},\varsigma^{\prime})$ between the metallic
states Eq.~(\ref{eq:general-metallic-nanotube-states}) is found by
inserting Eq.~(\ref{eq:linearized-upsilon}) (with
$\boldsymbol{\kk}=\kk_T\frac{\mathbf{T}}{|\mathbf{T}|}$) and
Eq.~(\ref{eq:general-metallic-nanotube-states}) into
Eq.~(\ref{eq:g-factor-graphene}):
\begin{align}
g(k,& \varsigma,\xi;k^{\prime},\varsigma^{\prime},\xi^{\prime})=
\nonumber \\
& \frac{1}{2}\left(\xi\xi^{\prime}
f_{n,m,\varsigma,\varsigma^{\prime}}+1\right)
-s_0\frac{\sqrt{3}a (\xi^{\prime}\kk_T^{\prime}+\xi\kk_T^{})}{4},
\label{eq:scattering-element-general-tube}
\end{align}
where we introduced
\begin{align}
f_{n,m,\varsigma,\varsigma^{\prime}}=\delta_{\varsigma,\varsigma^{\prime}}
&-\frac{n^2+m^2+4mn}{2(n^2+m^2+mn)}(1-\delta_{\varsigma,\varsigma^{\prime}})\nonumber \\
&+i\frac{\sqrt{3}\varsigma(m^2-n^2)}{2(n^2+m^2+mn)}(1-\delta_{\varsigma,\varsigma^{\prime}}).
\label{eq:def-af-f-afh-af-tube-index-generelt}
\end{align}
The $g$-factor in Eq.~(\ref{eq:scattering-element-general-tube})
has two terms: The first parentheses is the important wave vector
independent scalar-product of $\tiny{\left(\!\! \begin{array}{c} \alpha \\
\beta
\end{array}\!\! \right)}$ from
Eq.~(\ref{eq:general-metallic-nanotube-states}) and the second
term is a wave vector dependent correction term (of first order in
$s_0\sim 0.1$).

As we shall show in section \ref{sec:transresistivity-model}, only
backscattering contributes to the Coulomb drag in metallic tubes
and we therefore need to consider all possible backscattering
processes ($\xi=-\xi^{\prime}$) in any metallic tube.

\begin{figure}
\begin{center}
\epsfig{file=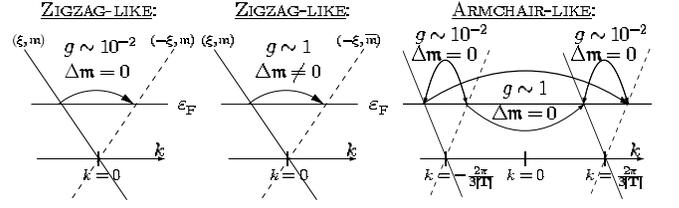,width=0.474\textwidth}
\caption{\footnotesize The possible backscattering processes in
\emph{any} metallic tube with a slightly raised Fermi level $\eF$.
(Left): Backscattering in a zigzag-like tube \emph{without}
crystal angular momentum change $\Delta\m=0$
(i.e.~$\varsigma=\varsigma^{\prime}$) and a small wave vector
$|k^{\prime}-k|\sim \frac{2\eF}{\h v_0}$ change, which is
suppressed by $g\lesssim 10^{-2}$ from
Eq.~(\ref{eq:g-i-anden-backscattering-metallic-varsigma-lig-varsigma-prime}).
(Center): Backscattering in a zigzag-like tube \emph{with} crystal
angular momentum change, which have $g\sim 1$ from
Eq.~(\ref{eq:g-for-varsigma-lig-minus-varsigma-prime-ca-1}). Here
$\overline{\m}$ denotes the opposite of $\m$ in the set
$\{\m_a,\m_b\}$.
(Right): Two types of backscattering in armchair-like tubes: $(i)$
A large wave vector transfer (for $\varsigma=-\varsigma^{\prime}$)
in between states with the same crystal angular momentum ($\m=0$)
and $g\sim 1$
(Eq.~(\ref{eq:g-for-varsigma-lig-minus-varsigma-prime-ca-1})) and
$(ii)$ A small wave vector transfer $q\simeq \frac{2\eF}{\h v_0}$
suppressed by $g\lesssim10^{-2}$. Note that the distance between
the points $\pm\frac{2\pi}{3|\mathbf{T}|}$ are not to scale
(i.e.~$\frac{2\eF}{\h v_0}\ll \frac{4\pi}{3|\mathbf{T}|}$) and
that the armchair-like bands are connected as in
Fig.~\ref{fig:two-types-of-tubes}.}
\label{fig:backscattering-metallic-tube-3-cases}
\end{center}
\end{figure}

Due to the double degeneracy of the zigzag-like bands
Eq.~(\ref{eq:zigzag-like-bands}) at the Fermi level, we must
consider backscattering both with and without crystal momentum
exchange (Fig.~\ref{fig:backscattering-metallic-tube-3-cases},
center and left panels, respectively).

If $\varsigma=\varsigma^{\prime}$ then
$\Delta\m\equiv\m^{\prime}-\m=0$ and from
Eq.~(\ref{eq:scattering-element-general-tube}) we have
\begin{align}
|g(k,\varsigma,\xi;k^{\prime},\varsigma,-\xi)|=
s_0\frac{\sqrt{3}a|k^{\prime}-k|}{4},
\label{eq:g-i-anden-backscattering-metallic-varsigma-lig-varsigma-prime}
\end{align}
which is of order $10^{-2}$ or less for scattering around the
Fermi level, i.e.~for $|k^{\prime}-k|\simeq\frac{2|\eF|}{\h v_0}$
the $g$-factor is $|g|=s_0\frac{|\eF|}{\gamma_0}\lesssim 10^{-2}$
for $|\eF|\lesssim 0.3\textrm{eV}$. If
$\varsigma=-\varsigma^{\prime}$ then $|\Delta\m|=|\m_a-\m_b|\neq0$
and for backscattering around the Fermi level the $g$-factor
squared is:\cite{Lunde-drag-in-MWCNT-masterthesis-2004}
\begin{align}
|g_{\e\sim\eF}|^2 &\simeq \frac{1}{4}
\bigg(1+\frac{n^2+m^2+4mn}{2(n^2+m^2+mn)}\bigg)^2\nonumber \\
&\hspace{8mm}+\frac{3}{16}\left(\frac{m^2-n^2}{n^2+m^2+mn}\right)^2,
\label{eq:g-for-varsigma-lig-minus-varsigma-prime-ca-1}
\end{align}
which is $\frac{3}{4}$ for $(n,0)$, $1$ for $(n,n)$ and in between
for all other tubes. \emph{So in a zigzag-like tube we have two
kinds of backscattering with small crystal wave vector exchange
$q\sim\frac{2\eF}{\h v_0}$ (and thereby large $V(q,\Delta\m)$):
Either $\Delta\m=0$ and $|g|\lesssim 10^{-2}$ or $\Delta\m\neq 0$
and $|g|\sim1$.} Note that the larger the $\Delta\m$ the smaller
$V(q,\Delta\m)$.
Even though $V(q,\Delta\m)$ 
is large the small $g$-factor suppresses the $\Delta\m=0$
backscattering.

Consider now armchair-like tubes where the bands crossing $\eF=0$
all have $\m=0$, so the small crystal wave vector transfer around
$\pm\frac{2\pi}{3|\mathbf{T}|}$ have
$\varsigma=\varsigma^{\prime}$ and therefore the $g$ factor is the
same as in
Eq.~(\ref{eq:g-i-anden-backscattering-metallic-varsigma-lig-varsigma-prime}),
i.e.~$|g|\lesssim 10^{-2}$ suppresses this kind of backscattering
(Fig.~\ref{fig:backscattering-metallic-tube-3-cases}(right)). If
we on the other hand have a large crystal wave vector transfer
backscattering
(Fig.~\ref{fig:backscattering-metallic-tube-3-cases}(right)), then
$\varsigma=-\varsigma^{\prime}$ and the $g$-factor of order 1 from
Eq.~(\ref{eq:g-for-varsigma-lig-minus-varsigma-prime-ca-1}) is
used. So the large crystal wave vector backscattering is most
important, since the Fourier transform does not grow enough to
compensate for the small $g$-factor.

Ando \emph{et
al.}~\cite{backscattering-Ando-JPJ-1998-1,backscattering-Ando-JPJ-1998-2}
have used the $\mathbf{k}\cdot\mathbf{p}$ approximation to
consider backscattering (from impurities) in metallic tubes and
found a result similar to Eq.~(\ref{eq:g-factor-graphene}), but
without the $s_0$ term. The small wave vector transfer
backscattering was found to be small in these papers.
Klesse~\cite{Klesse-PRB-2002-Coulomb-matrix-element-MWNT} has
found similar results for scattering in nanotubes, see also
[\onlinecite{pseudospin-Mceuen-PRL-1999}] for some experimental
evidence of lack of backscattering in metallic tubes compared to
semiconducting ones.

\subsection{Screening effects using RPA}

In appendix B, we derive the screened Coulomb interaction in the
random phase approximation (RPA) including the carbon nanotube
band structure with the result:
\begin{widetext}
\begin{align}
&\langle
k_1^{\prime}\m_1^{\prime}\xi_1^{\prime},k_2^{\prime}\m_2^{\prime}\xi_2^{\prime}|
V(\mathbf{r}_1,\mathbf{r}_2,\omega)|
k_1^{}\m_1^{}\xi_1^{},k_2^{}\m_2^{}\xi_2^{}\rangle=
\frac{1}{2\pi L}
g_1(k_1^{}\varsigma_1^{}\xi_1^{},k_1^{\prime}\varsigma_1^{\prime}\xi_1^{\prime})
g_2(k_2^{}\varsigma_2^{}\xi_2^{},k_2^{\prime}\varsigma_2^{\prime}\xi_2^{\prime})\nonumber\\
&\hspace{1.5cm}\times\hspace{-1mm}\sum_{G_{1},G_{2}}\sum_{u_{1},u_{2}}
\frac{V^0(k_{1}^{\prime}-k_{1}+G_{1},\m_{1}^{\prime}-\m_{1}^{}+\N_{1}^{}u_{1}^{},r_{1},r_{2})}{
\epsilon_{12}(k_{1}^{\prime}-k_{1},\m_{1}^{\prime}-\m_{1}^{},\om)}
%
\delta_{k_{1}^{}+k_{2}^{},k_{1}^{\prime}+k_{2}^{\prime}+G_{1}^{}+G_{2}^{}}
\delta_{\m_{1}^{\prime}+\m_{2}^{\prime}+\N_{1}^{}u_{1}^{},\m_{1}^{}+\m_{2}^{}+\N_{2}^{}u_{2}^{}}
\label{eq:endeligt-Coulomb-matrix-element}
\end{align}
\end{widetext}
where $\epsilon_{12}(q,\Delta\m,\om)$ is the dielectric function
disregarding the Umklapp processes (see
Eq.~(\ref{eq:dielectric-function-CNT})). Note that the effective
non-interacting polarization
$\chi_{\textrm{eff},i}^0(q,\Delta\m,\omega)$
Eq.~(\ref{eq:effective-polarizability-general}) entering the
dielectric function contains the $g$-factors. For armchair-like
tubes $\chi_{\textrm{eff},i}^0(q,\Delta\m,\omega)$ is given
explicitly in Eq.~(\ref{eq:armchair-like-polarizability-plus}) and
Eq.~(\ref{eq:armchair-like-polarizability-minus}). The bare
Coulomb interaction for cylindrical geometry is
\begin{align}
\hspace{-1mm}V^0(q,\Delta\m,r_{i},r_{j})=\frac{e^2}{\epsilon_0}
\textrm{I}_{\Delta\m}(qr_i)\textrm{K}_{\Delta\m}(qr_j) \quad
r_i\leq r_j,
\end{align}
where $\textrm{I}_{\Delta\m}(x)$ ($\textrm{K}_{\Delta\m}(x)$) is
the modified Bessel's functions of the first (second) kind of
order $\Delta\m$ and $\epsilon_0$ is the vacuum permittivity.


\section{The transresistance model}\label{sec:transresistivity-model}

The transresistance $R_{21}$ is now found for \emph{diffusive
na\-no\-tu\-bes using two coupled Boltzmann equations} (i.e.~Fermi
liquid theory) in linear response to the applied electric field
$E_1$ and for weak coupling between the tubes. The derivation is a
generalization of
ref.~[\onlinecite{Antti-smith-original-drag-PRB-93,karsten-ben-formel-PRB-1995,Audrius-drag-med-periodisk-modulation-PRB-2002,ben-Hu-drag-formel-med-flere-baand-PRB-98}]
(used to study bilayer systems) to the case of several general
bands. We only sketch the derivation and the details can be found
in Chap.3 of
Ref.~[\onlinecite{Lunde-drag-in-MWCNT-masterthesis-2004}]. In
order to simplify the notation we use $\nu$ as a collection of
band indices for the tube. A similar formula of $R_{21}$ can also
be found using the Kubo formula and doing perturbation theory to
second order in the intertube interaction (the first order DC
contribution is zero).\cite{karsten-kubo-formula-PRB-1995}

The coupled linearized Boltzmann equations for the non-equilibrium
distribution functions $f_i(k_i,\nu_i)$ ($i=1,2$ see
Fig.~\ref{fig:drag-setup-med-tubes}) are:
\begin{align}
\frac{e_1E_1}{\hbar} \frac{\p f^0_{}(\e_{k_1^{}\nu_1^{}})}{\p
k_1^{}} &= -
\frac{f_1^{}(k_1,\nu_1)-f_{}^{0}(\e_{k_1^{}\nu_1^{}})}{\tau_1^{}}
\label{eq:coupled-boltzmann-equation-1}\\
\frac{e_2E_2}{\hbar} \frac{\p f^0_{}(\e_{k_2^{}\nu_2^{}})}{\p
k_2^{}} &= -
\frac{f_2^{}(k_2,\nu_2)-f_{}^{0}(\e_{k_2^{}\nu_2^{}})}{\tau_2^{}}\nonumber\\
&\quad + S[f_1^{},f_{2}^{}=f_{}^{0}](k_2^{},\nu_2^{})
\label{eq:coupled-boltzmann-equation-2}
\end{align}
where a simple relaxation time approximation is used for the
impurity scattering~\cite{smith-boltzmann-bog}, $e_i$ is the
carrier charge in subsystem $i$ and $S[f_1^{},f_{2}^{}=f_{}^{0}]$
is the linearized collision integral coupling the two
subsystems/tubes. The assumption of weak intertube interaction and
small external electric field $E_1$ were used to linearize the
equations and to only include the lowest order terms and therefore
not have a collision integral on
Eq.~(\ref{eq:coupled-boltzmann-equation-1}). The linearized
collision integral is (using the
$H$-theorem~\cite{smith-boltzmann-bog}):
\begin{align}
& S[f_1^{},f_{2}^{}=f_{}^{0}](k_2^{},\nu_2^{})=
\label{eq:linearized-coll-integral}\\
& -\sum_{\sigma_1^{} \sigma_1^{\prime}\sigma_2^{\prime}}
\sum_{\nu_1^{} \nu_1^{\prime}\nu_2^{\prime}}
\ \sum_{k_1^{},k_1^{\prime},k_2^{\prime} \in \textrm{FBZ}} 
%
\!\!\!\!\!\! w(1^{\prime}2^{\prime};12)
f^0(\e_{k_1^{}\nu_1^{}})f^0(\e_{k_2^{}\nu_2^{}})\nonumber\\
&\!\times\!\left(1-f^0(\e_{k_1^{\prime}\nu_1^{\prime}})\right)\left(1-f^0(\e_{k_2^{\prime}\nu_2^{\prime}})\right)
\left[
\psi_1(k_1,\nu_1)-\psi_1(k_1^{\prime},\nu_1^{\prime})\right]\nonumber
\end{align}
where the deviation from equilibrium $\psi_i(k,\nu)$ was defined
though $f_i(k,\nu)-f_{}^{0}(\e_{k\nu}) \equiv
f_{}^{0}(\e_{k\nu})(1-f_{}^{0}(\e_{k\nu})) \psi_i(k,\nu)$ and
$w(1^{\prime}2^{\prime};12)$ is the transition rate for
electron-electron scattering between the tubes found from the
Fermis golden rule
$w(1^{\prime}2^{\prime};12)=\frac{2\pi}{\h} |\langle
k_1^{\prime}\nu_1^{\prime}k_2^{\prime}\nu_2^{\prime}|V_{12}(|\mathbf{r_1}-\mathbf{r_2}|)|k_1^{}\nu_1^{}k_2^{}\nu_2^{}\rangle|^2
%
\delta(\e_{k_1^{}\nu_1^{}}+\e_{k_2^{}\nu_2^{}}-\e_{k_1^{\prime}\nu_1^{\prime}}-\e_{k_2^{\prime}\nu_2^{\prime}})$
using the matrix element in
Eq.~(\ref{eq:endeligt-Coulomb-matrix-element}). To derive the
trans\-re\-sis\-tan\-ce $R_{21}=\frac{V_2}{I_1}$, 
we use the coupled Boltzmann equations
(\ref{eq:coupled-boltzmann-equation-1}) and
(\ref{eq:coupled-boltzmann-equation-2}) with
(\ref{eq:linearized-coll-integral}) and that $I_2=0$, since a
voltmeter is placed on subsystem
2.\cite{karsten-ben-formel-PRB-1995} After some
algebra~\cite{Lunde-drag-in-MWCNT-masterthesis-2004} we get:
\begin{align}
&R_{21}=\frac{\h^2}{\pi e_1 e_2 n_1 n_2 \kb T} \frac{L}{(2\pi)^2
r_1 r_2}
\sum_{G_1^{}G_2^{}}\delta_{G_1^{},G_2^{}}\nonumber\\
&\hspace{1.7cm}\times\frac{1}{(2\pi)^2}\!\!\!\sum_{\nu_1^{}\nu_1^{\prime}\nu_2^{}\nu_2^{\prime}}
\hspace{-0mm}|\J(\nu_1^{}\nu_1^{\prime},\nu_2^{}\nu_2^{\prime})|^2
\nonumber\\
&\hspace{-0mm}\times \int_{0}^{\infty}\!\! \frac{\dd q}{2\pi}
\int_{0}^{\infty}\!\!\!\! \dd \om
\frac{V_{12}(q,\nu_1^{},\nu_1^{\prime},\om)V_{12}^{\ast}(q+G_1^{},\nu_1^{},\nu_1^{\prime},\om)}{\sinh^2\big(
\frac{\h\om }{2 \kb T} \big)}\nonumber\\
&\hspace{2.4cm}\times F_{\nu_1^{}\nu_1^{\prime}}^{(1)}(q,\om)
F_{\nu_2^{}\nu_2^{\prime}}^{(2)}(q,\om),
\label{eq:rho-formel-general}
\end{align}
where $n_i$ is the carrier density,
$V_{12}(q,\nu_1^{},\nu_1^{\prime},\om)=\frac{V^0(q,\Delta\m,r_1,r_2)}{\epsilon_{12}(q,\Delta\m,\omega)}$
from Eq.~(\ref{eq:endeligt-Coulomb-matrix-element}),
$\J(\nu_1^{}\nu_1^{\prime},\nu_2^{}\nu_2^{\prime})$ are the
selection rules for the band indices such as crystal angular
momentum and/or parity (for armchair tubes) conservation and
$F_{\nu_i^{}\nu_i^{\prime}}^{(i)}(q,\om)$ is the available
$(q,\omega)$-phase space for scattering in the $i$:te tube given
by
\begin{align}
&F_{\nu_i^{}\nu_i^{\prime}}^{(i)}(q,\om)=
-\frac{e_i \tau_i}{\h^2 \mu_{\rm Tr}^{(i)}} \sum_{k_s}
\textrm{sign}(v_{k_s\nu_i}-v_{k_s+q\nu_i^{\prime}})
\label{eq:F-function-def-no-delta-function}\\
&\hspace{1.4cm}\times[f^0(\e_{k_s\nu_i^{}})-f^0(\e_{k_s+q\nu_i^{\prime}})]\
|g_i(k_s^{}\nu_i^{},k_s^{}+q\nu_i^{\prime})|^2, \nonumber
\end{align}
where the $k_s$ are the solutions to
$\e_{k\nu_i}-\e_{k+q\nu_i^{\prime}}-\h\omega=0$ in the FBZ of
subsystem $i$, $v_{k\nu}=\frac{1}{\h}\frac{\p\e_{k\nu}}{\p k}$ is
the velocity, $\s(x)$ gives the sign of $x$ (if $x=0$ then
$\s(x)=0$) and $\mu_{\rm Tr}^{(i)}$ is the transport mobility,
which is a single subsystem property. Note that the $F$-function
is periodic and odd in $q$.

Having stated this formula a few comments and interpretations are
in order.

Firstly, we note that \emph{only backscattering processes
contribute to the drag between metallic tubes} in the linearized
band models Eq.~(\ref{eq:zigzag-like-bands}) and
Eq.~(\ref{eq:armchair-like-bands}), since we only have two
velocities $\pm v_0=\pm\frac{\sqrt{3}\gamma_0a}{2\h}$ in the
metallic bands and therefore the sign-function of the velocity
difference before and after the scattering event in the
$F$-function Eq.~(\ref{eq:F-function-def-no-delta-function}) makes
only backscattering (i.e.~$v_{k_s\nu_i}=-v_{k_s+q\nu_i^{\prime}}$)
contribute to the $F$-function. In section \ref{subsec:g-factor}
we therefore analyzed the $g$-factors for all possible
backscattering processes in metallic tubes.
The interaction and $\sinh^{-2}\big(\frac{\h \om}{2\kb T}\big)$
are decreasing functions of $q$ and $\om$, respectively, so the
importance of the phase space (i.e.~the $F$-functions) in the
integral decreases from the origin.
It is worth to note that the forward scattering contribution which
for quadratic dispersion relation dominates at higher
temperatures,\cite{Pustilnik-small-q-important-in-drag-1D-PRL-2003}
thus plays no role here. If we included a curvature of the
dispersion relation for the nanotubes, we would get a correction
to the results presented here. However, there is one subtlety
hidden in this, because if we consider Coulomb drag between short
tubes, where the distribution functions are not relaxed to the
Galilean invariant form assumed in
[\onlinecite{Pustilnik-small-q-important-in-drag-1D-PRL-2003}],
but is instead given by a two-step distribution function, the
forward scattering does not contribute to the Coulomb drag as
shown in [\onlinecite{Drag-in-mesoscopic-systems-Asger-PRL-2001}].

Secondly, we have used a quantum number independent impurity
relaxation time $\tau_i$ for each subsystem in
Eq.~(\ref{eq:coupled-boltzmann-equation-1}) and
Eq.~(\ref{eq:coupled-boltzmann-equation-2}). The mobility
$\mu_{\rm Tr}^{(i)}$ can be shown to be proportional to $\tau_i$,
i.e.~$\mu_{\rm Tr}^{(i)}\propto \tau_i$, from a single subsystem
Boltzmann equation (like
Eq.~(\ref{eq:coupled-boltzmann-equation-1})). Therefore the
$F$-function Eq.~(\ref{eq:F-function-def-no-delta-function}) is
$\tau_i$ independent, so the \emph{trans\-re\-sis\-tan\-ce
$R_{21}$ is independent of the impurity relaxation times}. So in
the quasi-ballistic regime for large $\tau_i$ the transresistance
is still formally correct. However, there has been some work on
drag between ballistic 1D systems with free electron like
bands using Boltzmann equations, where almost identical transresistance formula is found.\cite{Gurevich-Drag-1D-Ballistiske-Blotzmann-JPCM-1998}

As a last comment, we note that Umklapp scattering is only
possible if the tubes are commensurable due to the
$\delta_{G_1,G_2}$ function in Eq.~(\ref{eq:rho-formel-general})
as also found in
ref.~[\onlinecite{Audrius-drag-med-periodisk-modulation-PRB-2002}].

%

\section{Electron-hole symmetry and Coulomb drag}\label{sec:Electron-hole-symmetry-and-Coulomb-drag}

All nanotubes have an inherited electron-hole symmetry from the
graphene band structure for $\eF=0$, which intuitively means that
there are as many electrons as holes for $\eF=0$ (for the precise
definition see
[\onlinecite{footnote-electron-hole-symmetry-definition}]; for a
recent measurement of electron-hole symmetry see
[\onlinecite{electron-hole-symmetry-in-semicond-CNT-eksperiment-Herrero-nature-2004}]).
So there will be an equal amount of momentum transfer to (from)
the electrons and holes and therefore no voltage difference will
arise, i.e.~$R_{21}=0$, if one of the subsystems has electron-hole
symmetry. Formally, the $F$-function can be seen to vanish at
electron-hole symmetry by using $f^0_{-\mu}(\e)=1-f^0_{\mu}(-\e)$
(after doing the sum over the band indices), where $\mu$ is the
chemical potential. This has also been used to show how $R_{21}$
can change
sign.\cite{Audrius-drag-med-periodisk-modulation-PRB-2002}

\begin{center}
\begin{figure}
\epsfig{file=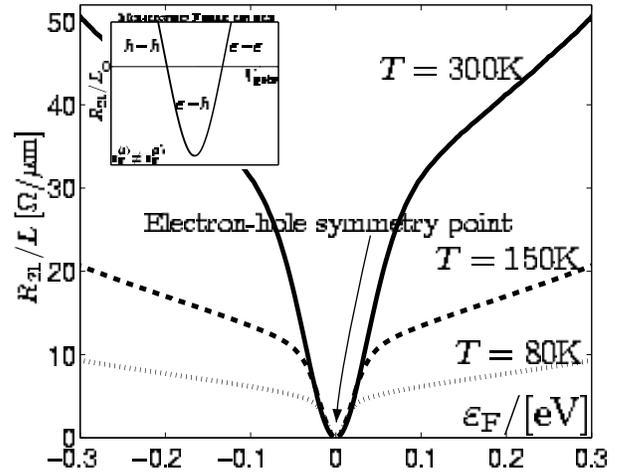,width=0.45\textwidth}
\caption{\footnotesize The transresistance per length
$\frac{R_{21}}{L}$ (in $\Omega/\mu \textrm{m}$) as a function of
the Fermi level $\eF$ (in eV) (e.g. a gate voltage). The
temperature is: $T=80K$ (dotted), $T=150K$ (dashed) and $T=300K$
(full line). The dip in $R_{21}$ at $\eF=0$ reflects the
electron-hole symmetry at this point. (Inset): A sketch of the
situation for misaligned Fermi levels (see text).}
\label{fig:rho-af-EF-armchair}
\end{figure}
\end{center}

Therefore, we predict that \emph{by varying the Fermi levels
(either by gate voltage or doping) a dip (or peak) in $R_{21}$
will appear due to the electron-hole symmetry for all kinds of
tube combinations}. For two concentric armchair tubes ($(5,5)$ in
$(10,10)$) the transresistance as a function of $\eF$ is shown on
Fig.~\ref{fig:rho-af-EF-armchair} (the range of $\eF$ is chosen to
correspond to typical
experiments\cite{gatevoltage-kruger-APL-2001,gatevoltage-doping-schonenberger-condmat-2001,gate-zhou-PRL-2000,doping-extreme-Javey-nature-2002,doping-lee-PRB-2000,doping-zhou-sceince-2000}).
Note that we use the simplification of having the same Fermi level
in the two tubes in the numerical calculation, but the situation
for misaligned Fermi levels is sketched in the inset of
Fig.~\ref{fig:rho-af-EF-armchair}. As indicated in the inset, we
have the following scenario for increasing gate voltage for
$\e^{(1)}_{\rm F}\neq \e^{(2)}_{\rm F}$: First hole-hole
($h\!-\!h$) scattering, then one subsystem passes electron-hole
symmetry, i.e.~$R_{21}=0$, afterwards $e\!-\!h$ scattering until
the other subsystem also passes though the electron-hole symmetry
point. The details of the calculation are given below in section
\ref{subsec:drag-between-armchair-tubes}.

\section{Coulomb drag between metallic tubes}\label{sec:Coulomb-drag-between-metallic-tubes}


\subsection{Drag between (real) armchair tubes}\label{subsec:drag-between-armchair-tubes}

\begin{figure}
\begin{center}
\epsfig{file=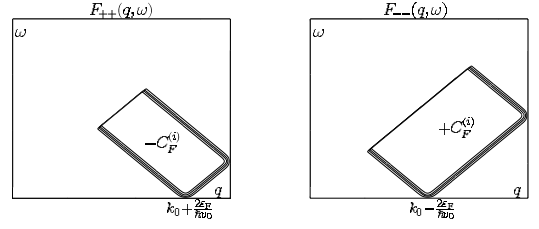,width=0.8\linewidth}
\caption{\footnotesize Contour plot of the $F$ functions for the
intraband scattering for $0<q<\frac{\pi}{|\mathbf{T}|}$, $\eF>0$
and the temperature $T=0.1\TF$. Note the smearing by the Fermi
functions due to the temperature on some edges and the sharp edge
at $\om=v_0 q$ from the step function $\theta(-\om+v_0 q)$.}
\label{fig:F-functions-intra-band-armchair-like}
\end{center}
\end{figure}

Let us begin by calculating the transresistance
Eq.~(\ref{eq:rho-formel-general}) between two concentric real
(i.e.~$(n,n)$) armchair nanotubes, which have $|\mathbf{T}|=a$
independent of $n$. The band index $\nu$ is in this case the index
$\Pi=\pm1$ from Eq.~(\ref{eq:armchair-like-bands}). To find the
$F_{\Pi\Pi^{\prime}}^{(i)}(q,\om)$ functions
Eq.~(\ref{eq:F-function-def-no-delta-function}) we need the
solutions of $\e_{k}^{\Pi}-\e_{k+q}^{\Pi^{\prime}}-\h\omega=0$
with the bands Eq.~(\ref{eq:armchair-like-bands}) and remembering
that $\e_{k}^{\Pi}$ should be made $\frac{2\pi}{|\mathbf{T}_i|}$
periodic by hand (in order to find two solutions and not only
one). The sign function only gives backscattering, which is
expressed by step functions. For intraband backscattering
$\Pi^{\prime}=\Pi$ we have $g\simeq 1$
(Eq.~(\ref{eq:g-for-varsigma-lig-minus-varsigma-prime-ca-1})) and
for interband backscattering $\Pi^{\prime}=-\Pi$ we have
$|g|^2=s_0^2\frac{3(aq)^2}{16}$
(Eq.~(\ref{eq:g-i-anden-backscattering-metallic-varsigma-lig-varsigma-prime}))
as found in section \ref{subsec:g-factor}. Therefore the
$F$-functions are\cite{Lunde-drag-in-MWCNT-masterthesis-2004} for
$0<q\leq\frac{\pi}{|\mathbf{T}_i|}$:
\begin{align}
&F_{--}^{(i)}(q,\om)=-C_F^{(i)}\theta(-\om+v_0 q)\\
&\hspace{19mm}\times
\bigg[-\bigg(f^0(\e_1)-f^0\Big(-\e_2-\frac{1}{2}k_0\h v_0
\Big)\bigg)\nonumber\\
&\hspace{25mm}+\bigg(f^0(\e_2)-f^0\Big(-\e_1-\frac{1}{2}k_0\h
v_0\Big)\bigg)\bigg]\nonumber
\end{align}
with $\e_1=\frac{\h}{2}(\om+v_0q-2v_0k_0)$ and
$\e_2=\frac{\h}{2}(\om-v_0q+v_0k_0)$,
\begin{align}
&F_{++}^{(i)}(q,\om)=-C_F^{(i)}\theta(-\om+v_0 q)\\
&\hspace{19mm}\times
\bigg[-\bigg(f^0(\tilde{\e}_1)-f^0\Big(-\tilde{\e}_2+\frac{1}{2}\h
v_0k_0\Big)\bigg)\nonumber\\
&\hspace{25mm}+\bigg(f^0(\tilde{\e}_2)-f^0\Big(-\tilde{\e}_1+\frac{1}{2}\h
v_0 k_0\Big)\bigg)\bigg]\nonumber
\end{align}
with $\tilde{\e}_1=\frac{\h}{2}(\om+v_0q-v_0k_0)$ and
$\tilde{\e}_2=\frac{\h}{2}(\om-v_0q+2v_0k_0)$,
\begin{align}
&F_{+-}^{(i)}(q,\om)=-C_F^{(i)}s_0^2\frac{3(qa)^2}{16}\theta(\om-v_0q+v_0k_0)\label{eq:F+-}\\
&\hspace{38mm}\times\theta(-\om-v_0q+2v_0k_0)\nonumber\\
&\hspace{7mm}\times\bigg[-\big(f^0(\e_1^{\prime})-f^0(-\e_2^{\prime})\big)+\big(f^0(\e_2^{\prime})-f^0(-\e_1^{\prime})\big)\bigg]\nonumber
\end{align}
and
\begin{align}
&F_{-+}^{(i)}(q,\om)=-C_F^{(i)}s_0^2\frac{3(qa)^2}{16}\theta(-\om-v_0q+v_0k_0)\label{eq:F-+}\\
&\hspace{7mm}\times\bigg[-\big(f^0(\e_1^{\prime})-f^0(-\e_2^{\prime})\big)+\big(f^0(\e_2^{\prime})-f^0(-\e_1^{\prime})\big)\bigg],\nonumber
\end{align}
where $\e_1^{\prime}=\frac{\h}{2}(\om+v_0q)$ and
$\e_2^{\prime}=\frac{\h}{2}(\om-v_0q)$ and we have calculated the
common single subsystem prefactor 
\begin{equation}
C_F^{(i)}\equiv \frac{\tilde{e}_i \tau_i}{\h^2 \mu_{\rm
Tr}^{(i)}}=\frac{2\eF+\h v_0 \frac{\pi}{|\mathbf{T}_i|}}{2(\h
v_0)^2}.
\end{equation}
It is important to note that the interband $F$-functions, $F_{+-}$
and $F_{-+}$, are heavily suppressed compared to the intraband
$F$-functions (shown in
Fig.~\ref{fig:F-functions-intra-band-armchair-like}) by
$|g|^2=s_0^2\frac{3(aq)^2}{16}$ of order $\lesssim10^{-4}$ for
backscattering around the Fermi level. Therefore, including the
tight-binding states in the Coulomb matrix element and not just in
the available phase space for scattering as in
Ref.~[\onlinecite{Lunde-Modena-drag-in-MWCNT-2004}] is a very
important effect.

\begin{figure}
\begin{center}
\begin{minipage}[c]{0.5\linewidth}
\epsfig{file=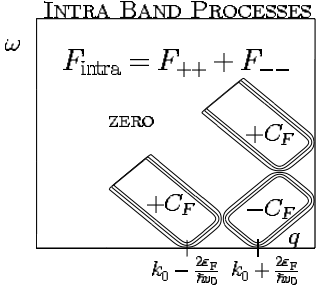,width=0.97\linewidth}
\end{minipage}\hfill
\begin{minipage}[c]{0.44\linewidth}
\caption{\footnotesize Contour plot of the $F_{\rm intra}$
function. $F_{\rm intra}$ gives the phase space for intraband
scattering in (real) armchair tubes. $F_{\rm intra}$ is seen for
$0<q<\frac{\pi}{a}$ and is odd in $q$ and should be repeated
periodically with $\frac{2\pi}{a}$ as a function of $q$.}
\label{fig:F-intra-armchair-like}
\end{minipage}
\end{center}
\end{figure}

In real armchair tubes the $\Pi$ index is a parity index in the
cylindrical
coordinate\cite{Vukovic-2002-PRB-bloch-states,Damnjanovic-1999-symmetri-og-potentialer-NT}
and therefore the Coulomb matrix element has the property:
\begin{multline}
\langle k_1^{\prime}\Pi_1^{\prime}k_2^{\prime}\Pi_2^{\prime}|
 V(\mathbf{r_1},\mathbf{r_2}|)|k_1^{}\Pi_1^{}k_2^{}\Pi_2^{}\rangle=\\
 \Pi_1^{\prime}\Pi_2^{\prime}\Pi_1^{}\Pi_2^{} \langle
k_1^{\prime}\Pi_1^{\prime}k_2^{\prime}\Pi_2^{\prime}|
 V(\mathbf{r_1},\mathbf{r_2}|)|k_1^{}\Pi_1^{}k_2^{}\Pi_2^{}\rangle,
\end{multline}
i.e.~\emph{the product of the parity is conserved in the
interaction}. Since both $\Pi=\pm1$ have $\m=0$ there is no
angular momentum selection rule, so the only selection rule $\J$
in Eq.~(\ref{eq:rho-formel-general}) is
$\J(\Pi_1^{},\Pi_1^{\prime},\Pi_2^{},\Pi_2^{\prime})=
\delta_{\Pi_1^{}\Pi_2^{},\Pi_1^{\prime}\Pi_2^{\prime}}$, which
reduces the number of terms by a factor of two. Since
$V(q,\Delta\m)$ is parity independent in
Eq.~(\ref{eq:rho-formel-general}), then the sum over band indices
for $|\mathbf{T}_1|=|\mathbf{T}_2|$ is:
\begin{align}
&\hspace{-2mm}\sum_{\Pi_1^{}\Pi_2^{}\Pi_1^{\prime}\Pi_2^{\prime}}
F^{}_{\Pi_1^{}\Pi_1^{\prime}}F^{}_{\Pi_2^{}\Pi_2^{\prime}}
\delta_{\Pi_1^{}\Pi_2^{},\Pi_1^{\prime}\Pi_2^{\prime}}^{}=\label{eq:F-inter+F-intra-def}\\
&\big(F_{++}+F_{--}\big)^2+\big(F_{+-}+F_{+-}\big)^2\equiv
(F_{\rm intra})^2+(F_{\rm inter})^2\nonumber,
\end{align}
which defines the inter and intraband $F$ functions. $(F_{\rm
inter})^2$ is of fourth order in $s_0q$ and therefore strongly
suppressed compared to $F_{\rm intra}$ even though $F_{\rm inter}$
has phase space for smaller $q$ and $\om$. $F_{\rm intra}(q,\om)$
is shown on Fig.~\ref{fig:F-intra-armchair-like}.

\begin{figure}
\begin{center}
\epsfig{file=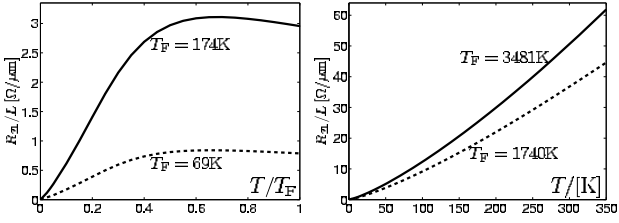,width=0.95\linewidth}
\caption{\footnotesize The transresistance per length
$\frac{R_{21}}{L}$ (in units of $\Omega/\mu \textrm{m}$) versus
temperature $T$ (or $T/\TF$ (left)). The curves are obtained from
a numerical integration of
eq.(\ref{eq:simple-drag-real-armchair-tubes}) for a $(5,5)$ in a
$(10,10)$ tube. Curves for four different Fermi levels $\eF$
(i.e.~gate voltages or dopings) are seen: $\eF=0.006$eV
($\TF=69$K) (left, dashed line), $\eF=0.015$eV ($\TF=174$K) (left,
full line), $\eF=0.15$eV ($\TF=1740$K) (right, dashed line) and
$\eF=0.3$eV (right, full line). Note the difference in magnitude
between the transresistances $R_{21}$.}
\label{fig:rho-af-TTF-armchair}
\end{center}
\end{figure}

We now have all the ingredients of the transresistance $R_{21}$:
\begin{align}
&\hspace{-1mm}\frac{R_{21}}{L}=\frac{\h^2}{\pi e^2 n_1n_2\kb
T}\frac{1}{2\pi r_1r_2}
\hspace{0mm} \int_{0}^{\infty}\!\! \frac{\dd q}{2\pi}
\int_{0}^{\infty}\!\!\!\! \dd \om
\frac{|V_{12}(q,0,\om)|^2}{\sinh^2\big(\frac{\h\om}{2\kb
T}\big)}\nonumber\\
&\hspace{21mm}\times \big[(F_{\rm intra}(q,\om))^2+(F_{\rm
inter}(q,\om))^2\big].
\label{eq:simple-drag-real-armchair-tubes}
\end{align}
A numerical integration yields $R_{21}$ as a function of $\eF$ and
the temperature $T$, shown on Figs.~\ref{fig:rho-af-EF-armchair}
and \ref{fig:rho-af-TTF-armchair}, respectively. The
transresistance per length $\frac{R_{21}}{L}$ is of the order a
few $\Omega$/$\mu$m. $R_{21}$ is seen to be linear in $T$ for
$T\lesssim0.4\TF$ as also found for free electron like
bands.\cite{Flensberg-drag-1D-duffusive-hcis-1996} For higher
temperatures the transresistance increases or decreases depending
on the Fermi level. Numerically, we find a factor of $10^6$
difference between the contribution to $R_{21}$ from
$F_{\textrm{inter}}$ and $F_{\textrm{intra}}$, so we can conclude
that \emph{the drag is due to the intraband backscattering
processes}. The largest contribution to the integral is around
$q=k_0\pm\frac{2\eF}{\h v_0}$ (see
Fig.~\ref{fig:F-intra-armchair-like}), which corresponds to
Umklapp scattering processes around the Fermi level,
e.g.~$k=k_0-\frac{\eF}{\h v_0}$ and $k^{\prime}=-k_0+\frac{\eF}{\h
v_0}$ so
$q=k^{\prime}-k+\frac{2\pi}{|\mathbf{T}_i|}=k_0+\frac{2\eF}{\h
v_0}$.

Note that screening induced by the substrate could change the
magnitude of the transresistance a small amount, which could be
modelled\cite{Egger-Luttinger-liquid-SWCNT-lang-udgave-Euro-phys-Jour-B-1998}
by introducing a new dielectric constant $\kappa=\epsilon_r
\epsilon_0$ instead of $\epsilon_0$ in
Eq.~(\ref{eq:bare-Coulomb-interaction-in-cylinders}) with
$\epsilon_r$ about 1 to 3.\cite{footnote-kappa-value}
%
%
For the present case, the magnitude of $R_{21}$ is changed
$\lesssim10\%$, when $\epsilon_r$ is increased from 1 to 3.

\begin{figure}
\begin{center}
\begin{minipage}[c]{0.54\linewidth}
\epsfig{file=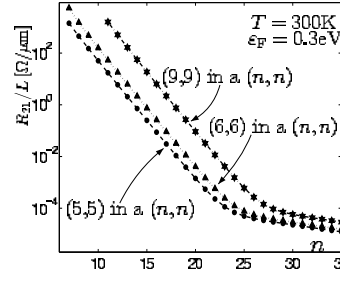,width=0.99\linewidth}
\end{minipage}\hfill
\begin{minipage}[c]{0.46\linewidth}
\caption{\footnotesize The transresistance per length
$\frac{R_{21}}{L}$ versus radius ($r\propto n$) for armchair
tubes. The different outer and inner armchair tubes are: A $(5,5)$
in a $(n,n)$ (dots), a $(6,6)$ in a $(n,n)$ (triangles) and a
$(9,9)$ in a $(n,n)$ (stars). The radius of the outer tube is:
$r=\frac{\sqrt{3}a}{2\pi}n$ for a $(n,n)$ tube. Here $T=300$K and
$\eF=0.3$eV is used. Note the logarithmic scale.}
\label{fig:rho-af-radius-armchair}
\end{minipage}
\end{center}
\end{figure}

The transresistance depends on the radii of the tubes only via the
bare Coulomb interaction
Eq.~(\ref{eq:bare-Coulomb-interaction-in-cylinders}).
Fig.~\ref{fig:rho-af-radius-armchair} shows that $R_{21}$
decreases exponentially (for $n\lesssim 25$) when keeping the
inner armchair tube at a fixed radius and increasing the outer
tube radius. For parallel 2DEG's $R_{21}$ was found to depend on
the separation $d$ as\cite{Antti-smith-original-drag-PRB-93}
$R_{21}\propto d^4$.


\subsection{Drag between armchair-like tubes}

For two general armchair-like tubes, we do not have a parity
selection rule and in general $|\mathbf{T}_1|\neq|\mathbf{T}_2|$
as seen in Table \ref{tab:examples-of-metallic-tubes}. Therefore
we have no selection rules, but all other terms than
$F_{\Pi\Pi}^{(1)}F_{\Pi^{\prime}\Pi^{\prime}}^{(2)}$ are of higher
order in $(s_0q)^2$ and therefore small, i.e.
\begin{align}
&\sum_{\Pi_1^{}\Pi_2^{}\Pi_1^{\prime}\Pi_2^{\prime}}
F^{(2)}_{\Pi_1^{}\Pi_1^{\prime}}F^{(2)}_{\Pi_2^{}\Pi_2^{\prime}}
\simeq\\
&F^{(1)}_{++}F^{(2)}_{++}+F^{(1)}_{--}F^{(2)}_{--}+F^{(1)}_{++}F^{(2)}_{--}+F^{(1)}_{--}F^{(2)}_{++}+\mathcal{O}\big((s_0q)^2\big)\nonumber
\end{align}
as for the (real) armchair tube case
Eq.~(\ref{eq:F-inter+F-intra-def}). The $F^{(i)}_{--}$ and
$F^{(i)}_{++}$ are the same as found in section
\ref{subsec:drag-between-armchair-tubes} and shown in
Fig.~\ref{fig:F-functions-intra-band-armchair-like} except that
$a$ is replaced by $|\mathbf{T}_i|$ (but not in the $g$-factor).

\begin{figure}
\begin{center}
\epsfig{file=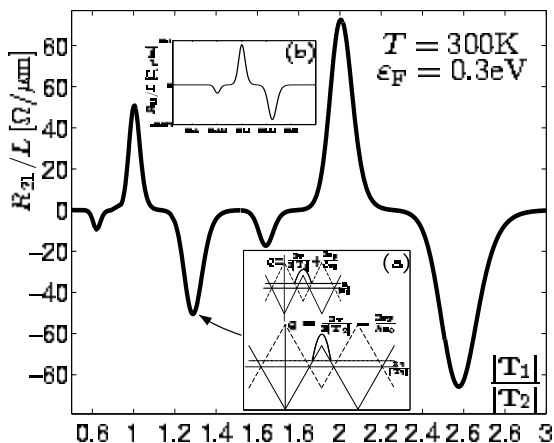,width=0.85\linewidth}
\caption{\footnotesize The transresistance per length
$\frac{R_{21}}{L}$ as a function of the ratio of the translational
vectors length $|\mathbf{T}_1|/|\mathbf{T}_2|$ for two
armchair-like tubes. The peaks corresponding to different
scattering processes are seen as explained in the text.
Numerically, we use $|\mathbf{T}_2|=a$, radii as for a $(5,5)$ in
an $(10,10)$ tube, $T=300$K and $\eF=0.3$eV. If the tubes have a
different radius, only the magnitude of the peak is changed, see
Fig.~\ref{fig:rho-af-radius-armchair}. (Inset (a)): The scattering
processes in tube 1 and 2 leading to the peak at
$|\mathbf{T}_1|/|\mathbf{T}_2|\simeq 1.28$. Note that the
backscattering processes are electron-like and hole-like,
respectively, so $R_{21}<0$. (Inset (b)): Peaks around
$|\mathbf{T}_1|/|\mathbf{T}_2|=1/2$. Note the difference in
scale.} \label{fig:rho-af-T-vektor-armchair-like}
\end{center}
\end{figure}

Since $|\mathbf{T}_1|$ and $|\mathbf{T}_2|$ are different (in
general), it is harder to conserve (crystal) momentum near the
Fermi level for the dominant backscattering process with momentum
transfer $q\simeq k_0^{(i)}\pm \frac{2\eF}{\h v_0}$ with
$k_0^{(i)}=\frac{2\pi}{3|\mathbf{T}_i|}$. However, for some values
of $|\mathbf{T}_1|$ and $|\mathbf{T}_2|$ it is possible to
conserve momentum near the Fermi level, which gives rise to peaks
in $R_{21}$ e.g.~at $\frac{|\mathbf{T}_1|}{|\mathbf{T}_2|}=1$ as
seen in Fig.~\ref{fig:rho-af-T-vektor-armchair-like}. The peaks on
both sides of $\frac{|\mathbf{T}_1|}{|\mathbf{T}_2|}=1$ are
\begin{equation}
\left|\frac{|\mathbf{T}_2|-|\mathbf{T}_1|}{|\mathbf{T}_1||\mathbf{T}_2|}\right|=\frac{6\eF}{\pi
\h v_0} \label{eq:resonance-in-T-vektor}
\end{equation}
corresponding to $k_0^{(1)}\pm \frac{2\eF}{\h v_0}=k_0^{(2)}\mp
\frac{2\eF}{\h v_0}$ (see inset (a) in
Fig.~\ref{fig:rho-af-T-vektor-armchair-like}). These peaks have
$R_{21}<0$, since they correspond to a resonance between a
electron-like and a hole-like backscattering in the sense that a
hole-like (electron-like) backscattering takes place in a
hole-like (electron-like) band with $\s(v_k)=-\s(k)$
($\s(v_k)=\s(k)$) in the FBZ. The peaks around
$\frac{|\mathbf{T}_1|}{|\mathbf{T}_2|}=\frac{1}{2}$ and $2$ are
found in the same way by taking the backscattering processes
$q\simeq 2k_0^{(i)}\pm \frac{2\eF}{\h v_0}$ into account.
If the radii of the tubes are different, then the magnitude of
$R_{21}$ will change (see Fig.~\ref{fig:rho-af-radius-armchair}),
but the signs and positions of the peaks are the same. The peaks
are broadened by increasing temperature and the positions of the
peaks depend on $\eF$ as seen e.g.~from
Eq.~(\ref{eq:resonance-in-T-vektor}) (except for
$\frac{|\mathbf{T}_1|}{|\mathbf{T}_2|}=\frac{1}{2}$, 1 and $2$).
The situation of varying $|\mathbf{T}_1|$ and $|\mathbf{T}_2|$ is
similar to varying the densities in the parallel 2D
systems.\cite{karsten-ben-formel-PRB-1995}
Note that if we have a tube configuration corresponding to a
negative dip in Fig.~\ref{fig:rho-af-T-vektor-armchair-like}
($R_{21}<0$), then this tube configuration will have a peak
instead of a dip as a function of the gate voltage.
%

Summarizing, Coulomb drag between armchair-like tubes is strongly
dependent on the magnitude of the translational vectors
$|\mathbf{T}_1|$ and $|\mathbf{T}_2|$ and can lead to both
negative and positive transresistance.

\subsection{Drag between zigzag-like tubes}

Consider the drag between two zigzag-like tubes, where the $\nu$
index is $\xi=\pm1$ and $\m\in\{\m_a,\m_b\}$ from
Eq.~(\ref{eq:zigzag-like-bands}). The backscattering $F$-function,
$F_{\xi\m,-\xi\m^{\prime}}^{(i)}$, has a form similar to
$F_{\Pi,-\Pi}^{(i)}$ for armchair-like tubes (Eq.~(\ref{eq:F+-})
and (\ref{eq:F-+})), where the important part is the
backscattering around the small $q\simeq \frac{2\eF}{\h v_0}$.
This backscattering can be both with ($\Delta\m\neq0$) and without
($\Delta\m=0$) exchange of crystal angular momentum with the
$g$-factors
\begin{equation}
|g(\Delta\m=0)|^2\propto (s_0aq)^2 \quad \textrm{and}\quad
|g(\Delta\m\neq0)|^2\simeq 1
\end{equation}
found in section \ref{subsec:g-factor}. Since there is crystal
angular momentum conservation\cite{footnote-commensurable}
%
%
it depends on the combination of the zigzag-like tubes (and their
$\m_a$ and $\m_b$) whether the $\Delta\m\neq0$ backscattering is
possible or not, so we have \emph{two very different cases}:

1. If $\Delta\m\neq0$ is not possible, then only $\Delta\m=0$
backscattering for $q\simeq\frac{2\eF}{\h v_0}$ is present, but
this is strongly suppressed by the small $g$-factor and so is the
drag. So in this case the small wave vector transfer forward
scattering (for non-linearized bands) could become important, but
in any case the effect is small. An example is the drag between
two (real) metallic zigzag tubes (see table
\ref{tab:examples-of-metallic-tubes}).

2. If $\Delta\m\neq0$ is possible, then this process is the
dominant, even though there is a small suppression (compared to
the $g$-factor) from having $\Delta\m\neq 0$ in the Fourier
transform $V_{12}(q,\Delta\m,\om)$, which is smaller the larger
$\Delta\m$. An example is a $(12,15)$ in a $(15,18)$, which has an
angular momentum exchange of $\Delta\m=\pm1$.

Furthermore, there are no peaks in $R_{21}$ as a function of
$\frac{|\mathbf{T}_1|}{|\mathbf{T}_2|}$ as for the armchair-like
tubes, since the transferred crystal wave vector
$q\simeq\pm\frac{2\eF}{\h v_0}$ is independent of
$|\mathbf{T}_i|$.

From the same principles as used above, we find the drag between
zigzag-like and armchair-like tubes to be strongly suppressed.

\section{Comments on the drag between semiconducting tubes}\label{sec:Comment-on-the-drag-between-semiconducting-tubes}

If the Fermi level for a semiconducting tube is shifted into the
conduction (or valence) band, then the drag processes are within a
single band (i.e. $\Delta\m=0$) similar to a quadratic band for
small tubes, where there are few bands with large separation. Here
both the small $q$ forward scattering and the large $q$
backscattering processes will contribute to the drag. We can
calculate the $g$ factors in the same way as for the metallic
tubes and for intraband scattering they are of order one. However,
the magnitude of the backscattering momentum transfer around the
Fermi level has to be approximately the same in the two tubes in
order to satisfy momentum conservation. In general, this is
\emph{not} the case.

If we deal with \emph{larger tubes} more bands can come into play
and thereby more scattering possibilities appear than captured in
the single band quadratic model (see
ref.~[\onlinecite{Klesse-PRB-2002-Coulomb-matrix-element-MWNT}]
for a discussion on scattering in larger MWCNT's). This is also
the case of larger metallic tubes. Coulomb drag in the quadratic
model with more bands (with different angular momentum along the
tube) for tubes of semiconducting material are considered in
ref.~[\onlinecite{Qin-drag-in-rund-quantumwells-JPCM-1995}].


%
%
%
%
%

\section{Summary}\label{sec:summary}

We have considered the intershell resistance $R_{21}$ originating
from the intershell Coulomb interaction neglecting tunnelling,
i.e. in a Coulomb drag configuration.

For any tube combination we predict a dip or peak in $R_{21}$ as a
function of gate voltage, which should be experimentally
observable. The dip (or peak) is due to the \emph{electron-hole
symmetry} of the carbon nanotube band structure. Whether $R_{21}$
has a dip or peak depends on the sign of $R_{21}$, when both
systems have Fermi levels above the electron-hole symmetry point.

The \emph{order of magnitude} and \emph{sign} of $R_{21}$ were
found to depend crucially on the chirality and Fermi level
mismatching of the two tubes. The magnitude of $R_{21}$ can reach
$\sim 50 \Omega/\mu$m under favorable circumstances.
The origin of the drastic change in magnitude between different
chiralities is the suppressed backscattering due to the Coulomb
matrix element between Bloch states combined with the mismatching
of wave vector and crystal angular momentum conservation near the
Fermi level. $R_{21}$ was found to be linear in temperature for
low temperatures (compared to $\TF$), just as for a single
quadratic band. To facilitate the analysis, we classified
\emph{all} metallic tubes in two categories: zigzag-like or
armchair-like, and described their crystal angular momentum
properties.

Throughout the paper, we use Fermi liquid theory to describe the
Coulomb drag in the MWCNT's, which gives a benchmark result for
comparison to future experiments and Luttinger liquid theories of
drag in MWCNT's. The effects considered in this paper should be
helpful in interpreting future measurements of the intershell
resistance.


\section*{Acknowledgements}

We thank Mads Brandbyge for several very useful discussions on the
carbon nanotube band structure and Reinhold Egger for a discussion
on the problem on the intershell resistance. Furthermore, Laurits
Højgaard Olesen is acknowledged for his help on numerical
problems. Jesper Nygård, Birte Rasmussen and Peter Bøggild have
given several useful comments on the experimental realization of
Coulomb drag in multiwall carbon nanotubes. Anna A. Jensen and
Morten H. Larsen are acknowledged for discussions of an algebraic
nature. A helpful comment from T. Vukovi$\acute{\textrm{c}}$
isalso appreciated.

\appendix
\section{Energy band structure of the carbon nanotubes}\label{sec:energy-bands-of-CNT-Appendix}

\renewcommand{\theequation}{A.\arabic{equation}} 
\setcounter{equation}{0}  

We will now give a rather detailed discussion of the band
structure of carbon nanotubes, since the intershell Coulomb
interaction matrix element turns out to depend critically on the
Bloch states of the two tubes due to the two atom primitive unit
cell (of a graphite layer) as seen in
section~\ref{sec:Coulomb-int}.

The carbon nanotube lattice can be thought of as a wrapping (i.e.
a conformal mapping) of a graphite layer into a tube. The wrapping
is preformed such that the chiral vector
$\mathbf{C}=n\mathbf{a}_1+m\mathbf{a}_2$ becomes the
circumferential of the $(n,m)$
 nanotube and this determines the lattice completely.\cite{Hamnda-PRL-1992-CNT-bands-periodic-boundary-conditions,Dresselhaus-Saito-PRB-1992-CNT-bandstructure-by-periodic-boundary-conditions}
(Here $\mathbf{a}_1=\frac{a}{2}(\sqrt{3},-1)$ and
$\mathbf{a}_2=\frac{a}{2}(\sqrt{3},1)$ are graphene lattice
vectors and $a=|\mathbf{a}_i|=\sqrt{3}\acc$, where $\acc$ is the
inter atomic distance).



Any $(n,m)$ nanotube  has three symmetries: A discrete
translational symmetry along the tube, a discrete rotational
symmetry around the tube axis and a helical symmetry (i.e.~a screw
operation). These symmetries give rise to the three corresponding
quantum numbers: $k$ (crystal wave vector along the tube), $\m$
(the crystal angular momentum component along the tube) and
$\kappa$ (helical quantum number). Only two of these symmetries
(quantum numbers) are needed to label the eigenstates, since the
symmetries are not independent.\cite{Mintmire-1995-carbon}
Conventionally translational symmetry is used to label the states,
but this does not use the smallest possible unit cell and can
therefore give many bands in the first Brillouin zone (FBZ) with
the same angular momentum.

Any carbon nanotube can be generated from a primitive two atom
unit cell using only discrete rotations and discrete screw
operations and thereby giving (generalized) Bloch states $|\kappa
\m\rangle$.\cite{White-1993-RPB,Mintmire-1995-carbon}
%
%
%
The advantage of using this method is that each energy band (as a
function of $\kappa$) has its own crystal angular momentum $\m$.
The discrete rotational symmetry is generated by the vector
$\mathbf{C}_{\N}$ along $\mathbf{C}$ giving the smallest possible
rotation leaving the lattice invariant, i.e.
\begin{equation}
\mathbf{C}_{\N}=\frac{n}{\N}\mathbf{a}_1+\frac{m}{\N}\mathbf{a}_2,
\quad \textrm{where} \quad \N=\gcd(n,m),
\end{equation}
i.e.~$\N$ is the greatest common divisor of $n$ and $m$. So a
given $(n,m)$ tube has crystal angular momentum
$\m\in\{0,1,\ldots,\N-1\}$. The disadvantage of using the symmetry
adapted Bloch states $|\kappa\m\rangle$ is that $\kappa$ is in the
direction the generator $\mathbf{H}$ of the helical symmetry,
which is general is different for different chiral vectors.

If we instead use the (often much) larger translational unit cell
the states can be labelled by $k\in
]-\frac{\pi}{\mathbf{T}},\frac{\pi}{\mathbf{T}}]$, where
$\mathbf{T}$ generates the translational symmetry (the
translational vector) and is given by~\cite{Dresselhaus-book-1998}
\begin{equation}
\mathbf{T}=\frac{(2m+n)\mathbf{a}_1-(2n+m)\mathbf{a}_2}{\gcd(2m+n,2n+m)}
\label{eq:T-vector-def}.
\end{equation}
Since we do not use the primitive unit cell in this case, but a
larger translational unit cell, we get a smaller FBZ and thereby
more bands in the FBZ than there are crystal angular momentum
quantum numbers.

\begin{figure}
\begin{center}
\epsfig{file=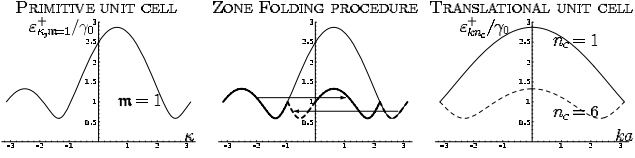,width=0.47\textwidth}
\caption{\footnotesize (Left): The $\m=1$ band for a $(5,5)$ tube
in the FBZ of the primitive unit cell as a function of
$\kappa\in]-\pi,\pi]$. (Center): The $\m=1$ band is pushed into
the smaller FBZ of the translational unit cell by using
$\kappa=\frac{ka}{2}+\frac{n_c\pi}{5}$ and that $n_c=1$ and
$n_c=6$, since $\m=1$. Note that the band is symmetrical around
$\frac{\pi}{5}$, since $\m=1$. (Right): The band structure for the
translational unit cell. Both bands have crystal angular momentum
$\m=1$, but indices $n_c=1$ and $n_c=6$.}
\label{fig:zone-folding-armchair}
\end{center}
\end{figure}

The conventional way to obtain the band structure for a isolated
singlewall $(n,m)$ nanotube using the translational unit cell is
to apply periodic boundary conditions on the two dimensional
graphene tight-binding
state\cite{Wallace-grafit-lag-baand-struktur-PR-1947}
$\psi_{\mathbf{k}}(\mathbf{r})$ along the circumferential
$\mathbf{C}$ of the
tube~\cite{Hamnda-PRL-1992-CNT-bands-periodic-boundary-conditions,Dresselhaus-Saito-PRB-1992-CNT-bandstructure-by-periodic-boundary-conditions},
i.e.
\begin{equation}
\psi_{\mathbf{k}}(\mathbf{r}+\mathbf{C})=e^{i
\mathbf{k}\cdot\mathbf{C}}\psi_{\mathbf{k}}(\mathbf{r})=\psi_{\mathbf{k}}(\mathbf{r})\Rightarrow
\mathbf{k}\cdot\mathbf{C}=2\pi n_c,
\label{eq:periodic-boundary-condition}
\end{equation}
where $n_c$ is an integer in $\{0,1,2,\ldots,\mathcal{N}-1\}$ with
$\mathcal{N}=\frac{2(n^2+m^2+nm)}{\gcd(2m+n,2n+m)}\geq \N$ being
the number of (two atomic) graphene unit cells in a translational
unit cell.\cite{Dresselhaus-book-1998} Thereby the $n_c$ labels
the bands (as a function of $k$) using the translational unit
cell. One disadvantage of using this larger translational unit
cell is, that $n_c$ \emph{is not the crystal angular momentum},
but only related to the actual physical crystal angular momentum
$\m$ by:
\begin{equation}
n_c=\m \ (\textrm{mod}\ \N). \label{eq:n-c-og-m-connection}
\end{equation}
Furthermore, we can connect the description of the band structure
using the primitive unit cell and the translational unit cell by
$\kappa=\mathbf{k}\cdot\mathbf{H}$, i.e. $\kappa$ depends on both
$k$ and $n_c$.\cite{Mintmire-1995-carbon} An example is given in
Fig.~\ref{fig:zone-folding-armchair}.

To do a tight-binding calculation for
graphene~\cite{Wallace-grafit-lag-baand-struktur-PR-1947} it is
essential that the unit cell of graphene has two atoms, so the
tight-binding state (Wannier decomposition of the eigenstate) has
two components~\cite{Ashcroft-Mermin-book-1976}:
\begin{equation}
\psi_{\mathbf{k}}(\mathbf{r})=
\frac{1}{\sqrt{N}}\sum_{\mathbf{R}}e^{i\mathbf{k}\cdot\mathbf{R}}
(\alpha_{\mathbf{k}}\Psi(\mathbf{r}-\mathbf{R})+\beta_{\mathbf{k}}\Psi(\mathbf{r}-\mathbf{R}-\mathbf{d})),
\label{eq:Bloch-state-graphene}
\end{equation}
where $\Psi$ is a $2p_z$ orbital (Wannier function) localized at
each atom, $\mathbf{R}=n_1\mathbf{a}_1+n_2\mathbf{a}_2$
($n_1,n_2\in\mathbb{Z}$) are lattice vectors, $N$ is the number of
unit cells in the layer,
$\mathbf{d}=\frac{1}{3}(\mathbf{a}_1+\mathbf{a}_2)$ is the vector
between the two atoms in the unit cell and $\alpha_{\mathbf{k}}$,
$\beta_{\mathbf{k}}$ are functions to be determined by the
tight-binding calculation. To find the energy we insert
$\psi_{\mathbf{k}}(\mathbf{r})$ in
$H\psi_{\mathbf{k}}(\mathbf{r})=\e_{\mathbf{k}}\psi_{\mathbf{k}}(\mathbf{r})$
and obtain a $2\times2$ matrix equation:
\begin{equation}
\hspace{-0.3mm}\left(
\begin{array}{cc}
H_{11} & H_{12}\\ H_{21} & H_{22}
\end{array} \right)
\left(
\begin{array}{c}
\alpha_{\mathbf{k}} \\ \beta_{\mathbf{k}} \end{array} \right)=
\e_{\mathbf{k}} \left(
\begin{array}{cc}
S_{11} & S_{12}\\ S_{21} & S_{22}
\end{array} \right)
\left(
\begin{array}{c}
\alpha_{\mathbf{k}} \\ \beta_{\mathbf{k}} \end{array} \right),
\label{eq:schrodinger-matrix-tight-binding}
\end{equation}
where $H_{ij}$, $S_{ij}$ are the overlap integrals with and
without the Hamiltonian found in the nearest neighbor
tight-binding approximation to be:
\begin{align}
H_{12}&=-\gamma_0
\big(1+e^{-i\mathbf{k}\cdot\mathbf{a}_1}+e^{-i\mathbf{k}\cdot\mathbf{a}_2}\big)
\equiv -\gamma_0 \Upsilon(\mathbf{k}),
\label{eq:H-AB-matrix-element}\\
S_{12}&=s_0 \Upsilon(\mathbf{k}), \quad S_{ii}=1 \quad
\textrm{and} \quad H_{ii}=\e_0,
\end{align}
where the value of the overlap integral is $\gamma_0\simeq
3\textrm{eV}$, the overlap of the orbitals
are~\cite{Dresselhaus-book-1998} $s_0\sim0.1$ and $\e_0$ is the
energy of the orbital, which is set to zero. Here the convention
$\gamma_0,s_0>0$ is used. In the literature a range of different
values is used (e.g.~$\gamma_0\simeq
2.5-3.1$).\cite{Brown-overlapsintegraler-eksperiment-PRB-2000,Reich-tight-binding-overlapsintegral-PRB-2002,Wilder-overlap-eksperiment-nature-1998,Odom-overlap-eksperiment-nature-1998}
By a diagonalization of
Eq.~(\ref{eq:schrodinger-matrix-tight-binding}) we find (for
$\Upsilon(\mathbf{k})\neq0$):
\begin{equation}
\e^{\pm}_{\mathbf{k}}=\pm\gamma_0 |\Upsilon(\mathbf{k})|, \quad
\left( \begin{array}{c} \alpha_{\mathbf{k}} \\ \beta_{\mathbf{k}}
\end{array} \right)_{\pm}=
\frac{1}{\sqrt{2}} \left( \begin{array}{c} \mp \frac{\Upsilon(\mathbf{k})}{|\Upsilon(\mathbf{k})|} \\
1
\end{array} \right), \label{eq:bandstructure-af-grafit-lag-s0=0}
\end{equation}
where we have neglected $s_0$ in the energy (but not in the
eigenstate).
By inserting $\mathbf{k}$ decomposed along the tube ($\mathbf{T}$)
and around the tube ($\mathbf{C}$):
$\mathbf{k}=k\frac{\mathbf{T}}{|\mathbf{T}|}+k_c\frac{\mathbf{C}}{|\mathbf{C}|}$
with $k_c=\frac{2\pi}{|\mathbf{C}|}n_c$, one obtains the band
structure for nanotube labelled by $k$ and $n_c$. Essentially the
same tight-binding calculation can be done using only the helical
and rotational symmetry as in~[\onlinecite{White-1993-RPB}] and
the same result is found, when we use
$\kappa=\mathbf{k}\cdot\mathbf{H}$ and
Eq.~(\ref{eq:n-c-og-m-connection}) to convert between the quantum
numbers\cite{footnote-Mads-Brandbyge-CNT-bands-program}
%
%
(see Fig.\ref{fig:zone-folding-armchair}).

The Fermi level is at $\eF=0$, since half of the states ($2p_z$
orbitals) are filled. By doping and/or a gate voltage the Fermi
level can be moved about $\sim\pm
0.4\textrm{eV}$.\cite{gatevoltage-kruger-APL-2001,gatevoltage-doping-schonenberger-condmat-2001,gate-zhou-PRL-2000,doping-extreme-Javey-nature-2002,doping-lee-PRB-2000,doping-zhou-sceince-2000}
Furthermore note that graphene has electron-hole
symmetry\cite{footnote-electron-hole-symmetry-definition}
%
%
for $\eF=0$ and therefore so does any $(n,m)$ carbon nanotube.

\begin{table}
  \centering
   \begin{tabular}{
   |c||c|c|c|c|c|c|}
    \hline
    \ Chirality \ & \ \ Type \ \ & \ $\m_a$ \ & \ $\m_b$ \ & \ $\N$ \ &\ \ \ $|\mathbf{T}|/a$ \ \ \ & \ \ \ $D/a$ \ \ \ \\
    \hline
    \hline
    $(n,n)$    & AL & 0 & 0 & $n$ & 1            & $\frac{\sqrt{3}n}{\pi}$   \\
    $(7,4)$    & AL & 0 & 0 &  1  & $\sqrt{31}$  & $\frac{\sqrt{93}}{\pi}$   \\
    $(15,6)$   & AL & 0 & 0 &  3  & $\sqrt{13}$  & $\frac{3\sqrt{39}}{\pi}$  \\
    $(8,23)$   & AL & 0 & 0 &  1  & $\sqrt{259}$ & $\frac{\sqrt{777}}{\pi}$  \\
    $(10,25)$  & AL & 0 & 0 &  5  & $\sqrt{13}$ & $\frac{5\sqrt{39}}{\pi}$   \\
    \hline
    $(n,0)$    & ZL & $\frac{2n}{3}$ & $\frac{n}{3}$ & $n$ & $\sqrt{3}$   & $\frac{n}{\pi}$          \\
    $(9,6)$    & ZL & 2              & 1             & 3   & $\sqrt{57}$  & $\frac{3\sqrt{19}}{\pi}$ \\
    $(6,21)$   & ZL & 2              & 1             & 3   & $\sqrt{201}$ & $\frac{3\sqrt{67}}{\pi}$ \\
    $(18,12)$  & ZL & 4              & 2             & 6   & $\sqrt{57}$  & $\frac{6\sqrt{19}}{\pi}$ \\
    $(12,24)$  & ZL & 4              & 8             & 12  & $\sqrt{21}$  & $\frac{12\sqrt{7}}{\pi}$ \\
    \hline
   \end{tabular}
  \caption{Examples of armchair-like (AL) and zigzag-like (ZL) metallic tubes, i.e.~all kinds of metallic tubes.
  For the AL tubes the difference in the length of the translational vector $\mathbf{T}$
  and the diameter $D$ is seen and for the ZL tubes we note the variety of the crystal angular momentum $\m_a=\frac{2n+m}{3}\ (\textrm{mod}\ \N)$
  and $\m_b=\frac{2m+n}{3}\ (\textrm{mod}\ \N)$ of the bands crossing the Fermi level ($\eF=0$).
  Numerically, it turns out, that $|\m_a-\m_b|=1$ for most of the ZL tubes, but there are other cases
  such as the $(12,24)$ tube. Remember that $\N=\gcd(n,m)$ and $a=|\mathbf{a}_i|$.}
\label{tab:examples-of-metallic-tubes}
\end{table}

\subsection{The linearized band structure}\label{subsec:The-linearized-band-structure}

We are only interested in the transport properties of nanotubes
and therefore expand $\Upsilon(\mathbf{k})$ around the Fermi level
$\eF=0$, i.e.~around the two
zeros\cite{footnote-zero-in-graphene-FBZ}
%
%
of $\Upsilon(\mathbf{k})$,
$\mathbf{K}_{\varsigma}=\frac{2\pi}{a}(\frac{1}{\sqrt{3}},\varsigma\frac{1}{3})$,
($\varsigma=\pm1$) and obtain
\begin{equation}
\Upsilon(\mathbf{K}_{\varsigma}+\boldsymbol{\kk})\simeq
\frac{\sqrt{3}a}{2} (i\kk_x+\varsigma\kk_y),
\label{eq:linearized-upsilon}
\end{equation}
where we have introduced the deviation from
$\mathbf{K}_{\varsigma}$ by $\boldsymbol{\kk}\equiv
\mathbf{k}-\mathbf{K}_{\varsigma}$. Note that
|$\Upsilon(\mathbf{K}_{\varsigma}+\boldsymbol{\kk})|\simeq
\frac{\sqrt{3}a}{2} |\boldsymbol{\kk}|$ used in
Eq.~(\ref{eq:bandstructure-af-grafit-lag-s0=0}). Furthermore, note
that we do not expand $\Upsilon$ around each individual $\eF\neq0$
used, but around $\eF=0$, since this preserves the electron-hole
symmetry of the band structure. By inserting
$\boldsymbol{\kk}=\kk_T\frac{\mathbf{T}}{|\mathbf{T}|}+\kk_C\frac{\mathbf{C}}{|\mathbf{C}|}$
into the periodic boundary condition
Eq.~(\ref{eq:periodic-boundary-condition}) the energy is found to
be
\begin{align}
& \e^{\pm}_{\kk_T,n_c}=\label{eq:energi-mulige-near-EF} \\
& \ \pm\frac{2\h
v_0}{D}\sqrt{\left(\frac{\kk_TD}{2}\right)^2+\left(n_c-\frac{(n+m)+\varsigma\frac{1}{3}(m-n)}{2}\right)^2},\nonumber
\end{align}
where  $D=\frac{a\sqrt{n^2+m^2+mn}}{\pi}$ is the diameter,
$v_0=\frac{\sqrt{3}\gamma_0 a}{2\h}$ is the value of the velocity
in all metallic tubes.

\subsection{Unified picture of metallic tubes: armchair-like and zigzag-like tubes}

We will now show using the linearized $\Upsilon$
Eq.~(\ref{eq:linearized-upsilon}) that \emph{all metallic tubes
are either zigzag-like or armchair-like} and define the precise
meaning of this. If $\frac{n-m}{3}\in \mathbb{Z}$ the $(n,m)$ tube
is metallic and has four crossings of the Fermi level found from
Eq.~(\ref{eq:energi-mulige-near-EF}) to be: Two (the $\pm$ in
Eq.~(\ref{eq:energi-mulige-near-EF})) for
$n_c^{\varsigma=+1}=\frac{2m+n}{3}$ and two ($\pm$) for
$n_c^{\varsigma=-1}=\frac{2n+m}{3}$ (i.e.~$\kk_C=0$). This gives
the energy and eigenstates for the bands crossing the Fermi level:
\begin{align}
\e_{\kk_T^{},n_c^{\varsigma}}^{\xi} &=\xi\h v_0 \kk_T \quad
\textrm{and} \label{eq:general-metallic-nanotube-bands-A}\\
\left(\!\!\begin{array}{c} \alpha \\ \beta
\end{array}\!\!\right)_{\xi,\varsigma}&=
\frac{1}{\sqrt{2}} \left(\!\begin{array}{c} -\xi
\frac{i(m-n)-\varsigma\sqrt{3}(n+m)}{2\sqrt{n^2+m^2+mn}}
\\ 1
\end{array}\!\right),
\label{eq:general-metallic-nanotube-states-A}
\end{align}
where $\xi=\pm1$. This is found by inserting the linearized
$\Upsilon$ Eq.~(\ref{eq:linearized-upsilon}) into
Eq.~(\ref{eq:bandstructure-af-grafit-lag-s0=0}) and using
$\kk_C=0$. By doing this, we get $\s (\kk_T)$ in $\alpha$ and $\e
\propto \pm |\kk_T|$, but we require continuity of the states
(across the $\Upsilon=0$ point where
Eq.~(\ref{eq:bandstructure-af-grafit-lag-s0=0}) was not valid) and
remove the sign-function and thereby also the absolute value,
i.e.~the metallic linear bands cross the Fermi level ($\eF=0$).
Note that $\alpha$ and $\beta$ are independent of $\kk_T$ and
thereby $k$ (to first order in $k$), which turns out to be
important in the Coulomb matrix element. The energy bands cross
the Fermi level ($\eF=0$) at $\kk_T=0$ and since
$k=\mathbf{k}\cdot\frac{\mathbf{T}}{|\mathbf{T}|}=\kk_T+
\mathbf{K}_{\varsigma}\cdot\frac{\mathbf{T}}{|\mathbf{T}|}$ the
crossing of $\eF=0$ as a function of $k$ is at
\begin{equation}
\mathbf{K}_{\varsigma}\cdot\frac{\mathbf{T}}{|\mathbf{T}|}=
%
%
\left\{
\begin{array}{rl}
\frac{-2\pi n}{\gcd(2n+m,2m+n)|\mathbf{T}|} & \textrm{for} \ \varsigma=+1\\
\frac{2\pi m}{\gcd(2n+m,2m+n)|\mathbf{T}|} & \textrm{for} \
\varsigma=-1
\end{array} \right., \label{eq:crossing-af-fermi-level-lig-nul-generelt}
\end{equation}
which are either both at $k=0$ (double degenerate,
$n_c^{\varsigma=\pm1}$) or $k=\pm\frac{2\pi}{3|\mathbf{T}|}$
(non-degenerate) for $k$ in the FBZ,
$-\frac{\pi}{|\mathbf{T}|}\leq k \leq \frac{\pi}{|\mathbf{T}|}$
(see [\onlinecite{Lunde-drag-in-MWCNT-masterthesis-2004}] for
details). Furthermore, we have the following connection between
the crossing of $\eF=0$ and the crystal angular momentum of the
bands crossing\cite{footnote-algebra-mangler}:
%
If the bands are crossing $\eF=0$ at $k=0$, then the two doubly
degenerate crosses have different nonzero angular momentum
$\m_a=\frac{2n+m}{3}\ (\textrm{mod}\ \N)$, $\m_b=\frac{2m+n}{3}\
(\textrm{mod}\ \N)$ and $\m_a\neq \m_b$. If on the other hand the
crossing is at $k=\pm\frac{2\pi}{3|\mathbf{T}|}$, then both
crosses have $\m_a=\m_b=0$. \emph{This makes it possible to divide
all metallic tubes into either armchair-like or zigzag-like tubes}
(see figure \ref{fig:two-types-of-tubes}) with the following bands
crossing the Fermi level ($\eF=0$):
\begin{align}
\hspace{-2.6mm}\textrm{Zigzag-like:}& \quad \e_{k\m}^{\xi}=\xi \h
v_0 k, \quad \m\in \{\m_a,
\m_b\} \label{eq:zigzag-like-bands-A}\\
\hspace{-2.6mm}\textrm{Armchair-like:}&  \quad \e_{k}^{\Pi}=-\Pi
\h v_0 (|k|-k_0), \ \ \textrm{($\m=0$)},
\label{eq:armchair-like-bands-A}
\end{align}
where $k_0=\frac{2\pi}{3|\mathbf{T}|}$, $\xi=\pm1$, $\Pi=\pm1$ and
$k\in]-\frac{\pi}{|\mathbf{T}|},\frac{\pi}{|\mathbf{T}|}]$. The
translational vector $\mathbf{T}$ (and $|\mathbf{T}|$) is
different for different metallic tubes independent of the type.
Note that the armchair-like bands are in general not connected in
the way modelled by Eq.~(\ref{eq:armchair-like-bands-A}) (consider
e.g. a $(7,4)$ tube), but since they have the same angular
momentum $\m=0$ we connect the bands in this way for convenience.
For scattering between the bands we will however consider the
bands as four bands as we will see in section
\ref{sec:Coulomb-int}. Examples of zigzag-like and armchair-like
tubes are found in table \ref{tab:examples-of-metallic-tubes}.

For a (real) armchair $(n,n)$ tube the $\Pi$ index in
Eq.~(\ref{eq:armchair-like-bands-A}) is the parity in the angular
coordinate in cylindrical
coordinates~\cite{Vukovic-2002-PRB-bloch-states,Damnjanovic-1999-symmetri-og-potentialer-NT}
and the states are $\tiny{\left(\!\! \begin{array}{c} \alpha \\
\beta
\end{array}\!\! \right)}=\frac{1}{\sqrt{2}}\tiny{\left(\! \begin{array}{c} \Pi \\
1
\end{array}\! \right)}$ to all orders in $k$ (in the nearest neighbor tight-binding
approximation). Results similar to the ones obtained from the
linearized $\Upsilon$ Eq.~(\ref{eq:linearized-upsilon}) can by
found by using the $\mathbf{k}\cdot\mathbf{p}$
approximation,\cite{Ando-electronic-states-SWCNT-JPSJ-1993}
however this does not reveal the crystal angular momentum.

\section{Screening in the RPA approach including the band
structure}\label{sec:RPA-in-tubes}

\renewcommand{\theequation}{B.\arabic{equation}} 
\setcounter{equation}{0}  

Here we calculate the screened Coulomb potential in the random
phase approximation (RPA) in order to include both static and
dynamical screening effects in the Coulomb drag, which have been
seen to be important perviously for bilayer
systems.\cite{Antti-smith-original-drag-PRB-93,karsten-ben-PRL-plasmon-1994,karsten-ben-formel-PRB-1995}

The Dyson equation for the screened potential in real and
frequency space is:
\begin{align}
V(\mathbf{r}_1,\mathbf{r}_2,\omega)= &V^0(|\mathbf{r}_1-\mathbf{r}_2|)+\nonumber\\
\int \! \dd \mathbf{r} \! \int \! \dd \mathbf{r}^{\prime}
\ &V^0(|\mathbf{r}_1-\mathbf{r}|)
\chi^0(\mathbf{r},\mathbf{r}^{\prime},\omega)V(\mathbf{r}^{\prime},\mathbf{r}_2,\omega),
\label{eq:Dyson-screened-pot}
\end{align}
where the non-interacting polarizability is
\begin{equation}
\chi^0(\mathbf{r},t,\mathbf{r}^{\prime},t^{\prime})=-i\theta(t-t^{\prime})\langle[\hat{\rho}(\mathbf{r},t),\hat{\rho}(\mathbf{r}^{\prime},t^{\prime})]\rangle_0,
\end{equation}
where $\hat{\rho}(\mathbf{r},t)$ is the density operator in the
interaction picture and the average $\langle\cdots\rangle_0$ is
taken for non-interacting particles. By writing the density
operator by the help of a complete set of quantum states
$\{\varphi_{\eta}(\mathbf{r})\}$ we find the polarizability to be
\begin{align}
\chi^0(\mathbf{r},&\mathbf{r}^{\prime},\omega)=\\
& \sum_{\eta \eta^{\prime}}
\frac{f^0(\e_{\eta})-f^0(\e_{\eta^{\prime}})}{\e_{\eta}-\e_{\eta^{\prime}}-\omega+i0^+}\
\varphi_{\eta}^{\ast}(\mathbf{r})
\varphi_{\eta^{\prime}}^{\ast}(\mathbf{r}^{\prime})
\varphi_{\eta^{\prime}}(\mathbf{r})
\varphi_{\eta}(\mathbf{r}^{\prime})\nonumber\\
\equiv& \sum_{\eta \eta^{\prime}}
\tilde{\chi}^0_{\eta,\eta^{\prime}}(\omega) \
\varphi_{\eta}^{\ast}(\mathbf{r})
\varphi_{\eta^{\prime}}^{\ast}(\mathbf{r}^{\prime})
\varphi_{\eta^{\prime}}(\mathbf{r})
\varphi_{\eta}(\mathbf{r}^{\prime}),\nonumber
\end{align}
where $0^+$ is a positive infinitesimal, $f^0(\e)$ is the Fermi
function and $\tilde{\chi}^0_{\eta,\eta^{\prime}}(\omega)$ was
introduced. To find the Coulomb matrix element we insert the RPA
equation (\ref{eq:Dyson-screened-pot}) into
\begin{align}
&\langle1^{\prime}2^{\prime}|V(\mathbf{r}_1,\mathbf{r}_2,\omega)|12\rangle=\\
&\hspace{5mm}\int \! \dd \mathbf{r}_1 \! \int \! \dd \mathbf{r}_2
\varphi_{1^{\prime}}^{\ast}(\mathbf{r}_1)
\varphi_{2^{\prime}}^{\ast}(\mathbf{r}_2)
V(\mathbf{r}_1,\mathbf{r}_2,\omega) \varphi_{1}(\mathbf{r}_1)
\varphi_{2}(\mathbf{r}_2)\nonumber
\end{align}
and get
\begin{align}
&\langle1^{\prime}2^{\prime}|V(\mathbf{r}_1,\mathbf{r}_2,\omega)|12\rangle=
\langle1^{\prime}2^{\prime}|V^0(|\mathbf{r}_1-\mathbf{r}_2|)|12\rangle+
\label{eq:Dyson-general-matrix-elements}\\
&\sum_{\eta \eta^{\prime}}
\tilde{\chi}^0_{\eta,\eta^{\prime}}(\omega)
\langle1^{\prime}\eta|V^0(|\mathbf{r}_1-\mathbf{r}|)|1\eta^{\prime}\rangle
\langle\eta^{\prime}2^{\prime}|V(\mathbf{r}^{\prime},\mathbf{r}_2,\omega)|\eta2\rangle.\nonumber
\end{align}
This equation can be used for any set of quantum states and in
particular for the metallic states for nanotubes, so $\eta$ is the
set of indices $(i,k,\xi,\varsigma,\sigma)$, where $i=1,2$ is the
tube index, $\sigma$ is the spin and remember that $\varsigma$
determines the angular momentum $\m$. The screened and unscreened
matrix elements
Eq.~(\ref{eq:Coulomb-int-matrix-element-trans-unitcell-tight-bin-screened})
and
Eq.~(\ref{eq:Coulomb-int-matrix-element-trans-unitcell-tight-bin-unscreened})
can now be inserted into
Eq.~(\ref{eq:Dyson-general-matrix-elements}) to get the screened
matrix element. Doing this, we observe that
$g_1(k_1^{}\varsigma_1^{}\xi_1^{},k_1^{\prime}\varsigma_1^{\prime}\xi_1^{\prime})
g_2(k_2^{}\varsigma_2^{}\xi_2^{},k_2^{\prime}\varsigma_2^{\prime}\xi_2^{\prime})$
is a common factor, which simplifies the result. To simplify
further, we use that $g_i$ and $\tilde{\chi}^0_i$ are periodic in
the reciprocal lattice $G_i$ for subsystem $i$,
$g(\eta,\eta^{\prime})=g^{\ast}(\eta^{\prime},\eta)$ and introduce
$q_i\equiv k_i^{\prime}-k_i^{}$,
$\Delta\m_i\equiv\m_i^{\prime}-\m_i^{}$ and
\begin{widetext}
\begin{multline}
\mathcal{W}_{i_1i_2}^{}(q_1,\Delta\m_1,q_2,\Delta\m_2,\omega)\equiv\hspace{-3mm}
\sum_{G_{i_1},G_{i_2}} \sum_{u_{i_1},u_{i_2}}
V(q_1+G_{i_1},\Delta\m_1+\N_{i_1}^{}u_{i_1}^{},
q_2+G_{i_2},\Delta\m_2+\N_{i_2}^{}u_{i_2}^{},r_{i_1},r_{i_2}),
\end{multline}
where $i_1$, $i_2$ are tube indices. Equivalently we introduce
$\mathcal{W}^0_{i_1i_2}$ for the sum over $V^0$ (without the $g$'s
and the $\frac{1}{2\pi L}$ factor). So
Eq.~(\ref{eq:Dyson-general-matrix-elements}) becomes
\begin{align}
&\mathcal{W}_{i_1i_2}^{}(q_{i_1},\Delta\m_{i_1},q_{i_2},\Delta\m_{i_2},\omega)=
2\pi L
\mathcal{W}_{i_1i_2}^0(q_{i_1},\Delta\m_{i_1},q_{i_2},\Delta\m_{i_2})
+\sum_{G_{i_1}^{}u_{i_1}^{}}\sum_i
V^0(q_{i_1}^{}+G_{i_1}^{},\Delta\m_{i_1}^{}+\N_{i_1}^{}u^{}_{i_1},r_{i_1}^{},r_i^{})\nonumber\\
&\hspace{4.5cm}\times\chi_{\textrm{eff},i}^0(q_{i_1}^{}+G_{i_1}^{},\Delta\m_{i_1}^{}+\N_{i_1}^{}u^{}_{i_1},\omega)
\mathcal{W}_{ii_2}^{}(q_{i_1}+G_{i_1}^{},\Delta\m_{i_1}+\N_{i_1}^{}u_{i_1}^{},q_{i_2},\Delta\m_{i_2},\omega),
\label{eq:Dyson-med-W-matrix-struktur-i-reciprocal-lattice}
\end{align}
which has a matrix structure in the reciprocal lattice and in $i$
and the effective polarization is
\begin{align}
&\chi_{\textrm{eff},i}^0(q,\Delta\m,\omega)=
\frac{2}{2\pi L}\sum_{k\varsigma}\sum_{\xi\xi^{\prime}}
\tilde{\chi}^0_{i}(k\xi\varsigma,k+q^{}\xi^{\prime}\varsigma^{\prime},\omega)
|g_{i}(k^{},\xi^{},\varsigma_{}^{};k+q,\xi^{\prime},\varsigma_{}^{\prime})|^2,
\label{eq:effective-polarizability-general}
\end{align}
where $\varsigma_{}^{\prime}$ is chosen such that
$\m^{\prime}=\m+\Delta\m$. Note that $\tilde{\chi}^0$ is diagonal
in the tube index $i$, since we do not include tunnelling between
the tubes. In order to find the screened intershell Coulomb
interaction we truncate
Eq.~(\ref{eq:Dyson-med-W-matrix-struktur-i-reciprocal-lattice})
and only include the $G_{i_1}=0$ and $u_{i_1}=0$ term in the sum,
which gives us a $2\times2$ matrix equation (in $i$) to find
$\mathcal{W}_{12}$, and therefore the screened Coulomb matrix
element is:
\begin{align}
&\langle
k_1^{\prime}\m_1^{\prime}\xi_1^{\prime},k_2^{\prime}\m_2^{\prime}\xi_2^{\prime}|
V(\mathbf{r}_1,\mathbf{r}_2,\omega)|
k_1^{}\m_1^{}\xi_1^{},k_2^{}\m_2^{}\xi_2^{}\rangle=
\frac{1}{2\pi L}
g_1(k_1^{}\varsigma_1^{}\xi_1^{},k_1^{\prime}\varsigma_1^{\prime}\xi_1^{\prime})
g_2(k_2^{}\varsigma_2^{}\xi_2^{},k_2^{\prime}\varsigma_2^{\prime}\xi_2^{\prime})\nonumber\\
&\hspace{0cm}\times\hspace{-1mm}\sum_{G_{1},G_{2}}\sum_{u_{1},u_{2}}
\frac{V^0(k_{1}^{\prime}-k_{1}+G_{1},\m_{1}^{\prime}-\m_{1}^{}+\N_{1}^{}u_{1}^{},r_{1},r_{2})}{
\epsilon_{12}(k_{1}^{\prime}-k_{1},\m_{1}^{\prime}-\m_{1}^{},\om)}
%
\delta_{k_{1}^{}+k_{2}^{},k_{1}^{\prime}+k_{2}^{\prime}+G_{1}^{}+G_{2}^{}}
\delta_{\m_{1}^{\prime}+\m_{2}^{\prime}+\N_{1}^{}u_{1}^{},\m_{1}^{}+\m_{2}^{}+\N_{2}^{}u_{2}^{}}
\label{eq:endeligt-Coulomb-matrix-element-A}
\end{align}
with
\begin{align}
\epsilon_{12}(q,\Delta\m,\om)&=
\Big[1-\chi_{\textrm{eff},1}^0(q,\Delta\m,\omega)V^0(q,\Delta\m,r_{1},r_{1})\Big]
\Big[1-\chi_{\textrm{eff},2}^0(q,\Delta\m,\omega)V^0(q,\Delta\m,r_{2},r_{2})\Big]\nonumber\\
&\qquad-\chi_{\textrm{eff},1}^0(q,\Delta\m,\omega)\chi_{\textrm{eff},2}^0(q,\Delta\m,\omega)
\times
V^0(q,\Delta\m,r_{1},r_{2})V^0(q,\Delta\m,r_{2},r_{1}),
\label{eq:dielectric-function-CNT}
\end{align}
where we have neglected the reciprocal lattice vectors different
from zero and therefore used
$V^0(q_{i_1}^{},\Delta\m_{i_1}^{},r_1^{},r_2^{})
\mathcal{W}_{22}^0(q_{i_1},\Delta\m_{i_1},q_{i_2},\Delta\m_{i_2})-
V^0(q_{i_1}^{},\Delta\m_{i_1}^{},r_2^{},r_2^{})
\mathcal{W}_{12}^0(q_{i_1},\Delta\m_{i_1},q_{i_2},\Delta\m_{i_2})\simeq0$.
%
%
%
%
%
%
%
%
%

If we consider armchair-like tubes (only the linear bands from
Eq.~(\ref{eq:armchair-like-bands-A})), then all the crystal
angular momentum is zero and from the $g$-factor analysis in
section \ref{subsec:g-factor} the interband transition
($\Pi=1\leftrightarrow \Pi^{\prime}=-1$ in
Eq.~(\ref{eq:armchair-like-bands-A})) can safely be neglected and
for the intraband transition we have $g\sim1$. Therefore
\begin{align}
\chi_{\textrm{eff},i}^0(q,0,\omega)&= \frac{2}{2\pi
L}\sum_{k}\sum_{\Pi=\pm1}\tilde{\chi}^0_{i}(k\Pi,k+q^{}\Pi,\omega)
\equiv
\chi_{\textrm{eff},i}^0(q,\omega)^{\Pi=+1}+\chi_{\textrm{eff},i}^0(q,\omega)^{\Pi=-1},
\end{align}
and for $0\leq q\leq \frac{\pi}{|\mathbf{T}|}$ we find in the long
tube limit and for zero temperature ($T=0$)
\begin{align}
\chi_{\textrm{eff},i}^0&(q,\omega)^{\Pi=+1}=
\frac{2}{(2\pi)^2\h}\Bigg[
\theta\left(k_0-\frac{\eF}{\h v_0}-q\right)
\frac{v_0q\left(k_0+\frac{2\eF}{\h
v_0}\right)}{\omega^2-v_0^2q^2_{}} \label{eq:armchair-like-polarizability-plus}\\
&+\theta\left(q-k_0+\frac{\eF}{\h v_0}\right)
\Bigg\{\frac{2v_0q\left(q-\frac{\pi}{|\mathbf{T}|}\right)}{v_0^2q^2-\omega^2}+
\frac{1}{2v_0}\ln\left(\left|\frac{\omega^2-v_0^2\left(q-2k_0^{}+\frac{2\eF}{\h
v_0}\right)^2}{\omega^2-v_0^2q^2}\right|\right)\Bigg\}\nonumber\\
&+\theta\left(q-\frac{1}{2}k_0-\frac{\eF}{\h v_0}\right)
\frac{1}{2v_0}\ln\left(\left|\frac{\omega^2-v_0^2\left(q-k_0^{}-\frac{2\eF}{\h
v_0}\right)^2}{\omega^2-v_0^2q^2}\right|\right)
+\theta\left(\frac{1}{2}k_0+\frac{\eF}{\h v_0}-q\right)
\frac{2v_0q\left(\frac{1}{2}k_0+\frac{\eF}{\h
v_0}-q\right)}{v_0^2q^2_{}-\omega^2}
\Bigg]\nonumber
\end{align}
and
\begin{align}
\chi_{\textrm{eff},i}^0&(q,\omega)^{\Pi=-1}=
\frac{2}{(2\pi)^2\h}\Bigg[
\theta\left(k_0+\frac{\eF}{\h v_0}-q\right)
\frac{2v_0q\left(k_0+\frac{\eF}{\h
v_0}-q\right)}{v_0^2q^2_{}-\omega^2}\nonumber\\
&+\theta\left(q-k_0-\frac{\eF}{\h v_0}\right)
\frac{1}{2v_0}\ln\left(\left|\frac{v_0^2\left(q-2k_0^{}-\frac{2\eF}{\h
v_0}\right)^2-\omega^2}{v_0^2q^2-\omega^2}\right|\right)
+\theta\left(\frac{1}{2}k_0-\frac{\eF}{\h v_0}-q\right)
\frac{2v_0q\left(k_0+\frac{\eF}{\h
v_0}\right)}{\omega^2-v_0^2q^2_{}}\nonumber\\
&+\theta\left(q-\frac{1}{2}k_0+\frac{\eF}{\h v_0}\right)
\Bigg\{\frac{2v_0q\left(\frac{\pi}{|\mathbf{T}|}-q\right)}{\omega^2-v_0^2q^2}+
\frac{1}{2v_0}\ln\left(\left|\frac{v_0^2\left(q-k_0^{}+\frac{2\eF}{\h
v_0}\right)^2-\omega^2}{v_0^2q^2-\omega^2}\right|\right)\Bigg\}
\Bigg] \label{eq:armchair-like-polarizability-minus},
\end{align}
\end{widetext}
which for small $q$ and $\omega$ simplifies to the result
in~\cite{Que-2002-RPA-potential-SWNT}:
\begin{align}
\chi_{\textrm{eff},i}^0(q,\omega)^{\Pi=+1}&=\chi_{\textrm{eff},i}^0(q,\omega)^{\Pi=-1}\nonumber\\
&=\frac{4v_0q^2}{(2\pi)^2\h(\omega^2-(v_0q)^2)}.
\label{eq:eff-polarizability-static-and-small-q-limit-Que}
\end{align}
Note that in the static limit the effective polarizability is just
a constant. The zero temperature
approximation~\cite{Antti-smith-original-drag-PRB-93} of the
polarizability is good as long as $T$ is much smaller than $\TF$,
which is often the case for nanotubes ($\TF \sim 1000$K).
Including finite temperature in the polarizability could give a
plasmon enhanced drag as previously found for bilayer
systems\cite{karsten-ben-PRL-plasmon-1994,Hill-Flensberg-palsmon-peak-i-drag-PRL-1997}
at $T\simeq 0.5 \TF$.

For zigzag-like tubes the effective polarizability can be found in
the same way, but for the linear bands crossing the Fermi level
($\eF=0$) we can -- in contrast to the armchair-like case --  have
both $\Delta\m=0$ and $\Delta\m=\pm(\m_a-\m_b)$.

The unscreened Coulomb interaction $V^0(q,\Delta\m,r_{i},r_{j})$
can be found from the Poisson equation by Fourier transforming in
the cylindrical coordinate and in the coordinate along the tube,
i.e.~\cite{Lunde-drag-in-MWCNT-masterthesis-2004}
\begin{align}
\hspace{-3.5mm}V^0(q,\Delta\m,r_{i},r_{j})=\frac{e^2}{\epsilon_0}
\textrm{I}_{\Delta\m}(qr_i)\textrm{K}_{\Delta\m}(qr_j) \quad
r_i\leq r_j,\label{eq:bare-Coulomb-interaction-in-cylinders}
\end{align}
where $\textrm{I}_{\Delta\m}(x)$ ($\textrm{K}_{\Delta\m}(x)$) is
the modified Bessel's functions of the first (second) kind of
order $\Delta\m$ and $\epsilon_0$ is the vacuum permittivity. Note
that the small $q$ limit is (logarithmic) divergent only for the
potential with $\Delta\m=0$.

So we have all the ingredients in the screened Coulomb matrix
element between different shells using the tight-binding states of
the carbon nanotubes, which is used to model the Coulomb drag
between the shells.

\vspace{4mm}

\noindent $^{\ast}$E-mail:  \verb"lunan"@\verb"fys.ku.dk"


\begin{thebibliography}{111}
\expandafter\ifx\csname
natexlab\endcsname\relax\def\natexlab#1{#1}\fi
\expandafter\ifx\csname bibnamefont\endcsname\relax
  \def\bibnamefont#1{#1}\fi
\expandafter\ifx\csname bibfnamefont\endcsname\relax
  \def\bibfnamefont#1{#1}\fi
\expandafter\ifx\csname citenamefont\endcsname\relax
  \def\citenamefont#1{#1}\fi
\expandafter\ifx\csname url\endcsname\relax
  \def\url#1{\texttt{#1}}\fi
\expandafter\ifx\csname
urlprefix\endcsname\relax\def\urlprefix{URL }\fi
\providecommand{\bibinfo}[2]{#2}
\providecommand{\eprint}[2][]{\url{#2}}

\bibitem[{\citenamefont{Reich et~al.}(2003)\citenamefont{Reich, Thomsen, and
  Maultzsch}}]{Reich-et-al-Book-CNT-2003}
\bibinfo{author}{\bibfnamefont{S.}~\bibnamefont{Reich}},
  \bibinfo{author}{\bibfnamefont{C.}~\bibnamefont{Thomsen}}, \bibnamefont{and}
  \bibinfo{author}{\bibfnamefont{J.}~\bibnamefont{Maultzsch}},
  \emph{\bibinfo{title}{Carbon nanotubes}} (\bibinfo{publisher}{Wiley-vch},
  \bibinfo{year}{2003}), \bibinfo{edition}{1st} ed., ISBN
  \bibinfo{isbn}{3-527-40386-8}, \bibinfo{note}{for a nice and resent review.}

\bibitem[{\citenamefont{Bachtold
  et~al.}(2000{\natexlab{a}})\citenamefont{Bachtold, Fuhrer, Plyasunov, Forero,
  Anderson, Zettl, and
  McEuen}}]{Bachtold-ballistic-SWNT-diffusiv-MWNT-PRL-2000}
\bibinfo{author}{\bibfnamefont{A.}~\bibnamefont{Bachtold}},
  \bibinfo{author}{\bibfnamefont{M.~S.} \bibnamefont{Fuhrer}},
  \bibinfo{author}{\bibfnamefont{S.}~\bibnamefont{Plyasunov}},
  \bibinfo{author}{\bibfnamefont{M.}~\bibnamefont{Forero}},
  \bibinfo{author}{\bibfnamefont{E.~H.} \bibnamefont{Anderson}},
  \bibinfo{author}{\bibfnamefont{A.}~\bibnamefont{Zettl}}, \bibnamefont{and}
  \bibinfo{author}{\bibfnamefont{P.~L.} \bibnamefont{McEuen}},
  \bibinfo{journal}{Phys. Rev. Lett.} \textbf{\bibinfo{volume}{84}},
  \bibinfo{pages}{6082} (\bibinfo{year}{2000}{\natexlab{a}}).

\bibitem[{\citenamefont{Liang et~al.}(2001)\citenamefont{Liang, Bockrath,
  Bozovic, Hafner, Tinkham, and
  Park}}]{Liang-ballistic-experiment-SWNT-nature-2001}
\bibinfo{author}{\bibfnamefont{W.}~\bibnamefont{Liang}},
  \bibinfo{author}{\bibfnamefont{M.}~\bibnamefont{Bockrath}},
  \bibinfo{author}{\bibfnamefont{D.}~\bibnamefont{Bozovic}},
  \bibinfo{author}{\bibfnamefont{J.~H.} \bibnamefont{Hafner}},
  \bibinfo{author}{\bibfnamefont{M.}~\bibnamefont{Tinkham}}, \bibnamefont{and}
  \bibinfo{author}{\bibfnamefont{H.}~\bibnamefont{Park}},
  \bibinfo{journal}{Nature} \textbf{\bibinfo{volume}{441}},
  \bibinfo{pages}{665} (\bibinfo{year}{2001}).

\bibitem[{\citenamefont{Bachtold
  et~al.}(2000{\natexlab{b}})\citenamefont{Bachtold, Fuhrer, Plyasunov, Forero,
  Anderson, Zettl, and
  McEuen}}]{diffusive-MWCNT-ballistic-SWCNT-experiment-McEuen-PRL-2000}
\bibinfo{author}{\bibfnamefont{A.}~\bibnamefont{Bachtold}},
  \bibinfo{author}{\bibfnamefont{M.~S.} \bibnamefont{Fuhrer}},
  \bibinfo{author}{\bibfnamefont{S.}~\bibnamefont{Plyasunov}},
  \bibinfo{author}{\bibfnamefont{M.}~\bibnamefont{Forero}},
  \bibinfo{author}{\bibfnamefont{E.~H.} \bibnamefont{Anderson}},
  \bibinfo{author}{\bibfnamefont{A.}~\bibnamefont{Zettl}}, \bibnamefont{and}
  \bibinfo{author}{\bibfnamefont{P.~L.} \bibnamefont{McEuen}},
  \bibinfo{journal}{Phys. Rev. Lett.} \textbf{\bibinfo{volume}{84}},
  \bibinfo{pages}{6082} (\bibinfo{year}{2000}{\natexlab{b}}).

\bibitem[{\citenamefont{Bachtold et~al.}(1999)\citenamefont{Bachtold, Strunk,
  Salvetat, Bonard, Forr\'{o}, Nussbaumer, and
  Sch$\ddot{\textrm{o}}$nenberger}}]{Bachtold-Aharonov-Bohm-nanotubes-nature-1%
999}
\bibinfo{author}{\bibfnamefont{A.}~\bibnamefont{Bachtold}},
  \bibinfo{author}{\bibfnamefont{C.}~\bibnamefont{Strunk}},
  \bibinfo{author}{\bibfnamefont{J.-P.} \bibnamefont{Salvetat}},
  \bibinfo{author}{\bibfnamefont{J.-M.} \bibnamefont{Bonard}},
  \bibinfo{author}{\bibfnamefont{L.}~\bibnamefont{Forr\'{o}}},
  \bibinfo{author}{\bibfnamefont{T.}~\bibnamefont{Nussbaumer}},
  \bibnamefont{and}
  \bibinfo{author}{\bibfnamefont{C.}~\bibnamefont{Sch$\ddot{\textrm{o}}$nenber%
ger}}, \bibinfo{journal}{Nature} \textbf{\bibinfo{volume}{397}},
  \bibinfo{pages}{673} (\bibinfo{year}{1999}).

\bibitem[{\citenamefont{Sch$\ddot{\textrm{o}}$nenberger
  et~al.}(2001)\citenamefont{Sch$\ddot{\textrm{o}}$nenberger, Buitelaar,
  Kr$\ddot{\textrm{u}}$ger, Widmer, Nussbaumer, and
  Iqbal}}]{gatevoltage-doping-schonenberger-condmat-2001}
\bibinfo{author}{\bibfnamefont{C.}~\bibnamefont{Sch$\ddot{\textrm{o}}$nenberge%
r}}, \bibinfo{author}{\bibfnamefont{M.}~\bibnamefont{Buitelaar}},
  \bibinfo{author}{\bibfnamefont{M.}~\bibnamefont{Kr$\ddot{\textrm{u}}$ger}},
  \bibinfo{author}{\bibfnamefont{I.}~\bibnamefont{Widmer}},
  \bibinfo{author}{\bibfnamefont{T.}~\bibnamefont{Nussbaumer}},
  \bibnamefont{and} \bibinfo{author}{\bibfnamefont{M.}~\bibnamefont{Iqbal}},
  \bibinfo{journal}{Proceedings Moriond 2001, cond-mat/0106501}
  (\bibinfo{year}{2001}).

\bibitem[{\citenamefont{Martel et~al.}(1998)\citenamefont{Martel, Schmidt,
  Shea, Hertel, and Avouris}}]{Martel-diffusiv-experiment-SWNT-APL-1998}
\bibinfo{author}{\bibfnamefont{R.}~\bibnamefont{Martel}},
  \bibinfo{author}{\bibfnamefont{T.}~\bibnamefont{Schmidt}},
  \bibinfo{author}{\bibfnamefont{H.~R.} \bibnamefont{Shea}},
  \bibinfo{author}{\bibfnamefont{T.}~\bibnamefont{Hertel}}, \bibnamefont{and}
  \bibinfo{author}{\bibfnamefont{P.}~\bibnamefont{Avouris}},
  \bibinfo{journal}{Appl. Phys. Lett.} \textbf{\bibinfo{volume}{73}},
  \bibinfo{pages}{2447} (\bibinfo{year}{1998}).

\bibitem[{\citenamefont{Frank et~al.}(1998)\citenamefont{Frank, Poncharal,
  Wang, and de~Heer}}]{frank-Ballistic-MWNT-1998}
\bibinfo{author}{\bibfnamefont{S.}~\bibnamefont{Frank}},
  \bibinfo{author}{\bibfnamefont{P.}~\bibnamefont{Poncharal}},
  \bibinfo{author}{\bibfnamefont{Z.}~\bibnamefont{Wang}}, \bibnamefont{and}
  \bibinfo{author}{\bibfnamefont{W.~A.} \bibnamefont{de~Heer}},
  \bibinfo{journal}{Science} \textbf{\bibinfo{volume}{280}},
  \bibinfo{pages}{1744} (\bibinfo{year}{1998}).

\bibitem[{\citenamefont{Bockrath et~al.}(1997)\citenamefont{Bockrath, Cobden,
  McEuen, Chopra, Zettl, Thess, and
  Smalley}}]{Coulomb-blockade-in-CNT-science-Bockrath-1997}
\bibinfo{author}{\bibfnamefont{M.}~\bibnamefont{Bockrath}},
  \bibinfo{author}{\bibfnamefont{D.~H.} \bibnamefont{Cobden}},
  \bibinfo{author}{\bibfnamefont{P.~L.} \bibnamefont{McEuen}},
  \bibinfo{author}{\bibfnamefont{N.~G.} \bibnamefont{Chopra}},
  \bibinfo{author}{\bibfnamefont{A.}~\bibnamefont{Zettl}},
  \bibinfo{author}{\bibfnamefont{A.}~\bibnamefont{Thess}}, \bibnamefont{and}
  \bibinfo{author}{\bibfnamefont{R.~E.} \bibnamefont{Smalley}},
  \bibinfo{journal}{Science} \textbf{\bibinfo{volume}{275}},
  \bibinfo{pages}{1922} (\bibinfo{year}{1997}).

\bibitem[{\citenamefont{Tans et~al.}(1997)\citenamefont{Tans, Devoret, Dai,
  Thess, Smalley, Geerligs, and
  Dekker}}]{Coulomb-blockade-in-CNT-nature-Tans-1997}
\bibinfo{author}{\bibfnamefont{S.~J.} \bibnamefont{Tans}},
  \bibinfo{author}{\bibfnamefont{M.~H.} \bibnamefont{Devoret}},
  \bibinfo{author}{\bibfnamefont{H.}~\bibnamefont{Dai}},
  \bibinfo{author}{\bibfnamefont{A.}~\bibnamefont{Thess}},
  \bibinfo{author}{\bibfnamefont{R.~E.} \bibnamefont{Smalley}},
  \bibinfo{author}{\bibfnamefont{L.~J.} \bibnamefont{Geerligs}},
  \bibnamefont{and} \bibinfo{author}{\bibfnamefont{C.}~\bibnamefont{Dekker}},
  \bibinfo{journal}{Nature} \textbf{\bibinfo{volume}{386}},
  \bibinfo{pages}{474} (\bibinfo{year}{1997}).

\bibitem[{\citenamefont{Cobden and
  Nyg{\aa}rd}(2002)}]{Coulomb-blockade-in-CNT-nygaard-PRL-2002}
\bibinfo{author}{\bibfnamefont{D.~H.} \bibnamefont{Cobden}} \bibnamefont{and}
  \bibinfo{author}{\bibfnamefont{J.}~\bibnamefont{Nyg{\aa}rd}},
  \bibinfo{journal}{Phys. Rev. Lett.} \textbf{\bibinfo{volume}{89}},
  \bibinfo{pages}{046803} (\bibinfo{year}{2002}).

\bibitem[{\citenamefont{Nyg{\aa}rd et~al.}(2000)\citenamefont{Nyg{\aa}rd,
  Cobden, and Lindelof}}]{Kondo-in-CNT-nygaard-nature-2000}
\bibinfo{author}{\bibfnamefont{J.}~\bibnamefont{Nyg{\aa}rd}},
  \bibinfo{author}{\bibfnamefont{D.~H.} \bibnamefont{Cobden}},
  \bibnamefont{and} \bibinfo{author}{\bibfnamefont{P.~E.}
  \bibnamefont{Lindelof}}, \bibinfo{journal}{Nature}
  \textbf{\bibinfo{volume}{408}}, \bibinfo{pages}{342} (\bibinfo{year}{2000}).

\bibitem[{\citenamefont{Jarillo-Herrero
  et~al.}(2004)\citenamefont{Jarillo-Herrero, Sapmaz, Dekker, Kouwenhoven, and
  van~der
  Zant}}]{electron-hole-symmetry-in-semicond-CNT-eksperiment-Herrero-nature-20%
04}
\bibinfo{author}{\bibfnamefont{P.}~\bibnamefont{Jarillo-Herrero}},
  \bibinfo{author}{\bibfnamefont{S.}~\bibnamefont{Sapmaz}},
  \bibinfo{author}{\bibfnamefont{C.}~\bibnamefont{Dekker}},
  \bibinfo{author}{\bibfnamefont{L.~P.} \bibnamefont{Kouwenhoven}},
  \bibnamefont{and} \bibinfo{author}{\bibfnamefont{H.~S.~J.}
  \bibnamefont{van~der Zant}}, \bibinfo{journal}{Nature}
  \textbf{\bibinfo{volume}{429}}, \bibinfo{pages}{389} (\bibinfo{year}{2004}).

\bibitem[{\citenamefont{Nyg{\aa}rd et~al.}(1999)\citenamefont{Nyg{\aa}rd,
  D.H.Cobden, Bockrath, McEuen, and
  Lindelof}}]{review-transport-bl-a-Coulomb-blockade-nygaard-1999}
\bibinfo{author}{\bibfnamefont{J.}~\bibnamefont{Nyg{\aa}rd}},
  \bibinfo{author}{\bibnamefont{D.H.Cobden}},
  \bibinfo{author}{\bibfnamefont{M.}~\bibnamefont{Bockrath}},
  \bibinfo{author}{\bibfnamefont{P.~L.} \bibnamefont{McEuen}},
  \bibnamefont{and} \bibinfo{author}{\bibfnamefont{P.~E.}
  \bibnamefont{Lindelof}}, \bibinfo{journal}{Appl. Phys. A}
  \textbf{\bibinfo{volume}{69}}, \bibinfo{pages}{297} (\bibinfo{year}{1999}).

\bibitem[{\citenamefont{McEuen et~al.}(1999)\citenamefont{McEuen, Bockrath,
  Cobden, Yoon, and Louie}}]{pseudospin-Mceuen-PRL-1999}
\bibinfo{author}{\bibfnamefont{P.~L.} \bibnamefont{McEuen}},
  \bibinfo{author}{\bibfnamefont{M.}~\bibnamefont{Bockrath}},
  \bibinfo{author}{\bibfnamefont{D.~H.} \bibnamefont{Cobden}},
  \bibinfo{author}{\bibfnamefont{Y.-G.} \bibnamefont{Yoon}}, \bibnamefont{and}
  \bibinfo{author}{\bibfnamefont{S.~G.} \bibnamefont{Louie}},
  \bibinfo{journal}{Phys. Rev. Lett.} \textbf{\bibinfo{volume}{83}},
  \bibinfo{pages}{5098} (\bibinfo{year}{1999}).

\bibitem[{\citenamefont{Kong et~al.}(2001)\citenamefont{Kong, Yenilmez,
  Tombler, Kim, Dai, Laughlin, Liu, Jayanthi, and
  Wu}}]{Ballistic-4ee/h-good-contacts-Kong-PRL-2001}
\bibinfo{author}{\bibfnamefont{J.}~\bibnamefont{Kong}},
  \bibinfo{author}{\bibfnamefont{E.}~\bibnamefont{Yenilmez}},
  \bibinfo{author}{\bibfnamefont{T.~W.} \bibnamefont{Tombler}},
  \bibinfo{author}{\bibfnamefont{W.}~\bibnamefont{Kim}},
  \bibinfo{author}{\bibfnamefont{H.}~\bibnamefont{Dai}},
  \bibinfo{author}{\bibfnamefont{R.~B.} \bibnamefont{Laughlin}},
  \bibinfo{author}{\bibfnamefont{L.}~\bibnamefont{Liu}},
  \bibinfo{author}{\bibfnamefont{C.~S.} \bibnamefont{Jayanthi}},
  \bibnamefont{and} \bibinfo{author}{\bibfnamefont{S.~Y.} \bibnamefont{Wu}},
  \bibinfo{journal}{Phys. Rev. Lett.} \textbf{\bibinfo{volume}{87}},
  \bibinfo{pages}{106801} (\bibinfo{year}{2001}).

\bibitem[{\citenamefont{Javey et~al.}(2003)\citenamefont{Javey, Guo, Wang,
  Lundstrom, and Dai}}]{Ballistic-CNT-good-contacts-Pd-Javey-nature-2003}
\bibinfo{author}{\bibfnamefont{A.}~\bibnamefont{Javey}},
  \bibinfo{author}{\bibfnamefont{J.}~\bibnamefont{Guo}},
  \bibinfo{author}{\bibfnamefont{Q.}~\bibnamefont{Wang}},
  \bibinfo{author}{\bibfnamefont{M.}~\bibnamefont{Lundstrom}},
  \bibnamefont{and} \bibinfo{author}{\bibfnamefont{H.}~\bibnamefont{Dai}},
  \bibinfo{journal}{Nature} \textbf{\bibinfo{volume}{424}},
  \bibinfo{pages}{654} (\bibinfo{year}{2003}).

\bibitem[{\citenamefont{Mann et~al.}(2003)\citenamefont{Mann, Javey, Kong,
  Wang, and Dai}}]{Ballistic-CNT-good-contacts-Pd-Mann-nanolett-2003}
\bibinfo{author}{\bibfnamefont{D.}~\bibnamefont{Mann}},
  \bibinfo{author}{\bibfnamefont{A.}~\bibnamefont{Javey}},
  \bibinfo{author}{\bibfnamefont{J.}~\bibnamefont{Kong}},
  \bibinfo{author}{\bibfnamefont{Q.}~\bibnamefont{Wang}}, \bibnamefont{and}
  \bibinfo{author}{\bibfnamefont{H.}~\bibnamefont{Dai}}, \bibinfo{journal}{Nano
  Lett.} \textbf{\bibinfo{volume}{3}}, \bibinfo{pages}{1541}
  (\bibinfo{year}{2003}).

\bibitem[{\citenamefont{Egger and
  Gogolin}(1997)}]{Egger-luttinger-liquids-teori-SWCNT-prl-1997}
\bibinfo{author}{\bibfnamefont{R.}~\bibnamefont{Egger}} \bibnamefont{and}
  \bibinfo{author}{\bibfnamefont{A.~O.} \bibnamefont{Gogolin}},
  \bibinfo{journal}{Phys. Rev. Lett.} \textbf{\bibinfo{volume}{79}},
  \bibinfo{pages}{5082} (\bibinfo{year}{1997}).

\bibitem[{\citenamefont{Egger and
  Gogolin}(1998)}]{Egger-Luttinger-liquid-SWCNT-lang-udgave-Euro-phys-Jour-B-1%
998}
\bibinfo{author}{\bibfnamefont{R.}~\bibnamefont{Egger}} \bibnamefont{and}
  \bibinfo{author}{\bibfnamefont{A.~O.} \bibnamefont{Gogolin}},
  \bibinfo{journal}{Eur. Phys. J. B} \textbf{\bibinfo{volume}{3}},
  \bibinfo{pages}{281} (\bibinfo{year}{1998}).

\bibitem[{\citenamefont{Bockrath et~al.}(1999)\citenamefont{Bockrath, Cobden,
  Lu, Rinzler, Smalley, Balents, and
  McEuen}}]{Luttinger-liquids-experiment-nature-mceuen-1999}
\bibinfo{author}{\bibfnamefont{M.}~\bibnamefont{Bockrath}},
  \bibinfo{author}{\bibfnamefont{D.~H.} \bibnamefont{Cobden}},
  \bibinfo{author}{\bibfnamefont{J.}~\bibnamefont{Lu}},
  \bibinfo{author}{\bibfnamefont{A.~G.} \bibnamefont{Rinzler}},
  \bibinfo{author}{\bibfnamefont{R.~E.} \bibnamefont{Smalley}},
  \bibinfo{author}{\bibfnamefont{L.}~\bibnamefont{Balents}}, \bibnamefont{and}
  \bibinfo{author}{\bibfnamefont{P.~L.} \bibnamefont{McEuen}},
  \bibinfo{journal}{Nature} \textbf{\bibinfo{volume}{397}},
  \bibinfo{pages}{598} (\bibinfo{year}{1999}).

\bibitem[{\citenamefont{Ishii et~al.}(2003)\citenamefont{Ishii, Kataura,
  Shiozawa, Yoshioka, Otsubo, Takayama, Miyahara, Suzuki, Achiba, Nakatake
  et~al.}}]{Luttinger-liquids-experiment-in-CNT-nature-ishil-2003}
\bibinfo{author}{\bibfnamefont{H.}~\bibnamefont{Ishii}},
  \bibinfo{author}{\bibfnamefont{H.}~\bibnamefont{Kataura}},
  \bibinfo{author}{\bibfnamefont{H.}~\bibnamefont{Shiozawa}},
  \bibinfo{author}{\bibfnamefont{H.}~\bibnamefont{Yoshioka}},
  \bibinfo{author}{\bibfnamefont{H.}~\bibnamefont{Otsubo}},
  \bibinfo{author}{\bibfnamefont{Y.}~\bibnamefont{Takayama}},
  \bibinfo{author}{\bibfnamefont{T.}~\bibnamefont{Miyahara}},
  \bibinfo{author}{\bibfnamefont{S.}~\bibnamefont{Suzuki}},
  \bibinfo{author}{\bibfnamefont{Y.}~\bibnamefont{Achiba}},
  \bibinfo{author}{\bibfnamefont{M.}~\bibnamefont{Nakatake}},
  \bibnamefont{et~al.}, \bibinfo{journal}{Nature}
  \textbf{\bibinfo{volume}{426}}, \bibinfo{pages}{540} (\bibinfo{year}{2003}).

\bibitem[{\citenamefont{Kasumov et~al.}(2003)\citenamefont{Kasumov, Kociak,
  Ferrier, Deblock, Gu\'{e}ron, Reulet, Khodos, St\'{e}phan, and
  Bouchiat}}]{Luttinger-liquid-effects-could-be-Coulomb-blockade-effect-superl%
edning-Kasumov-PRB-2003}
\bibinfo{author}{\bibfnamefont{A.}~\bibnamefont{Kasumov}},
  \bibinfo{author}{\bibfnamefont{M.}~\bibnamefont{Kociak}},
  \bibinfo{author}{\bibfnamefont{M.}~\bibnamefont{Ferrier}},
  \bibinfo{author}{\bibfnamefont{R.}~\bibnamefont{Deblock}},
  \bibinfo{author}{\bibfnamefont{S.}~\bibnamefont{Gu\'{e}ron}},
  \bibinfo{author}{\bibfnamefont{B.}~\bibnamefont{Reulet}},
  \bibinfo{author}{\bibfnamefont{I.}~\bibnamefont{Khodos}},
  \bibinfo{author}{\bibfnamefont{O.}~\bibnamefont{St\'{e}phan}},
  \bibnamefont{and} \bibinfo{author}{\bibfnamefont{H.}~\bibnamefont{Bouchiat}},
  \bibinfo{journal}{Phys. Rev. B} \textbf{\bibinfo{volume}{68}},
  \bibinfo{pages}{214521} (\bibinfo{year}{2003}).

\bibitem[{\citenamefont{Tarkiainen et~al.}(2001)\citenamefont{Tarkiainen,
  Ahlskog, Penttil$\ddot{\textrm{a}}$, Roschier, Hakonen, Paalanen, and
  Sonin}}]{luttinger-liquids-kontra-fermi-liquids-MWNT-Tarkiainen-PRB}
\bibinfo{author}{\bibfnamefont{R.}~\bibnamefont{Tarkiainen}},
  \bibinfo{author}{\bibfnamefont{M.}~\bibnamefont{Ahlskog}},
  \bibinfo{author}{\bibfnamefont{J.}~\bibnamefont{Penttil$\ddot{\textrm{a}}$}},
  \bibinfo{author}{\bibfnamefont{L.}~\bibnamefont{Roschier}},
  \bibinfo{author}{\bibfnamefont{P.}~\bibnamefont{Hakonen}},
  \bibinfo{author}{\bibfnamefont{M.}~\bibnamefont{Paalanen}}, \bibnamefont{and}
  \bibinfo{author}{\bibfnamefont{E.}~\bibnamefont{Sonin}},
  \bibinfo{journal}{Phys. Rev. B} \textbf{\bibinfo{volume}{64}},
  \bibinfo{pages}{195412} (\bibinfo{year}{2001}), \bibinfo{note}{and
  ref.~therein.}

\bibitem[{\citenamefont{Krsti\'{c} et~al.}(2003)\citenamefont{Krsti\'{c},
  Blumentritt, Muster, Roth, and
  Rubio}}]{luttinger-liquids-absent-in-doped-MWNT-PRB-Krstic-2003}
\bibinfo{author}{\bibfnamefont{V.}~\bibnamefont{Krsti\'{c}}},
  \bibinfo{author}{\bibfnamefont{S.}~\bibnamefont{Blumentritt}},
  \bibinfo{author}{\bibfnamefont{J.}~\bibnamefont{Muster}},
  \bibinfo{author}{\bibfnamefont{S.}~\bibnamefont{Roth}}, \bibnamefont{and}
  \bibinfo{author}{\bibfnamefont{A.}~\bibnamefont{Rubio}},
  \bibinfo{journal}{Phys. Rev. B} \textbf{\bibinfo{volume}{67}},
  \bibinfo{pages}{041401(R)} (\bibinfo{year}{2003}).

\bibitem[{\citenamefont{Kang et~al.}(2003)\citenamefont{Kang, Lu, Kong, Hu, Yi,
  Wang, Zhang, Pan, and
  Xie}}]{luttinger-liquids-kontra-fermi-liquids-MWNT-PRB-Xie}
\bibinfo{author}{\bibfnamefont{N.}~\bibnamefont{Kang}},
  \bibinfo{author}{\bibfnamefont{L.}~\bibnamefont{Lu}},
  \bibinfo{author}{\bibfnamefont{W.~J.} \bibnamefont{Kong}},
  \bibinfo{author}{\bibfnamefont{J.~S.} \bibnamefont{Hu}},
  \bibinfo{author}{\bibfnamefont{W.}~\bibnamefont{Yi}},
  \bibinfo{author}{\bibfnamefont{Y.~P.} \bibnamefont{Wang}},
  \bibinfo{author}{\bibfnamefont{D.~L.} \bibnamefont{Zhang}},
  \bibinfo{author}{\bibfnamefont{Z.~W.} \bibnamefont{Pan}}, \bibnamefont{and}
  \bibinfo{author}{\bibfnamefont{S.~S.} \bibnamefont{Xie}},
  \bibinfo{journal}{Phys. Rev. B} \textbf{\bibinfo{volume}{67}},
  \bibinfo{pages}{033404} (\bibinfo{year}{2003}).

\bibitem[{\citenamefont{Egger}(1999)}]{Egger-Luttinger-liquids-in-MWNT-PRL-199%
9}
\bibinfo{author}{\bibfnamefont{R.}~\bibnamefont{Egger}},
  \bibinfo{journal}{Phys. Rev. Lett.} \textbf{\bibinfo{volume}{83}},
  \bibinfo{pages}{5547} (\bibinfo{year}{1999}).

\bibitem[{\citenamefont{Hunger et~al.}(2004)\citenamefont{Hunger, Lengeler, and
  Appenzeller}}]{Luttinger-liquid-and-Contact-barriers-Hunger-PRB-2004}
\bibinfo{author}{\bibfnamefont{T.}~\bibnamefont{Hunger}},
  \bibinfo{author}{\bibfnamefont{B.}~\bibnamefont{Lengeler}}, \bibnamefont{and}
  \bibinfo{author}{\bibfnamefont{J.}~\bibnamefont{Appenzeller}},
  \bibinfo{journal}{Phys. Rev. B} \textbf{\bibinfo{volume}{69}},
  \bibinfo{pages}{195406} (\bibinfo{year}{2004}).

\bibitem[{\citenamefont{Yoon et~al.}(2002)\citenamefont{Yoon, Delaney, and
  Louie}}]{no-intershell-tunnelling-MWCNT-theory-Yoon-PRB-2002}
\bibinfo{author}{\bibfnamefont{Y.-G.} \bibnamefont{Yoon}},
  \bibinfo{author}{\bibfnamefont{P.}~\bibnamefont{Delaney}}, \bibnamefont{and}
  \bibinfo{author}{\bibfnamefont{S.~G.} \bibnamefont{Louie}},
  \bibinfo{journal}{Phys. Rev. B} \textbf{\bibinfo{volume}{66}},
  \bibinfo{pages}{073407} (\bibinfo{year}{2002}).

\bibitem[{\citenamefont{Maarouf et~al.}(2000)\citenamefont{Maarouf, Kane, and
  Mele}}]{Maarouf-small-tunnelling-between-tubes-PRB-2000}
\bibinfo{author}{\bibfnamefont{A.~A.} \bibnamefont{Maarouf}},
  \bibinfo{author}{\bibfnamefont{C.~L.} \bibnamefont{Kane}}, \bibnamefont{and}
  \bibinfo{author}{\bibfnamefont{E.~J.} \bibnamefont{Mele}},
  \bibinfo{journal}{Phys. Rev. B} \textbf{\bibinfo{volume}{61}},
  \bibinfo{pages}{11156} (\bibinfo{year}{2000}), \bibinfo{note}{and
  ref.~therein.}

\bibitem[{\citenamefont{Collins et~al.}(2001)\citenamefont{Collins, Arnold, and
  Avouris}}]{collins-shell-remove-science-2001}
\bibinfo{author}{\bibfnamefont{P.~G.} \bibnamefont{Collins}},
  \bibinfo{author}{\bibfnamefont{M.~S.} \bibnamefont{Arnold}},
  \bibnamefont{and} \bibinfo{author}{\bibfnamefont{P.}~\bibnamefont{Avouris}},
  \bibinfo{journal}{Science} \textbf{\bibinfo{volume}{292}},
  \bibinfo{pages}{706} (\bibinfo{year}{2001}).

\bibitem[{\citenamefont{Cumings et~al.}(2000)\citenamefont{Cumings, Collins,
  and Zettl}}]{cumings-shell-remove-nature-2000}
\bibinfo{author}{\bibfnamefont{J.}~\bibnamefont{Cumings}},
  \bibinfo{author}{\bibfnamefont{P.~G.} \bibnamefont{Collins}},
  \bibnamefont{and} \bibinfo{author}{\bibfnamefont{A.}~\bibnamefont{Zettl}},
  \bibinfo{journal}{Nature} \textbf{\bibinfo{volume}{406}},
  \bibinfo{pages}{586} (\bibinfo{year}{2000}).

\bibitem[{\citenamefont{Dohn}(2003)}]{Dohn-shell-remove-masterthesis-2003}
\bibinfo{author}{\bibfnamefont{S.}~\bibnamefont{Dohn}}, Master's thesis,
  \bibinfo{school}{MIC--Department of Micro and Nanotechnology}
  (\bibinfo{year}{2003}).

\bibitem[{\citenamefont{Collins and
  Avouris}(2002)}]{Colling-shell-remove-Appl-Phys-2002}
\bibinfo{author}{\bibfnamefont{P.~G.} \bibnamefont{Collins}} \bibnamefont{and}
  \bibinfo{author}{\bibfnamefont{P.}~\bibnamefont{Avouris}},
  \bibinfo{journal}{Appl. Phys. A-Mater.} \textbf{\bibinfo{volume}{74}},
  \bibinfo{pages}{329} (\bibinfo{year}{2002}).

\bibitem[{\citenamefont{Li and Wang}(2004)}]{Li-shell-remove-China-2004}
\bibinfo{author}{\bibfnamefont{Q.~H.} \bibnamefont{Li}} \bibnamefont{and}
  \bibinfo{author}{\bibfnamefont{T.}~\bibnamefont{Wang}},
  \bibinfo{journal}{Sci. China Ser. E.} \textbf{\bibinfo{volume}{47}},
  \bibinfo{pages}{1} (\bibinfo{year}{2004}).

\bibitem[{\citenamefont{Cumings and
  Zettl}(2000)}]{cuming-oscillator-in-MWCNT-eksperiment-2000-science}
\bibinfo{author}{\bibfnamefont{J.}~\bibnamefont{Cumings}} \bibnamefont{and}
  \bibinfo{author}{\bibfnamefont{A.}~\bibnamefont{Zettl}},
  \bibinfo{journal}{Science} \textbf{\bibinfo{volume}{289}},
  \bibinfo{pages}{602} (\bibinfo{year}{2000}).

\bibitem[{\citenamefont{Roche et~al.}(2001{\natexlab{a}})\citenamefont{Roche,
  Triozon, Rubio, and Mayou}}]{Roche-intershell-tunnelling-PRB-2001}
\bibinfo{author}{\bibfnamefont{S.}~\bibnamefont{Roche}},
  \bibinfo{author}{\bibfnamefont{F.}~\bibnamefont{Triozon}},
  \bibinfo{author}{\bibfnamefont{A.}~\bibnamefont{Rubio}}, \bibnamefont{and}
  \bibinfo{author}{\bibfnamefont{D.}~\bibnamefont{Mayou}},
  \bibinfo{journal}{Phys. Rev. B} \textbf{\bibinfo{volume}{64}},
  \bibinfo{pages}{121401} (\bibinfo{year}{2001}{\natexlab{a}}).

\bibitem[{\citenamefont{Roche et~al.}(2001{\natexlab{b}})\citenamefont{Roche,
  Triozon, Rubio, and Mayou}}]{Roche-intershell-tunnelling-Phys-lett-A-2001}
\bibinfo{author}{\bibfnamefont{S.}~\bibnamefont{Roche}},
  \bibinfo{author}{\bibfnamefont{F.}~\bibnamefont{Triozon}},
  \bibinfo{author}{\bibfnamefont{A.}~\bibnamefont{Rubio}}, \bibnamefont{and}
  \bibinfo{author}{\bibfnamefont{D.}~\bibnamefont{Mayou}},
  \bibinfo{journal}{Phys. Lett. A} \textbf{\bibinfo{volume}{285}},
  \bibinfo{pages}{94} (\bibinfo{year}{2001}{\natexlab{b}}).

\bibitem[{\citenamefont{Triozon et~al.}(2004)\citenamefont{Triozon, Roche,
  Rubio, and Mayou}}]{Intershell-conductance-MWCNT-Roche-PRB-2004}
\bibinfo{author}{\bibfnamefont{F.}~\bibnamefont{Triozon}},
  \bibinfo{author}{\bibfnamefont{S.}~\bibnamefont{Roche}},
  \bibinfo{author}{\bibfnamefont{A.}~\bibnamefont{Rubio}}, \bibnamefont{and}
  \bibinfo{author}{\bibfnamefont{D.}~\bibnamefont{Mayou}},
  \bibinfo{journal}{Phys. Rev. B} \textbf{\bibinfo{volume}{69}},
  \bibinfo{pages}{121410(R)} (\bibinfo{year}{2004}).

\bibitem[{\citenamefont{Ahn et~al.}(2004)\citenamefont{Ahn, Kim, Wiersig, and
  Chang}}]{Intershell-tunneling-MWCNT-Ahn-Physica-E-2004}
\bibinfo{author}{\bibfnamefont{K.-H.} \bibnamefont{Ahn}},
  \bibinfo{author}{\bibfnamefont{Y.-H.} \bibnamefont{Kim}},
  \bibinfo{author}{\bibfnamefont{J.}~\bibnamefont{Wiersig}}, \bibnamefont{and}
  \bibinfo{author}{\bibfnamefont{K.~J.} \bibnamefont{Chang}},
  \bibinfo{journal}{Physica E} \textbf{\bibinfo{volume}{22}},
  \bibinfo{pages}{666} (\bibinfo{year}{2004}).

\bibitem[{\citenamefont{Hansson and
  Stafstr$\ddot{\textrm{o}}$m}(2003)}]{hansson-intershell-resistance-PRB-2003}
\bibinfo{author}{\bibfnamefont{A.}~\bibnamefont{Hansson}} \bibnamefont{and}
  \bibinfo{author}{\bibfnamefont{S.}~\bibnamefont{Stafstr$\ddot{\textrm{o}}$m}%
}, \bibinfo{journal}{Phys. Rev. B} \textbf{\bibinfo{volume}{67}},
  \bibinfo{pages}{075406} (\bibinfo{year}{2003}).

\bibitem[{\citenamefont{Sanvito et~al.}(2000)\citenamefont{Sanvito, Kwon,
  Tom\'{a}nek, and Lambert}}]{sanvito-intershell-resistance-PRL-2003}
\bibinfo{author}{\bibfnamefont{S.}~\bibnamefont{Sanvito}},
  \bibinfo{author}{\bibfnamefont{Y.-K.} \bibnamefont{Kwon}},
  \bibinfo{author}{\bibfnamefont{D.}~\bibnamefont{Tom\'{a}nek}},
  \bibnamefont{and} \bibinfo{author}{\bibfnamefont{C.~J.}
  \bibnamefont{Lambert}}, \bibinfo{journal}{Phys. Rev. Lett.}
  \textbf{\bibinfo{volume}{84}}, \bibinfo{pages}{1974} (\bibinfo{year}{2000}).

\bibitem[{\citenamefont{Miyamoto et~al.}(2002)\citenamefont{Miyamoto, Saito,
  and Tom\'{a}nek}}]{Tomanek-intershell-resistance-PRB-2002}
\bibinfo{author}{\bibfnamefont{Y.}~\bibnamefont{Miyamoto}},
  \bibinfo{author}{\bibfnamefont{S.}~\bibnamefont{Saito}}, \bibnamefont{and}
  \bibinfo{author}{\bibfnamefont{D.}~\bibnamefont{Tom\'{a}nek}},
  \bibinfo{journal}{Phys. Rev. B} \textbf{\bibinfo{volume}{65}},
  \bibinfo{pages}{041402(R)} (\bibinfo{year}{2002}).

\bibitem[{\citenamefont{Uryu}(2004)}]{Uryu-matrix-elementer-mellem-tubes-PRB-2%
004}
\bibinfo{author}{\bibfnamefont{S.}~\bibnamefont{Uryu}}, \bibinfo{journal}{Phys.
  Rev. B} \textbf{\bibinfo{volume}{69}}, \bibinfo{pages}{075402}
  (\bibinfo{year}{2004}).

\bibitem[{\citenamefont{Kim and
  Chang}(2002)}]{Kim-telescoping-MWCNT-teori-PRB-2002}
\bibinfo{author}{\bibfnamefont{D.-H.} \bibnamefont{Kim}} \bibnamefont{and}
  \bibinfo{author}{\bibfnamefont{K.~J.} \bibnamefont{Chang}},
  \bibinfo{journal}{Phys. Rev. B} \textbf{\bibinfo{volume}{66}},
  \bibinfo{pages}{155402} (\bibinfo{year}{2002}).

\bibitem[{\citenamefont{Bourlon et~al.}(2004)\citenamefont{Bourlon, Miko,
  Forro, Glattli, and
  Bachtold}}]{Bacdtold-intershell-resistance-measurement-condmat-2004}
\bibinfo{author}{\bibfnamefont{B.}~\bibnamefont{Bourlon}},
  \bibinfo{author}{\bibfnamefont{C.}~\bibnamefont{Miko}},
  \bibinfo{author}{\bibfnamefont{L.}~\bibnamefont{Forro}},
  \bibinfo{author}{\bibfnamefont{D.}~\bibnamefont{Glattli}}, \bibnamefont{and}
  \bibinfo{author}{\bibfnamefont{A.}~\bibnamefont{Bachtold}}
  (\bibinfo{year}{2004}), \bibinfo{note}{condmat/0407672}.

\bibitem[{\citenamefont{Pogrebinskii}(1977)}]{Pogrebinskii-original-drag-forsl%
ag-1977}
\bibinfo{author}{\bibfnamefont{M.~B.} \bibnamefont{Pogrebinskii}},
  \bibinfo{journal}{Fiz. Tekh. Poluprovodn. {\bf 11} 637 (1977) [Sov. Phys.
  Semicond. {\bf 11} 372 (1977)]}  (\bibinfo{year}{1977}).

\bibitem[{\citenamefont{Price}(1983)}]{Price-original-drag-forslag-1983}
\bibinfo{author}{\bibfnamefont{P.~J.} \bibnamefont{Price}},
  \bibinfo{journal}{Physica \textbf{B} and \textbf{C}}
  \textbf{\bibinfo{volume}{117}}, \bibinfo{pages}{750} (\bibinfo{year}{1983}).

\bibitem[{foo({\natexlab{a}})}]{footnote-other-interactions-than-Coluomb}
\bibinfo{note}{Other $e\!-\!e$ interactions such as phonon mediated interaction
  could give an important contribution as seen in bilayer
  systems.\cite{Boensager-phonon-drag-PRB-1998} This could be included in
  future studies.}

\bibitem[{\citenamefont{Zheng and
  Jiang}(2002)}]{MWCNT-as-GHz-oscillatorer-PRL-2002}
\bibinfo{author}{\bibfnamefont{Q.}~\bibnamefont{Zheng}} \bibnamefont{and}
  \bibinfo{author}{\bibfnamefont{Q.}~\bibnamefont{Jiang}},
  \bibinfo{journal}{Phys. Rev. Lett.} \textbf{\bibinfo{volume}{88}},
  \bibinfo{pages}{045503} (\bibinfo{year}{2002}), \bibinfo{note}{and citations
  of this paper.}

\bibitem[{pri()}]{private-comm-nygard}
\bibinfo{note}{J. Nyg{\aa}rd, P. B{\o}gglid and P. Hakonen, private comm.}

\bibitem[{\citenamefont{Sugai et~al.}(2003)\citenamefont{Sugai, Yoshida,
  Shimada, Okazaki, Shinohara, and Bandow}}]{Sugai-gro-DWCNT-NL-2003}
\bibinfo{author}{\bibfnamefont{T.}~\bibnamefont{Sugai}},
  \bibinfo{author}{\bibfnamefont{H.}~\bibnamefont{Yoshida}},
  \bibinfo{author}{\bibfnamefont{T.}~\bibnamefont{Shimada}},
  \bibinfo{author}{\bibfnamefont{T.}~\bibnamefont{Okazaki}},
  \bibinfo{author}{\bibfnamefont{H.}~\bibnamefont{Shinohara}},
  \bibnamefont{and} \bibinfo{author}{\bibfnamefont{S.}~\bibnamefont{Bandow}},
  \bibinfo{journal}{Nano Lett.} \textbf{\bibinfo{volume}{3}},
  \bibinfo{pages}{769} (\bibinfo{year}{2003}).

\bibitem[{\citenamefont{Rojo}(1999)}]{Rojo-review-af-drag-mellem-2DEG-1999}
\bibinfo{author}{\bibfnamefont{A.~G.} \bibnamefont{Rojo}}, \bibinfo{journal}{J.
  Phys.: Condens. Matter} \textbf{\bibinfo{volume}{11}}, \bibinfo{pages}{R31}
  (\bibinfo{year}{1999}), \bibinfo{note}{for a good review.}

\bibitem[{\citenamefont{Gramila et~al.}(1991)\citenamefont{Gramila, Eisenstein,
  MacDonald, Pfeiffer, and
  West}}]{Gramila-first-experiment-drag-mellem-2DEG-PRL-1991}
\bibinfo{author}{\bibfnamefont{T.~J.} \bibnamefont{Gramila}},
  \bibinfo{author}{\bibfnamefont{J.~P.} \bibnamefont{Eisenstein}},
  \bibinfo{author}{\bibfnamefont{A.~H.} \bibnamefont{MacDonald}},
  \bibinfo{author}{\bibfnamefont{L.~N.} \bibnamefont{Pfeiffer}},
  \bibnamefont{and} \bibinfo{author}{\bibfnamefont{K.~W.} \bibnamefont{West}},
  \bibinfo{journal}{Phys. Rev. Lett.} \textbf{\bibinfo{volume}{66}},
  \bibinfo{pages}{1216} (\bibinfo{year}{1991}).

\bibitem[{\citenamefont{Jauho and
  Smith}(1993)}]{Antti-smith-original-drag-PRB-93}
\bibinfo{author}{\bibfnamefont{A.-P.} \bibnamefont{Jauho}} \bibnamefont{and}
  \bibinfo{author}{\bibfnamefont{H.}~\bibnamefont{Smith}},
  \bibinfo{journal}{Phys. Rev. B} \textbf{\bibinfo{volume}{47}},
  \bibinfo{pages}{4420} (\bibinfo{year}{1993}).

\bibitem[{\citenamefont{Flensberg and Hu}(1994)}]{karsten-ben-PRL-plasmon-1994}
\bibinfo{author}{\bibfnamefont{K.}~\bibnamefont{Flensberg}} \bibnamefont{and}
  \bibinfo{author}{\bibfnamefont{B.~Y.-K.} \bibnamefont{Hu}},
  \bibinfo{journal}{Phys. Rev. Lett.} \textbf{\bibinfo{volume}{73}},
  \bibinfo{pages}{3572} (\bibinfo{year}{1994}).

\bibitem[{\citenamefont{Hill et~al.}(1997)\citenamefont{Hill, Nicholls,
  Linfield, Pepper, Ritchie, Jones, Hu, and
  Flensberg}}]{Hill-Flensberg-palsmon-peak-i-drag-PRL-1997}
\bibinfo{author}{\bibfnamefont{N.~P.~R.} \bibnamefont{Hill}},
  \bibinfo{author}{\bibfnamefont{J.~T.} \bibnamefont{Nicholls}},
  \bibinfo{author}{\bibfnamefont{E.~H.} \bibnamefont{Linfield}},
  \bibinfo{author}{\bibfnamefont{M.}~\bibnamefont{Pepper}},
  \bibinfo{author}{\bibfnamefont{D.~A.} \bibnamefont{Ritchie}},
  \bibinfo{author}{\bibfnamefont{G.~A.~C.} \bibnamefont{Jones}},
  \bibinfo{author}{\bibfnamefont{B.~Y.-K.} \bibnamefont{Hu}}, \bibnamefont{and}
  \bibinfo{author}{\bibfnamefont{K.}~\bibnamefont{Flensberg}},
  \bibinfo{journal}{Phys. Rev. Lett.} \textbf{\bibinfo{volume}{78}},
  \bibinfo{pages}{2204} (\bibinfo{year}{1997}).

\bibitem[{\citenamefont{B{\o}nsager et~al.}(1998)\citenamefont{B{\o}nsager,
  Flensberg, Hu, and MacDonald}}]{Boensager-phonon-drag-PRB-1998}
\bibinfo{author}{\bibfnamefont{M.~C.} \bibnamefont{B{\o}nsager}},
  \bibinfo{author}{\bibfnamefont{K.}~\bibnamefont{Flensberg}},
  \bibinfo{author}{\bibfnamefont{B.~Y.-K.} \bibnamefont{Hu}}, \bibnamefont{and}
  \bibinfo{author}{\bibfnamefont{A.~H.} \bibnamefont{MacDonald}},
  \bibinfo{journal}{Phys. Rev. B} \textbf{\bibinfo{volume}{57}},
  \bibinfo{pages}{7085} (\bibinfo{year}{1998}).

\bibitem[{\citenamefont{Gornyi et~al.}(2004)\citenamefont{Gornyi, Mirlin, and
  von Oppen}}]{Gornyi-drag-i-lave-B-felter-condmat-2004}
\bibinfo{author}{\bibfnamefont{I.}~\bibnamefont{Gornyi}},
  \bibinfo{author}{\bibfnamefont{A.}~\bibnamefont{Mirlin}}, \bibnamefont{and}
  \bibinfo{author}{\bibfnamefont{F.}~\bibnamefont{von Oppen}},
  \bibinfo{journal}{cond-mat/0406176}  (\bibinfo{year}{2004}).

\bibitem[{\citenamefont{Komnik and
  Egger}(1998)}]{crossed-nanotubbes-egger-PRL-1998}
\bibinfo{author}{\bibfnamefont{A.}~\bibnamefont{Komnik}} \bibnamefont{and}
  \bibinfo{author}{\bibfnamefont{R.}~\bibnamefont{Egger}},
  \bibinfo{journal}{Phys. Rev. Lett.} \textbf{\bibinfo{volume}{80}},
  \bibinfo{pages}{2881} (\bibinfo{year}{1998}).

\bibitem[{\citenamefont{Flensberg}(1998)}]{crossed-luttinger-liquids-Flensberg%
-PRL-1998}
\bibinfo{author}{\bibfnamefont{K.}~\bibnamefont{Flensberg}},
  \bibinfo{journal}{Phys. Rev. Lett.} \textbf{\bibinfo{volume}{81}},
  \bibinfo{pages}{184} (\bibinfo{year}{1998}).

\bibitem[{\citenamefont{Nazarov and
  Averin}(1998)}]{Luttinger-drag-Nazarov-PRL-1998}
\bibinfo{author}{\bibfnamefont{Y.~V.} \bibnamefont{Nazarov}} \bibnamefont{and}
  \bibinfo{author}{\bibfnamefont{D.~V.} \bibnamefont{Averin}},
  \bibinfo{journal}{Phys. Rev. Lett.} \textbf{\bibinfo{volume}{81}},
  \bibinfo{pages}{653} (\bibinfo{year}{1998}).

\bibitem[{\citenamefont{Ponomarenko and
  Averin}(2000)}]{Luttinger-drag-Ponomarenko-PRL-2000}
\bibinfo{author}{\bibfnamefont{V.~V.} \bibnamefont{Ponomarenko}}
  \bibnamefont{and} \bibinfo{author}{\bibfnamefont{D.~V.}
  \bibnamefont{Averin}}, \bibinfo{journal}{Phys. Rev. Lett.}
  \textbf{\bibinfo{volume}{85}}, \bibinfo{pages}{4928} (\bibinfo{year}{2000}).

\bibitem[{\citenamefont{Klesse~R}(2000)}]{Luttinger-drag-Klesse--PRB-2000}
\bibinfo{author}{\bibfnamefont{S.~A.} \bibnamefont{Klesse~R}},
  \bibinfo{journal}{Phys. Rev. B} \textbf{\bibinfo{volume}{62}},
  \bibinfo{pages}{16912} (\bibinfo{year}{2000}).

\bibitem[{\citenamefont{Komnik and
  Egger}(2000)}]{crossed-nanotubbes-egger-EPJB-2000}
\bibinfo{author}{\bibfnamefont{A.}~\bibnamefont{Komnik}} \bibnamefont{and}
  \bibinfo{author}{\bibfnamefont{R.}~\bibnamefont{Egger}},
  \bibinfo{journal}{Eur. Phys. J. B} \textbf{\bibinfo{volume}{19}},
  \bibinfo{pages}{271} (\bibinfo{year}{2000}).

\bibitem[{\citenamefont{Trauzettel et~al.}(2002)\citenamefont{Trauzettel,
  Egger, and Grabert}}]{Luttinger-drag-Trauzettel-RPL-2002}
\bibinfo{author}{\bibfnamefont{B.}~\bibnamefont{Trauzettel}},
  \bibinfo{author}{\bibfnamefont{R.}~\bibnamefont{Egger}}, \bibnamefont{and}
  \bibinfo{author}{\bibfnamefont{H.}~\bibnamefont{Grabert}},
  \bibinfo{journal}{Phys. Rev. Lett.} \textbf{\bibinfo{volume}{88}},
  \bibinfo{pages}{116401} (\bibinfo{year}{2002}).

\bibitem[{\citenamefont{Gao et~al.}(2004)\citenamefont{Gao, Komnik, Egger,
  Glattli, and
  Bachtold}}]{crossed-SWCNT-Luttinger-liquid-experiment-Egger-Bachtold-PRL-200%
4}
\bibinfo{author}{\bibfnamefont{B.}~\bibnamefont{Gao}},
  \bibinfo{author}{\bibfnamefont{A.}~\bibnamefont{Komnik}},
  \bibinfo{author}{\bibfnamefont{R.}~\bibnamefont{Egger}},
  \bibinfo{author}{\bibfnamefont{D.~C.} \bibnamefont{Glattli}},
  \bibnamefont{and} \bibinfo{author}{\bibfnamefont{A.}~\bibnamefont{Bachtold}},
  \bibinfo{journal}{Phys. Rev. Lett.} \textbf{\bibinfo{volume}{92}},
  \bibinfo{pages}{216804} (\bibinfo{year}{2004}).

\bibitem[{\citenamefont{Ando and
  Nakanishi}(1998)}]{backscattering-Ando-JPJ-1998-1}
\bibinfo{author}{\bibfnamefont{T.}~\bibnamefont{Ando}} \bibnamefont{and}
  \bibinfo{author}{\bibfnamefont{T.}~\bibnamefont{Nakanishi}},
  \bibinfo{journal}{J. Phys. soc. Jpn.} \textbf{\bibinfo{volume}{67}},
  \bibinfo{pages}{1704} (\bibinfo{year}{1998}).

\bibitem[{\citenamefont{Ando et~al.}(1998)\citenamefont{Ando, Nakanishi, and
  Saito}}]{backscattering-Ando-JPJ-1998-2}
\bibinfo{author}{\bibfnamefont{T.}~\bibnamefont{Ando}},
  \bibinfo{author}{\bibfnamefont{T.}~\bibnamefont{Nakanishi}},
  \bibnamefont{and} \bibinfo{author}{\bibfnamefont{R.}~\bibnamefont{Saito}},
  \bibinfo{journal}{J. Phys. soc. Jpn.} \textbf{\bibinfo{volume}{67}},
  \bibinfo{pages}{2857} (\bibinfo{year}{1998}).

\bibitem[{\citenamefont{Klesse}(2002)}]{Klesse-PRB-2002-Coulomb-matrix-element%
-MWNT}
\bibinfo{author}{\bibfnamefont{R.}~\bibnamefont{Klesse}},
  \bibinfo{journal}{Phys. Rev. B} \textbf{\bibinfo{volume}{66}},
  \bibinfo{pages}{085409} (\bibinfo{year}{2002}).

\bibitem[{\citenamefont{White et~al.}(1993)\citenamefont{White, Robertson, and
  Mintmire}}]{White-1993-RPB}
\bibinfo{author}{\bibfnamefont{C.~T.} \bibnamefont{White}},
  \bibinfo{author}{\bibfnamefont{D.~H.} \bibnamefont{Robertson}},
  \bibnamefont{and} \bibinfo{author}{\bibfnamefont{J.~W.}
  \bibnamefont{Mintmire}}, \bibinfo{journal}{Phys. Rev. B}
  \textbf{\bibinfo{volume}{47}}, \bibinfo{pages}{5485} (\bibinfo{year}{1993}).

\bibitem[{\citenamefont{Mintmire and White}(1995)}]{Mintmire-1995-carbon}
\bibinfo{author}{\bibfnamefont{J.~W.} \bibnamefont{Mintmire}} \bibnamefont{and}
  \bibinfo{author}{\bibfnamefont{C.~T.} \bibnamefont{White}},
  \bibinfo{journal}{Carbon} \textbf{\bibinfo{volume}{33}}, \bibinfo{pages}{893}
  (\bibinfo{year}{1995}).

\bibitem[{\citenamefont{Vukovi$\acute{\textrm{c}}$
  et~al.}(2002)\citenamefont{Vukovi$\acute{\textrm{c}}$, Milo$\breve{\rm
  s}$evi$\acute{\textrm{c}}$, and
  Damnjanovi$\acute{\textrm{c}}$}}]{Vukovic-2002-PRB-bloch-states}
\bibinfo{author}{\bibfnamefont{T.}~\bibnamefont{Vukovi$\acute{\textrm{c}}$}},
  \bibinfo{author}{\bibfnamefont{I.}~\bibnamefont{Milo$\breve{\rm
  s}$evi$\acute{\textrm{c}}$}}, \bibnamefont{and}
  \bibinfo{author}{\bibfnamefont{M.}~\bibnamefont{Damnjanovi$\acute{\textrm{c}%
}$}}, \bibinfo{journal}{Phys. Rev. B}
\textbf{\bibinfo{volume}{65}},
  \bibinfo{pages}{045418} (\bibinfo{year}{2002}).

\bibitem[{\citenamefont{Damnjanovi$\acute{\textrm{c}}$
  et~al.}(1999)\citenamefont{Damnjanovi$\acute{\textrm{c}}$, Milo$\breve{\rm
  s}$evi$\acute{\textrm{c}}$, Vukovi$\acute{\textrm{c}}$, and
  Sredanovi$\acute{\textrm{c}}$}}]{Damnjanovic-1999-symmetri-og-potentialer-NT}
\bibinfo{author}{\bibfnamefont{M.}~\bibnamefont{Damnjanovi$\acute{\textrm{c}}$%
}},
\bibinfo{author}{\bibfnamefont{I.}~\bibnamefont{Milo$\breve{\rm
  s}$evi$\acute{\textrm{c}}$}},
  \bibinfo{author}{\bibfnamefont{T.}~\bibnamefont{Vukovi$\acute{\textrm{c}}$}},
  \bibnamefont{and}
  \bibinfo{author}{\bibfnamefont{R.}~\bibnamefont{Sredanovi$\acute{\textrm{c}}%
$}}, \bibinfo{journal}{Phys. Rev. B}
\textbf{\bibinfo{volume}{60}},
  \bibinfo{pages}{2728} (\bibinfo{year}{1999}).

\bibitem[{foo({\natexlab{b}})}]{footnote-exchange-interaction}
\bibinfo{note}{We do not have exchange interaction between the tubes, since the
  inter tube tunnelling were neglected. Therefore we can use product states
  instead of anti-symmetrized wave functions.}

\bibitem[{\citenamefont{Lee et~al.}(2000)\citenamefont{Lee, Kim, Fischer,
  Lefebvre, Radosavljevi$\acute{\textrm{c}}$, Hone, and
  Johnson}}]{doping-lee-PRB-2000}
\bibinfo{author}{\bibfnamefont{R.~S.} \bibnamefont{Lee}},
  \bibinfo{author}{\bibfnamefont{H.~J.} \bibnamefont{Kim}},
  \bibinfo{author}{\bibfnamefont{J.~E.} \bibnamefont{Fischer}},
  \bibinfo{author}{\bibfnamefont{J.}~\bibnamefont{Lefebvre}},
  \bibinfo{author}{\bibfnamefont{M.}~\bibnamefont{Radosavljevi$\acute{\textrm{%
c}}$}}, \bibinfo{author}{\bibfnamefont{J.}~\bibnamefont{Hone}},
  \bibnamefont{and} \bibinfo{author}{\bibfnamefont{A.~T.}
  \bibnamefont{Johnson}}, \bibinfo{journal}{Phys. Rev. B}
  \textbf{\bibinfo{volume}{61}}, \bibinfo{pages}{4526} (\bibinfo{year}{2000}).

\bibitem[{\citenamefont{Zhou et~al.}(2000{\natexlab{a}})\citenamefont{Zhou,
  Kong, Yenilmez, and Dai}}]{doping-zhou-sceince-2000}
\bibinfo{author}{\bibfnamefont{C.}~\bibnamefont{Zhou}},
  \bibinfo{author}{\bibfnamefont{J.}~\bibnamefont{Kong}},
  \bibinfo{author}{\bibfnamefont{E.}~\bibnamefont{Yenilmez}}, \bibnamefont{and}
  \bibinfo{author}{\bibfnamefont{H.}~\bibnamefont{Dai}},
  \bibinfo{journal}{Science} \textbf{\bibinfo{volume}{290}},
  \bibinfo{pages}{1552} (\bibinfo{year}{2000}{\natexlab{a}}).

\bibitem[{foo({\natexlab{c}})}]{footnote-2D-til-3D}
\bibinfo{note}{Note that in doing the wrapping of the graphene band structure
  we are going from a two to three dimensional description and thereby the
  radial part of the wave function is $|\psi|\propto \frac{\delta(r-r^c)}{r}$,
  where $r^c$ is the radius of the tube.}

\bibitem[{foo({\natexlab{d}})}]{footnote-explicit-periodisk}
\bibinfo{note}{The matrix elements
  (\ref{eq:Coulomb-int-matrix-element-trans-unitcell-tight-bin-screened}) and
  (\ref{eq:Coulomb-int-matrix-element-trans-unitcell-tight-bin-unscreened})
  could also have been obtained using the helical symmetry states
  $|\kappa\m\rangle$, which has the advantage of making the $g$-factor an
  explicitly periodic function of reciprocal lattice vectors whereas using the
  translational unit cell this periodicity has to be imposed by
  hand.\cite{Lunde-drag-in-MWCNT-masterthesis-2004} The states are given
  explicitly in Ref.~[\onlinecite{White-1993-RPB,Mintmire-1995-carbon}].}

\bibitem[{\citenamefont{Lunde}(2004)}]{Lunde-drag-in-MWCNT-masterthesis-2004}
\bibinfo{author}{\bibfnamefont{A.~M.} \bibnamefont{Lunde}}, Master's thesis,
  \bibinfo{school}{Niels Bohr Institute and MIC--Department of Micro and
  Nanotechnology, see www.fys.ku.dk/flensberg/students.htm}
  (\bibinfo{year}{2004}).

\bibitem[{\citenamefont{Flensberg and Hu}(1995)}]{karsten-ben-formel-PRB-1995}
\bibinfo{author}{\bibfnamefont{K.}~\bibnamefont{Flensberg}} \bibnamefont{and}
  \bibinfo{author}{\bibfnamefont{B.~Y.-K.} \bibnamefont{Hu}},
  \bibinfo{journal}{Phys. Rev. B} \textbf{\bibinfo{volume}{52}},
  \bibinfo{pages}{14796} (\bibinfo{year}{1995}).

\bibitem[{\citenamefont{Alkauskas et~al.}(2002)\citenamefont{Alkauskas,
  Flensberg, Hu, and Jauho}}]{Audrius-drag-med-periodisk-modulation-PRB-2002}
\bibinfo{author}{\bibfnamefont{A.}~\bibnamefont{Alkauskas}},
  \bibinfo{author}{\bibfnamefont{K.}~\bibnamefont{Flensberg}},
  \bibinfo{author}{\bibfnamefont{B.~Y.-K.} \bibnamefont{Hu}}, \bibnamefont{and}
  \bibinfo{author}{\bibfnamefont{A.-P.} \bibnamefont{Jauho}},
  \bibinfo{journal}{Phys. Rev. B} \textbf{\bibinfo{volume}{66}},
  \bibinfo{pages}{201304(R)} (\bibinfo{year}{2002}).

\bibitem[{\citenamefont{Hu}(1998)}]{ben-Hu-drag-formel-med-flere-baand-PRB-98}
\bibinfo{author}{\bibfnamefont{B.~Y.-K.} \bibnamefont{Hu}},
  \bibinfo{journal}{Phys. Rev. B} \textbf{\bibinfo{volume}{57}},
  \bibinfo{pages}{12345} (\bibinfo{year}{1998}).

\bibitem[{\citenamefont{Flensberg et~al.}(1995)\citenamefont{Flensberg, Hu,
  Jauho, and Kinaret}}]{karsten-kubo-formula-PRB-1995}
\bibinfo{author}{\bibfnamefont{K.}~\bibnamefont{Flensberg}},
  \bibinfo{author}{\bibfnamefont{B.~Y.-K.} \bibnamefont{Hu}},
  \bibinfo{author}{\bibfnamefont{A.-P.} \bibnamefont{Jauho}}, \bibnamefont{and}
  \bibinfo{author}{\bibfnamefont{J.~M.} \bibnamefont{Kinaret}},
  \bibinfo{journal}{Phys. Rev. B} \textbf{\bibinfo{volume}{52}},
  \bibinfo{pages}{14761} (\bibinfo{year}{1995}).

\bibitem[{\citenamefont{Smith and Jensen}(1989)}]{smith-boltzmann-bog}
\bibinfo{author}{\bibfnamefont{H.}~\bibnamefont{Smith}} \bibnamefont{and}
  \bibinfo{author}{\bibfnamefont{H.~H.} \bibnamefont{Jensen}},
  \emph{\bibinfo{title}{Transport Phenomena}} (\bibinfo{publisher}{Clarendon
  Press, Oxford}, \bibinfo{year}{1989}).

\bibitem[{\citenamefont{Pustilnik et~al.}(2003)\citenamefont{Pustilnik,
  Mishchenko, Glazman, and
  Andreev}}]{Pustilnik-small-q-important-in-drag-1D-PRL-2003}
\bibinfo{author}{\bibfnamefont{M.}~\bibnamefont{Pustilnik}},
  \bibinfo{author}{\bibfnamefont{E.~G.} \bibnamefont{Mishchenko}},
  \bibinfo{author}{\bibfnamefont{L.~I.} \bibnamefont{Glazman}},
  \bibnamefont{and} \bibinfo{author}{\bibfnamefont{A.~V.}
  \bibnamefont{Andreev}}, \bibinfo{journal}{Phys. Rev. Lett.}
  \textbf{\bibinfo{volume}{91}}, \bibinfo{pages}{126805}
  (\bibinfo{year}{2003}).

\bibitem[{\citenamefont{Mortensen et~al.}(2001)\citenamefont{Mortensen,
  Flensberg, and Jauho}}]{Drag-in-mesoscopic-systems-Asger-PRL-2001}
\bibinfo{author}{\bibfnamefont{N.~A.} \bibnamefont{Mortensen}},
  \bibinfo{author}{\bibfnamefont{K.}~\bibnamefont{Flensberg}},
  \bibnamefont{and} \bibinfo{author}{\bibfnamefont{A.-P.} \bibnamefont{Jauho}},
  \bibinfo{journal}{Phys. Rev. Lett.} \textbf{\bibinfo{volume}{86}},
  \bibinfo{pages}{1841} (\bibinfo{year}{2001}).

\bibitem[{\citenamefont{Gurevich et~al.}(1998)\citenamefont{Gurevich, Pevzner,
  and Fenton}}]{Gurevich-Drag-1D-Ballistiske-Blotzmann-JPCM-1998}
\bibinfo{author}{\bibfnamefont{V.~L.} \bibnamefont{Gurevich}},
  \bibinfo{author}{\bibfnamefont{V.~B.} \bibnamefont{Pevzner}},
  \bibnamefont{and} \bibinfo{author}{\bibfnamefont{E.~W.}
  \bibnamefont{Fenton}}, \bibinfo{journal}{J. Phys.: Condens. Matter}
  \textbf{\bibinfo{volume}{10}}, \bibinfo{pages}{2551} (\bibinfo{year}{1998}).

\bibitem[{foo({\natexlab{e}})}]{footnote-electron-hole-symmetry-definition}
\bibinfo{note}{The definition of electron-hole symmetry can be given as
  follows: For each occupied $k$ state (at $T=0$) with energy $\e_k$ and
  velocity $v_k$ there exists one and only one empty state $k^{\prime}$ with
  the same velocity $v_{k^{\prime}}=v_k$ and the opposite energy with respect
  to the Fermi level: $\eF-\e_{k^{\prime}}=-(\eF-\e_k)$. In this way there is a
  one to one correspondence between the electrons (i.e. the filled states) and
  holes (i.e. the empty states).}

\bibitem[{\citenamefont{Kr$\ddot{\textrm{u}}$ger
  et~al.}(2001)\citenamefont{Kr$\ddot{\textrm{u}}$ger, Buitelaar, Nussbaumer,
  Sch$\ddot{\textrm{o}}$nenberger, and
  Forr$\acute{\textrm{o}}$}}]{gatevoltage-kruger-APL-2001}
\bibinfo{author}{\bibfnamefont{M.}~\bibnamefont{Kr$\ddot{\textrm{u}}$ger}},
  \bibinfo{author}{\bibfnamefont{M.~R.} \bibnamefont{Buitelaar}},
  \bibinfo{author}{\bibfnamefont{T.}~\bibnamefont{Nussbaumer}},
  \bibinfo{author}{\bibfnamefont{C.}~\bibnamefont{Sch$\ddot{\textrm{o}}$nenber%
ger}}, \bibnamefont{and}
  \bibinfo{author}{\bibfnamefont{L.}~\bibnamefont{Forr$\acute{\textrm{o}}$}},
  \bibinfo{journal}{Appl. Phys. Lett.} \textbf{\bibinfo{volume}{78}},
  \bibinfo{pages}{1291} (\bibinfo{year}{2001}).

\bibitem[{\citenamefont{Zhou et~al.}(2000{\natexlab{b}})\citenamefont{Zhou,
  Kong, and Dai}}]{gate-zhou-PRL-2000}
\bibinfo{author}{\bibfnamefont{C.}~\bibnamefont{Zhou}},
  \bibinfo{author}{\bibfnamefont{J.}~\bibnamefont{Kong}}, \bibnamefont{and}
  \bibinfo{author}{\bibfnamefont{H.}~\bibnamefont{Dai}},
  \bibinfo{journal}{Phys. Rev. Lett.} \textbf{\bibinfo{volume}{84}},
  \bibinfo{pages}{5604} (\bibinfo{year}{2000}{\natexlab{b}}).

\bibitem[{\citenamefont{Javey et~al.}(2002)\citenamefont{Javey, Kim, Brink,
  Wang, Ural, Guo, Mcintyre, Mceuen, Lundstrom, and
  Dai}}]{doping-extreme-Javey-nature-2002}
\bibinfo{author}{\bibfnamefont{A.}~\bibnamefont{Javey}},
  \bibinfo{author}{\bibfnamefont{H.}~\bibnamefont{Kim}},
  \bibinfo{author}{\bibfnamefont{M.}~\bibnamefont{Brink}},
  \bibinfo{author}{\bibfnamefont{Q.}~\bibnamefont{Wang}},
  \bibinfo{author}{\bibfnamefont{A.}~\bibnamefont{Ural}},
  \bibinfo{author}{\bibfnamefont{J.}~\bibnamefont{Guo}},
  \bibinfo{author}{\bibfnamefont{P.}~\bibnamefont{Mcintyre}},
  \bibinfo{author}{\bibfnamefont{P.}~\bibnamefont{Mceuen}},
  \bibinfo{author}{\bibfnamefont{M.}~\bibnamefont{Lundstrom}},
  \bibnamefont{and} \bibinfo{author}{\bibfnamefont{H.}~\bibnamefont{Dai}},
  \bibinfo{journal}{Nature Materials} \textbf{\bibinfo{volume}{1}},
  \bibinfo{pages}{241} (\bibinfo{year}{2002}).

\bibitem[{\citenamefont{Lunde and
  Jauho}(2004)}]{Lunde-Modena-drag-in-MWCNT-2004}
\bibinfo{author}{\bibfnamefont{A.~M.} \bibnamefont{Lunde}} \bibnamefont{and}
  \bibinfo{author}{\bibfnamefont{A.-P.} \bibnamefont{Jauho}},
  \bibinfo{journal}{Semicond. Sci. Technol.} \textbf{\bibinfo{volume}{19}},
  \bibinfo{pages}{S433} (\bibinfo{year}{2004}).

\bibitem[{\citenamefont{Hu and
  Flensberg}(1996)}]{Flensberg-drag-1D-duffusive-hcis-1996}
\bibinfo{author}{\bibfnamefont{B.~Y.-K.} \bibnamefont{Hu}} \bibnamefont{and}
  \bibinfo{author}{\bibfnamefont{K.}~\bibnamefont{Flensberg}},
  \bibinfo{journal}{Hot Carriers in Semicondoctors (HCIS-9), eds. K. Hess et
  al., 421, Plenum Press}  (\bibinfo{year}{1996}).

\bibitem[{foo({\natexlab{f}})}]{footnote-kappa-value}
\bibinfo{note}{In\cite{Egger-Luttinger-liquid-SWCNT-lang-udgave-Euro-phys-Jour%
-B-1998} $\epsilon_r=1.4$ is used.}

\bibitem[{foo({\natexlab{g}})}]{footnote-commensurable}
\bibinfo{note}{Note that the crystal angular momentum conservation depend on
  wether the tubes are commensurable or
  not.\cite{Lunde-drag-in-MWCNT-masterthesis-2004}}.

\bibitem[{\citenamefont{Qin}(1995)}]{Qin-drag-in-rund-quantumwells-JPCM-1995}
\bibinfo{author}{\bibfnamefont{G.}~\bibnamefont{Qin}}, \bibinfo{journal}{J.
  Phys.: Condens. Matter} \textbf{\bibinfo{volume}{7}}, \bibinfo{pages}{9785}
  (\bibinfo{year}{1995}).

\bibitem[{\citenamefont{Hamada et~al.}(1992)\citenamefont{Hamada, Sawada, and
  Oshiyama}}]{Hamnda-PRL-1992-CNT-bands-periodic-boundary-conditions}
\bibinfo{author}{\bibfnamefont{N.}~\bibnamefont{Hamada}},
  \bibinfo{author}{\bibfnamefont{S.-I.} \bibnamefont{Sawada}},
  \bibnamefont{and} \bibinfo{author}{\bibfnamefont{A.}~\bibnamefont{Oshiyama}},
  \bibinfo{journal}{Phys. Rev. Lett.} \textbf{\bibinfo{volume}{68}},
  \bibinfo{pages}{1579} (\bibinfo{year}{1992}).

\bibitem[{\citenamefont{Saito et~al.}(1992)\citenamefont{Saito, Fujita,
  Dresselhaus, and
  Dresselhaus}}]{Dresselhaus-Saito-PRB-1992-CNT-bandstructure-by-periodic-boun%
dary-conditions}
\bibinfo{author}{\bibfnamefont{R.}~\bibnamefont{Saito}},
  \bibinfo{author}{\bibfnamefont{M.}~\bibnamefont{Fujita}},
  \bibinfo{author}{\bibfnamefont{G.}~\bibnamefont{Dresselhaus}},
  \bibnamefont{and} \bibinfo{author}{\bibfnamefont{M.~S.}
  \bibnamefont{Dresselhaus}}, \bibinfo{journal}{Phys. Rev. B}
  \textbf{\bibinfo{volume}{46}}, \bibinfo{pages}{1804} (\bibinfo{year}{1992}).

\bibitem[{\citenamefont{Saito et~al.}(1998)\citenamefont{Saito, Dresselhaus,
  and Dresselhaus}}]{Dresselhaus-book-1998}
\bibinfo{author}{\bibfnamefont{R.}~\bibnamefont{Saito}},
  \bibinfo{author}{\bibfnamefont{G.}~\bibnamefont{Dresselhaus}},
  \bibnamefont{and} \bibinfo{author}{\bibfnamefont{M.~S.}
  \bibnamefont{Dresselhaus}}, \emph{\bibinfo{title}{Physical Properties of
  Carbon Nanotubes}} (\bibinfo{publisher}{Imperial College press},
  \bibinfo{year}{1998}), \bibinfo{edition}{3rd} ed.

\bibitem[{\citenamefont{Wallace}(1947)}]{Wallace-grafit-lag-baand-struktur-PR-%
1947}
\bibinfo{author}{\bibfnamefont{P.~R.} \bibnamefont{Wallace}},
  \bibinfo{journal}{Phys. Rev.} \textbf{\bibinfo{volume}{71}},
  \bibinfo{pages}{622} (\bibinfo{year}{1947}).

\bibitem[{\citenamefont{Ashcroft and Mermin}(1976)}]{Ashcroft-Mermin-book-1976}
\bibinfo{author}{\bibfnamefont{N.}~\bibnamefont{Ashcroft}} \bibnamefont{and}
  \bibinfo{author}{\bibfnamefont{N.}~\bibnamefont{Mermin}},
  \emph{\bibinfo{title}{Solid State Physics}} (\bibinfo{publisher}{Harcourt
  College Publishers}, \bibinfo{year}{1976}), \bibinfo{edition}{college} ed.

\bibitem[{\citenamefont{Brown et~al.}(2000)\citenamefont{Brown, Corio, Marucci,
  Dresselhaus, and
  Dresselhaus}}]{Brown-overlapsintegraler-eksperiment-PRB-2000}
\bibinfo{author}{\bibfnamefont{S.~D.~M.} \bibnamefont{Brown}},
  \bibinfo{author}{\bibfnamefont{P.}~\bibnamefont{Corio}},
  \bibinfo{author}{\bibfnamefont{A.}~\bibnamefont{Marucci}},
  \bibinfo{author}{\bibfnamefont{M.~A.}~\bibnamefont{Pimenta}},
  \bibinfo{author}{\bibfnamefont{M.~S.} \bibnamefont{Dresselhaus}},
  \bibnamefont{and}
  \bibinfo{author}{\bibfnamefont{G.}~\bibnamefont{Dresselhaus}},
  \bibinfo{journal}{Phys. Rev. B} \textbf{\bibinfo{volume}{61}},
  \bibinfo{pages}{7734} (\bibinfo{year}{2000}).

\bibitem[{\citenamefont{Reich et~al.}(2002)\citenamefont{Reich, Maultzsch,
  Thomsen, and
  Ordej$\acute{\textrm{o}}$n}}]{Reich-tight-binding-overlapsintegral-PRB-2002}
\bibinfo{author}{\bibfnamefont{S.}~\bibnamefont{Reich}},
  \bibinfo{author}{\bibfnamefont{J.}~\bibnamefont{Maultzsch}},
  \bibinfo{author}{\bibfnamefont{C.}~\bibnamefont{Thomsen}}, \bibnamefont{and}
  \bibinfo{author}{\bibfnamefont{P.}~\bibnamefont{Ordej$\acute{\textrm{o}}$n}},
  \bibinfo{journal}{Phys. Rev. B} \textbf{\bibinfo{volume}{66}},
  \bibinfo{pages}{035412} (\bibinfo{year}{2002}).

\bibitem[{\citenamefont{Wilder et~al.}(1998)\citenamefont{Wilder, Venema,
  Rinzler, Smalley, and Dekker}}]{Wilder-overlap-eksperiment-nature-1998}
\bibinfo{author}{\bibfnamefont{J.~W.~G.} \bibnamefont{Wilder}},
  \bibinfo{author}{\bibfnamefont{L.~C.} \bibnamefont{Venema}},
  \bibinfo{author}{\bibfnamefont{A.~G.} \bibnamefont{Rinzler}},
  \bibinfo{author}{\bibfnamefont{R.~E.} \bibnamefont{Smalley}},
  \bibnamefont{and} \bibinfo{author}{\bibfnamefont{C.}~\bibnamefont{Dekker}},
  \bibinfo{journal}{Nature} \textbf{\bibinfo{volume}{391}}, \bibinfo{pages}{59}
  (\bibinfo{year}{1998}).

\bibitem[{\citenamefont{Odom et~al.}(1998)\citenamefont{Odom, Huang, Kim, and
  Lieber}}]{Odom-overlap-eksperiment-nature-1998}
\bibinfo{author}{\bibfnamefont{T.~W.} \bibnamefont{Odom}},
  \bibinfo{author}{\bibfnamefont{J.-L.} \bibnamefont{Huang}},
  \bibinfo{author}{\bibfnamefont{P.}~\bibnamefont{Kim}}, \bibnamefont{and}
  \bibinfo{author}{\bibfnamefont{C.~M.} \bibnamefont{Lieber}},
  \bibinfo{journal}{Nature} \textbf{\bibinfo{volume}{391}}, \bibinfo{pages}{62}
  (\bibinfo{year}{1998}).

\bibitem[{foo({\natexlab{h}})}]{footnote-Mads-Brandbyge-CNT-bands-program}
\bibinfo{note}{To get a feeling for the connection between the description of
  the band structure using the translational and primitive unit cell one can
  use the \emph{Mathematica} program: Atomic and Electronic Structure of Carbon
  nanotubes, by M. Brandbyge, found at
  http://library.wolfram.com/infocenter/MathSource/384/}.

\bibitem[{foo({\natexlab{i}})}]{footnote-zero-in-graphene-FBZ}
\bibinfo{note}{Note that graphene has six zeros all at the FBZ boundary and
  these are given in three pairs connected by reciprocal lattice vectors of
  graphene.}

\bibitem[{foo({\natexlab{j}})}]{footnote-algebra-mangler}
\bibinfo{note}{We only have numerical proof for this property, found by trying
  all combinations of $n,m\leq 100$.}

\bibitem[{\citenamefont{Ajiki and
  Ando}(1993)}]{Ando-electronic-states-SWCNT-JPSJ-1993}
\bibinfo{author}{\bibfnamefont{H.}~\bibnamefont{Ajiki}} \bibnamefont{and}
  \bibinfo{author}{\bibfnamefont{T.}~\bibnamefont{Ando}}, \bibinfo{journal}{J.
  Phys. soc. Jpn.} \textbf{\bibinfo{volume}{62}}, \bibinfo{pages}{1255}
  (\bibinfo{year}{1993}).

\bibitem[{\citenamefont{Que}(2002)}]{Que-2002-RPA-potential-SWNT}
\bibinfo{author}{\bibfnamefont{W.}~\bibnamefont{Que}}, \bibinfo{journal}{Phys.
  Rev. B} \textbf{\bibinfo{volume}{66}}, \bibinfo{pages}{193405}
  (\bibinfo{year}{2002}).

\end{thebibliography}

\end{document}